\documentclass[12pt]{article}
\pdfoutput=1
\usepackage{amssymb,amsmath,dsfont}
\usepackage[normalem]{ulem}
\usepackage{slashed}
\usepackage{epsfig}
\usepackage{epstopdf}
\usepackage{latexsym}
\usepackage{graphicx}
\usepackage{booktabs}
\usepackage{bbm}
\usepackage{xspace}
\usepackage{cancel}
\usepackage[utf8]{inputenc}
\usepackage[babel]{csquotes}
\usepackage{enumitem}
\usepackage[numbers,compress]{natbib}
\usepackage[T1]{fontenc}
\usepackage[margin=20pt,small]{caption}
\usepackage{subfig}
\usepackage[toc]{appendix}
\usepackage{color}
\usepackage{datetime}
\usepackage[
colorlinks=false,
linkcolor=darkblue,  
urlcolor=blue,    
filecolor=blue,     
citecolor=red,
linktocpage=true,
pdfstartview=FitV,
bookmarksopen=true,
hidelinks
]{hyperref}
\numberwithin{equation}{section}

\ifpdf
\fi
\newcommand*{\boxedcolor}{red}
\makeatletter
\renewcommand{\boxed}[1]{\textcolor{\boxedcolor}{%
		\fbox{\normalcolor\m@th$\displaystyle#1$}}}
\makeatother

\newcommand{\bxx}[1]{\begin{#1}}
\newcommand{\be}{\bxx{equation}}
\newcommand{\ee}{\end{equation}}
\newcommand{\vev}[1]{\ensuremath{\langle #1 \rangle}\xspace}

\definecolor{cardinal}{rgb}{0.6,0,0}
\definecolor{darkgreen}{rgb}{0,0.5,0}
\definecolor{golden}{rgb}{0.92, 0.7, 0}
\definecolor{midnight}{rgb}{0, 0, 0.5}
\definecolor{darkblue}{rgb}{0.2, 0, 0.8}

\newcommand{\id}{{\mathds 1}}

\def\sl{\frak{sl}}

\begin{document}  
	
	\begin{titlepage}
		
		\medskip
		\begin{center} 
			{\Large \bf Bootstrapping boundary-localized interactions}

			\bigskip
			\bigskip
			\bigskip
			
			{\bf Connor Behan$^1$, Lorenzo Di Pietro$^{2,3}$, Edoardo Lauria$^{4}$ and Balt C. van Rees$^4$\\ }
			\bigskip
			\bigskip
            	${}^{1}$
            	Mathematical Institute, University of Oxford, Andrew Wiles Building, Radcliffe Observatory Quarter, Woodstock Road, Oxford, OX2 6GG, UK\\
           		${}^{2}$
            	Dipartimento di Fisica, Universit\`{a} di Trieste, Strada Costiera 11, I-34151 Trieste, Italy\\
            	${}^{3}$
            	INFN, Sezione di Trieste, Via Valerio 2, I-34127 Trieste, Italy\\
				${}^{4}$
				CPHT, CNRS, Institut Polytechnique de Paris, France
			\vskip 5mm				
			\texttt{behan@maths.ox.ac.uk,~ldipietro@units.it,\\edoardo.lauria@polytechnique.edu,~balt.van-rees@polytechnique.edu} \\
		\end{center}
		
		\bigskip
		\bigskip
		
		\begin{abstract}
			\noindent We study conformal boundary conditions for the theory of a single real scalar to investigate whether the known Dirichlet and Neumann conditions are the only possibilities. For this free bulk theory there are strong restrictions on the possible boundary dynamics. In particular, we find that the bulk-to-boundary operator expansion of the bulk field involves at most a `shadow pair' of boundary fields, irrespective of the conformal boundary condition. We numerically analyze the four-point crossing equations for this shadow pair in the case of a three-dimensional boundary (so a four-dimensional scalar field) and find that large ranges of parameter space are excluded. However a `kink' in the numerical bounds obeys all our consistency checks and might be an indication of a new conformal boundary condition.
		\end{abstract}

		\noindent

	\end{titlepage}
	
	
	\setcounter{tocdepth}{2}	

\tableofcontents
\newpage
\section{Introduction and summary}
\label{sec:intro}

The classification of conformal boundary conditions for a bulk CFT is a difficult problem. Besides well-known results for rational boundary conditions in rational CFTs (reviewed in \cite{Cardy:2004hm}), very little is known even for relatively simple theories. It is natural to ask whether a systematic approach is feasible -- one which does not rely on explicit constructions but leverages instead the modern conformal bootstrap methods \cite{Rattazzi:2008pe} (see \cite{Poland:2018epd} for a review and \cite{Liendo:2012hy} for a first application to BCFT which relied on results from \cite{McAvity:1993ue,McAvity:1995zd}). A promising methodology is to start from theories that are as simple as possible in the bulk. In this work we pursue precisely such a direction in the case where the bulk theory is a single real free scalar field.

In any spacetime dimension a free scalar can certainly have Dirichlet or Neumann conformal boundary conditions. The question we try to answer here is whether more general conformal boundary conditions are possible, for example by coupling the bulk scalar to new boundary degrees of freedom and flowing to the infrared. These putative boundary conditions should modify the behavior of the scalar near the boundary and produce non-trivial boundary correlators, analogous to those of an interacting one lower-dimensional CFT. We find numerical evidence for at least one such `exotic' boundary condition in four dimensions, and more generally very strong constraints on the space of potential conformal boundary conditions.\footnote{The existence of such strong constraints is remarkable. In some cases, for example Maxwell theory in four dimensions, it is known that the space of conformal boundary conditions is vast, since it includes the space of all CFTs with a $U(1)$ symmetry in three dimensions \cite{Witten:2003ya,Gaiotto:2008ak,Seiberg:2016gmd,DiPietro:2019hqe}.}

In exploring consistent boundary conditions for a free scalar theory we obtained a very special set of `shadow-related' crossing symmetry equations, as follows. First of all, the $\square \phi = 0$ equation of motion implies that the bulk-boundary expansion of the bulk field $\phi$ is restricted to contain at most two operators that we denote as $\widehat O_1$ and $\widehat O_2$; their dimensions are $\Delta_\phi$ and $\Delta_\phi + 1$, respectively. At most one of these two operators can vanish, and if so then we are in the Dirichlet or Neumann case and the two operators are immediately recognizable as the restriction of $\phi$ or $\partial_{\perp} \phi$ to the boundary. If they are both non-vanishing then the operators can be thought of as a `shadow pair' in the sense of Ferrara, Gatto, Grillo and Parisi \cite{Ferrara:1972xe,Ferrara:1972ay,Ferrara:1972uq,Ferrara:1973vz}. Their dimensions match this observation since $2 \Delta_\phi + 1 = d -1$, the dimension of the boundary, but the picture extends to their three-point functions: for a generic third defect operator $\widehat{\mathcal{O}}$ with dimension $\widehat{\Delta}$ and spin $l$ we find the relations
\begin{align}
\begin{split}\label{ssspinconstr0}
\frac{1}{\Gamma \left(\frac{l+\widehat{\Delta}}{2}\right) \Gamma \left(\frac{d+l-\widehat{\Delta}-2}{2} \right)}\hat{f}_{11{\widehat{\mathcal{O}}^{(l)}}}  & =  -\frac{b_2/b_1}{2\Gamma\left(\frac{d+l-\widehat{\Delta}-1}{2} \right)\Gamma \left(\frac{l+\widehat{\Delta}+1}{2}\right)} \hat{f}_{12{\widehat{\mathcal{O}}^{(l)}}}~,\\
\frac{1}{\Gamma \left(\frac{l+\widehat{\Delta}}{2}\right) \Gamma \left(\frac{d+l-\widehat{\Delta}}{2} \right)} \hat{f}_{22{\widehat{\mathcal{O}}^{(l)}}} & = -\frac{2 b_1/b_2}{\Gamma\left(\frac{d+l-\widehat{\Delta}-1}{2} \right)\Gamma \left(\frac{l+\widehat{\Delta}-1}{2} \right)} \hat{f}_{12{\widehat{\mathcal{O}}^{(l)}}}~,
\end{split}
\end{align}
where
\be
\frac{b_1}{b_2} =\sqrt{\frac{1+2^{d-2}a_{\phi^2}}{\left(d-2\right)(1-2^{d-2}a_{\phi^2})}}
\ee
and $a_{\phi^2}$ is the one-point function of the operator $\phi^2$  in the presence of the boundary. This relation between OPE coefficients agrees with the result obtained from applying a shadow transformation to the relevant three-point functions. In section \ref{sec:conformalbc} we derive this equation by demanding the absence of unphysical singularities in a three-point functions involving two bulk operators $\phi$.\footnote{In \cite{Lauria:2020emq} the same analysis was carried out for defects with a higher co-dimension in the free scalar theory, leading to similar shadow relations and a proof of triviality in many cases. For non-integer dimensions this setup can also be used to describe the long-range Ising model, where the relations can be found from the non-local equation of motion \cite{Paulos:2015jfa}. More details can be found in \cite{Behan:2017emf,Behan:2017dwr} and a first numerical analysis in this context was done in \cite{Behan:2018hfx}.} Note that the relations as written are still valid when the dimension of $\widehat{\mathcal{O}}$ is such that the gamma functions have poles; this is precisely when the operators are of `double-twist' type and the shadow transformation is singular.

The properties of the previous paragraph already lead to a remarkable bootstrap problem. Indeed, up to the special `double-twist' operators there is \emph{one} spectrum and set of OPE coefficients that needs to solve the \emph{five} different crossing symmetry equations corresponding to the possible four-point functions of $\widehat O_1$ and $\widehat O_2$.\footnote{The corresponding four-point functions \emph{should} be related by the integral transformation that implements the shadow symmetry. However it is not clear to us whether the conformal block decompositions of such shadow-transformed four-point functions are automatically consistent. For example, the integral transformation is sensitive to contact terms and it seems unlikely that it can be swapped with the sum over conformal blocks.} This is intriguing in itself, but in the numerical analysis we can actually impose three more constraints. The first one is related to the Ward identity for the displacement operator. The second one is that of locality of the BCFT setup, which translates to the absence of any vector operators of dimension $d$ in the $\widehat O_1 \times \widehat O_2$ OPE. Both of these are described in section \ref{ss:displ_flux}. The third one is imposed to separate local three-dimensional CFTs, which do not interest us here, from boundary conditions: this requires the scaling dimension of the first spin 2 operator to be strictly greater than 3.

We have numerically explored the system of crossing equations originating from the $\langle 1111 \rangle$, $\langle 1122 \rangle$ and $\langle 2222 \rangle$ four-point functions in four bulk dimensions subject to all the above conditions. It might be tempting to conjecture that no non-trivial conformal boundary conditions exist that meet such stringent criteria, but surprisingly this is not quite what we find. On the one hand, there does exist a large range of possible values of $a_{\phi^2}$ where the first spin 2 operator must have a dimension less than about 3.1. Since this value is likely to decrease even further when increasing computational power, it is natural to conjecture that it must converge to 3, and then no conformal boundary conditions would be possible in this range. On the other hand, for $a_{\phi^2}$ near its Neumann value we suddenly find room for interesting physics: as indicated in figure \ref{TvsA_4d}, the first spin 2 operator can have a dimension of nearly four without violating any other constraint. A corresponding kink in all the numerical plots points towards at least one possible exotic boundary condition for a free scalar in this neighborhood. We subject this point to a detailed analysis in section \ref{sec:numerics} and show that it passes all the obvious consistency checks for a proper local boundary condition. Using the `extremal functional' methods of \cite{Poland:2011,ElShowk:2013} we also estimate the dimensions of several low-lying operators and in particular conclude that the higher-spin symmetries of the bulk theory are broken. We leave for the future the interesting question of identifying a microscopic candidate for such a conformal boundary condition.

The rest of the paper is organized as follows. In section \ref{sec:conformalbc} we study the constraints that any conformal boundary condition of a free scalar field must obey, and in particular derive the shadow relations \eqref{ssspinconstr0}. In section \ref{sec:examples} we provide some examples, starting by reviewing the free boundary conditions and then using conformal perturbation theory to construct interacting ones (under the assumption that we can appropriately tune the parameters of the local degrees of freedom on the boundary). In section \ref{sec:crossingeq} we derive the set of crossing equations for the four-point functions involving $\widehat O_1$ and $\widehat O_2$, we organize them in a way that takes advantage of the exact relations, and we explain the approximations of the resulting `superblocks' that we use in our numerical implementation. In section \ref{sec:numerics} we present the numerical results in the case of $d=4$, showing plots that involve several different characteristic observables, and in particular we show the kink that we mentioned above. We finally discuss possible future directions in section \ref{sec:conclusions}. A summary of the conventions and various technical results that we used along the way are relegated to the appendices.

\section{Analytic constraints on the free scalar BCFT}
\label{sec:conformalbc}
Consider a free massless scalar field $\phi$ in $d>2$ dimensions with a planar boundary. We use the coordinate $y\geq 0$ for the direction orthogonal to the boundary, and $\vec{x}$ for the directions parallel to the boundary. We denote the  components of $x = (\vec{x},y)\in \mathbb{R}^{d-1} \times \mathbb{R}_+ $ as $x^\mu$, $\mu = 1,\dots, d$ with $x^d = y$, and those of $\vec{x}\in\mathbb{R}^{d-1}$ as $x^a$, $a=1,\dots, d-1$. We are interested in unitary boundary conditions that preserve the boundary conformal symmetry $SO(d,1)$.

\subsection{Two-point functions with the scalar field}\label{sec:two_point}

In this section we discuss the bulk-to-boundary OPE (bOPE) of the scalar field for a generic conformal boundary condition, and the constraints imposed by crossing symmetry on the bulk two-point function of the scalar field.

\subsubsection{Bulk-boundary two-point functions}\label{ss:BDcorr}
In this section we review the existence of two operators with protected scaling dimension in the bOPE of the free scalar field \cite{Liendo:2012hy,Dimofte:2012pd,Gaiotto:2014gha,Gliozzi:2015qsa,Giombi:2019enr}. In a BCFT the bulk-boundary two-point function of a scalar bulk operator $O$ of scaling dimension $\Delta_{O}$ and a scalar boundary operator $\widehat{O}$ of scaling dimension $\widehat{\Delta}_{\widehat{O}}$ is \cite{McAvity:1993ue,McAvity:1995zd,Liendo:2012hy}
\begin{align}\label{BD2pt}
\langle O(\vec{x},y)\widehat{O}(0)\rangle =\frac{b_{O\widehat{O}}}{y^{\Delta_{O}-\widehat{\Delta}_{\widehat{O}}}(|\vec{x}|^2+y^2)^{\widehat{\Delta}_{\widehat{O}}}}~.
\end{align}
The bOPE coefficient $b_{O\widehat{O}}$, which is real for Hermitian operators, is not determined by symmetry.
Specializing $O$ to be a free scalar $\phi$ of scaling dimension $\Delta_\phi=\frac{d}{2}-1$, the equation of motion $\square \phi=0$ gives 
\begin{align}\label{eom2ptBdry}
0=\langle \square\phi(\vec{x},y)\widehat{O}(0)\rangle =\left(\frac{d}{2}-\widehat{\Delta}_{\widehat{O}}\right)\left(\frac{d}{2}-1-\widehat{\Delta}_{\widehat{O}}\right)\frac{b_{\phi\widehat{O}}}{y^{\frac{d}{2}-\widehat{\Delta}_{\widehat{O}}+1}(|\vec{x}|^2+y^2)^{\widehat{\Delta}_{\widehat{O}}}}~.
\end{align}
Therefore the possible scaling dimensions for boundary primaries with $b_{\phi \hat{O}} \neq 0$ are
\begin{align}
\label{Deltai}
\quad \widehat{\Delta}_1= \frac{d}{2}-1~, \quad \widehat{\Delta}_2 = \frac{d}{2}~.
\end{align}
Without loss of generality, we can assume there is at most one boundary operator of dimension $\widehat{\Delta}_1$ with $b_{\phi \hat{O}} \neq 0$, that we denote as $\widehat{O}_1$, and similarly for $\widehat{\Delta}_2$, the corresponding operator being denoted as $\widehat{O}_2$. As observed in \cite{Giombi:2019enr}, the scaling dimensions of these operators add up to $d-1$, which suggests that the two operators might be thought of as a `shadow pair'. In the next subsection we will show that also their three-point functions are compatible with such a `shadow relation'. The bOPE of the free scalar is \cite{McAvity:1993ue,McAvity:1995zd,Liendo:2012hy}
\begin{align}
\begin{split}\label{BOPEphi}
\phi(\vec{x},y) 
= b_1 \, \mathfrak{C}_{\frac{d}{2}-1}[y, \vec \nabla^2]\,\widehat{{O}}_1 (\vec{x})+b_2\,y\,\mathfrak{C}_{\frac{d}{2}}[y, \vec \nabla^2]\,\widehat{{O}}_2 (\vec{x})~.
\end{split}
\end{align}
where we defined $b_i\equiv b_{\phi \widehat{O}_i}$, $i=1,2$. The explicit form of the differential operator $\mathfrak{C}_{\widehat{\Delta}_i}[y, \vec \nabla^2]$ is given in appendix \ref{app:conventions}.
The bOPE can be used to reconstruct bulk correlators starting from the boundary ones.

\subsubsection{Bulk two-point function}\label{ss:BBcorr}

Next, we consider the two-point function
\begin{align}\label{BB2pt}
\langle \phi(\vec{x}_1,y_1)\phi(\vec{x}_2,y_2)\rangle.
\end{align}
This correlator is not completely fixed by the symmetry as it depends on the cross-ratio
\begin{align}\label{xidef}
\xi \equiv \frac{(x_1-x_2)^2}{y_1 y_2}\equiv\frac{|\vec{x}_1-\vec{x}_2|^2+(y_1-y_2)^2}{y_1 y_2}.
\end{align}
We can compute \eqref{BB2pt} by plugging twice the bOPE \eqref{BOPEphi}, using the diagonal and unit-normalized boundary two-punt functions
\begin{equation}
\langle \widehat{O}_i(\vec{x}_1) \widehat{O}_j(\vec{x}_2)\rangle = \delta_{ij}\frac{1}{|\vec{x}_1 -\vec{x}_2|^{2\widehat{\Delta}_i}}~,~~i,j=1,2
\end{equation}
and resumming the contributions from the descendants. The resulting \emph{boundary channel} decomposition of \eqref{BB2pt} is 
\begin{align}
\begin{split}\label{2ptFunctionsDEF}
& \langle \phi(\vec{x}_1,y_1)\phi(\vec{x}_2,y_2)\rangle
\\ &~~~~~~ = \frac{1}{(y_1 y_2)^{\frac{d}{2}-1}}\left[\frac{b_1^2 }{2}  \left(\xi^{1-d/2}+(\xi +4)^{1-d/2}\right) + \frac{b_2^2}{2 (d-2)}  \left(\xi^{1-d/2}-(\xi +4)^{1-d/2}\right)\right]~.
\end{split}
\end{align}

An alternative way of computing the two-point function \eqref{BB2pt} is to invoke the bulk OPE $\phi\times \phi$, namely
\begin{align}\label{phiphiOPE}
\phi(x)\phi(0)=\frac{\mathbb{{\mathds 1}}}{(x^2)^{d/2-1}}+\phi^2(0)+\frac{c_{\phi\phi T}}{C_T}{ x}^\mu { x}^\nu T_{\mu\nu}(0)+\sum_{\ell=4,6,\dots}\frac{c_{\phi\phi \ell}}{C_{J_\ell}}{ x}^{\mu_1}\dots { x}^{\mu_\ell}J_\ell^{\mu_1\dots \mu_\ell}(0)+\dots
\end{align}
where $T_{\mu\nu}$ is the stress tensor, and the operators $J_\ell^{\mu_1\dots \mu_\ell}$ with $\ell\geq 4$ are the tower of higher-spin conserved currents present in the free scalar CFT. The OPE data involving the stress tensor are \cite{Osborn:1993cr}
\begin{align}\label{cphiphiT2}
c_{\phi\phi T}= -\frac{d (d-2)}{2(d-1)S_d}, \quad C_T= \frac{d }{(d-1) S_d^2}, \quad S_d\equiv\text{Vol}(S^{d-1})=\frac{2 \pi ^{d/2}}{\Gamma \left(\frac{d}{2}\right)}.
\end{align}
Plugging in \eqref{BB2pt}, we write the two-point function as a sum over bulk one-point functions and their derivatives. Boundary conformal invariance allows only for scalar bulk one-point functions \cite{McAvity:1993ue,McAvity:1995zd,Liendo:2012hy}, hence from the $\phi\times \phi$ bulk OPE the only non-trivial contributions are due to
\begin{align}\label{one_point}
\langle \phi^2(\vec{x},y)\rangle = \frac{a_{\phi^2}}{y^{d-2}}, \quad \langle \mathds 1\rangle =a_{\mathds 1} =1.
\end{align}
Resumming the contribution from bulk descendants we obtain the \emph{bulk channel} decomposition of \eqref{BB2pt}:
\begin{align}
\begin{split}\label{2ptFunctionsB}
\langle \phi(\vec{x}_1,y_1)\phi(\vec{x}_2,y_2)\rangle
=\frac{\xi^{1-\frac{d}{2}}}{(y_1 y_2)^{\frac{d}{2}-1}} \left[1 + a_{\phi^2}\,  2^{d-2} \xi ^{\frac{d-2}{2}} (\xi +4)^{1-\frac{d}{2}}\right]~.
\end{split}
\end{align}

Equating the two different decompositions \eqref{2ptFunctionsDEF} and \eqref{2ptFunctionsB} gives the \emph{bulk-to-boundary crossing} equation \cite{Liendo:2012hy}.
Since everything else in the equation is fixed, the only dynamical data are the one-point function coefficient $a_{\phi^2}$ on the l.h.s. and the bulk-to-boundary couplings $(b_1,b_2)$ on the r.h.s. The solution is \cite{Liendo:2012hy,Gliozzi:2015qsa}
\begin{align}\label{b1b2def}
b_1^2=1+2^{d-2}a_{\phi^2}~,\quad b_2^2=\left(d-2\right)(1-2^{d-2}a_{\phi^2})~.
\end{align}
This result tells us that in any boundary condition for a free scalar the parameter $a_{\phi^2}$ is constrained by unitarity to lie in an interval
\begin{align}\label{aphi2_range}
-\frac{1}{2^{d-2}} = a^{(D)}_{\phi^2} \leq a_{\phi^2}\leq a^{(N)}_{\phi^2} = \frac{1}{2^{d-2}}.
\end{align}
As we indicated above, the boundaries of the interval correspond to the Dirichlet ($b_1 = 0$) and Neumann ($b_2 = 0$) boundary condition. These elementary boundary conditions will be discussed in detail in section \ref{sec:examples}, but in the remainder of this section we will assume that $b_1 b_2 \neq 0$ because we would like to explore the possibility of more exotic boundary conditions.

\subsection{Three-point functions with the scalar field}\label{sec:three_pt}
In this section we consider three-point functions with two insertions of the free scalar $\phi$ and a generic boundary operator $\widehat{\mathcal{O}}$. Note that, by Lorentz invariance, these correlators can be non-vanishing only if the third operator transforms as a symmetric and traceless tensor of $SO(d-1)$. Without loss of generality we can place the boundary operator at infinity and consider
\begin{align}\label{phiphiOspinj}
\langle \phi(\vec{x}_1,y_1)\phi(\vec{x}_2,y_2){\widehat{\mathcal{O}}^{(l)}} (\theta,\infty)\rangle.
\end{align}
Following the standard procedure \cite{Costa:2011mg}, in the above expression we contracted the tensor indices with a boundary polarization vector $\theta^a$ as follows 
\begin{align}
{\widehat{\mathcal{O}}^{(l)}}(\theta,\infty)\equiv \theta^{a_1}\dots \theta^{a_l}\widehat{\mathcal{O}}_{a_1,\dots,a_l} (\infty),\quad \theta\cdot \theta=0.
\end{align}
We will show that the boundary channel expansion of this correlation function exhibits unphysical singularities, which can be removed only if special conditions are met. Therefore these conditions have to be satisfied in any conformal boundary condition of the free scalar.

\subsubsection{Boundary channel computation}

The bOPE \eqref{BOPEphi} allows to completely determine the correlator \eqref{phiphiOspinj} in terms of the three-point functions between the operators $\widehat{O}_i$, $i=1,2$, and ${\widehat{\mathcal{O}}^{(l)}}$. Conformal invariance fixes the latter three-point functions to take the form \cite{Osborn:1993cr,Costa:2011mg}
\begin{align}\label{3ptssl}
\langle \widehat{O}_i (\vec{x}_1)\widehat{O}_j (\vec{x}_2){\widehat{\mathcal{O}}^{(l)}} (\theta,\infty)\rangle=&\,\frac{\hat{f}_{ij{\widehat{\mathcal{O}}^{(l)}}}}{|\vec{x}_{12}|^{\widehat{\Delta}_i+\widehat{\Delta}_j-\widehat{\Delta}}}P^{(l)}_\parallel(\hat{x}_{12},{\theta})~,
\end{align}
where $\widehat{\Delta}$ denotes the scaling dimension of the operator $\widehat{\mathcal{O}}^{(l)}$ which carries $SO(d-1)$ spin $l$. The dependence on the polarization vector is through the following polynomial 
\begin{align}\label{parjpoly}
P^{(l)}_\parallel(\hat{x}_{12},{\theta})\equiv\left(- \hat{x}_{12}\cdot {\theta}\right)^l, \quad \hat{x}^a\equiv \frac{x^a}{|\vec{x}|}~.
\end{align}
By Bose symmetry
\begin{align}\label{Bosesymm}
\hat{f}_{ij{\widehat{\mathcal{O}}^{(l)}}}=(-1)^l {}\hat{f}_{ji{\widehat{\mathcal{O}}^{(l)}}}~,
\end{align}
which implies that only even spins $l$ are allowed in \eqref{3ptssl} if $i=j$. 

To compute \eqref{phiphiOspinj}, we act twice with the bOPE on the boundary three-point functions \eqref{3ptssl}. After some algebra to resum the contributions from the descendants, we obtain the following boundary channel expansion
\begin{align}\label{phiphiOgen}
\langle \phi(\vec{x}_1,y_1)&\phi(\vec{x}_2,y_2){{\widehat{\mathcal{O}}^{(l)}}}(\theta,\infty)\rangle=\frac{P^{(l)}_\parallel(\hat{x}_{12},{\theta})}{ |\vec{x}_{12}|^{d-2-\widehat{\Delta}}}\times\nonumber\\
&\left( b_1^2 {}\hat{f}_{11{\widehat{\mathcal{O}}^{(l)}}}\,\hat{\mathcal{F}}^{11}_{\widehat{\Delta},l}({w_+},{w_-})+{b_1 b_2{}\hat{f}_{12{\widehat{\mathcal{O}}^{(l)}}}}{}\hat{\mathcal{F}}^{12}_{\widehat{\Delta},l}({w_+},{w_-})+{b_2^2 {}\hat{f}_{22{\widehat{\mathcal{O}}^{(l)}}}}{}\hat{\mathcal{F}}^{22}_{\widehat{\Delta},l}({w_+},{w_-})\right).
\end{align}
This expression depends on two cross-ratios $w_\pm$, which we take as follows:
\begin{align}\label{wdef}
{w_\pm}=\frac{({y_1\pm y_2})^2}{|\vec{x}_{12}|^2}~.
\end{align}
The functions $\hat{\mathcal{F}}^{ij}_{\widehat{\Delta},l}({w_+},{w_-})$ are computed in Appendix \ref{app:blocks_defect} and their explicit expressions are given in  \eqref{phiphiOblocks_def}. In the next section we will study the analyticity properties of the correlator \eqref{phiphiOgen}.

\subsubsection{Constraints from analyticity of the bulk OPE}\label{ss:reg_OPE}
Next, we study the same three-point function using the bulk $\phi\times \phi$ OPE. Since the only singular term in this OPE is given by the identity operator, which does not contribute to the three-point function, we conclude that the three-point function must be free of singularities when the two bulk points coincide. In terms of the cross-ratios $w_\pm$, this limit corresponds to $w_+ \to \infty$ with any fixed $w_-$.

As we show in the appendix \ref{app:exactrelDer}, for generic values of their parameters the boundary blocks on the r.h.s of \eqref{phiphiOgen} become singular in this limit. These unphysical singularities can be removed if the OPE coefficients are related in the following way
\begin{align}
\begin{split}\label{ssspinconstr}
{}\hat{f}_{11{\widehat{\mathcal{O}}^{(l)}}}& =\kappa_1(\widehat{\Delta},l)\hat{f}_{12{\widehat{\mathcal{O}}^{(l)}}},\quad \kappa_1(\widehat{\Delta},l)\equiv -\frac{b_2 \Gamma \left(\frac{l+\widehat{\Delta}}{2}\right) \Gamma \left(\frac{d+l-\widehat{\Delta}-2}{2} \right)}{2b_1 \Gamma\left(\frac{d+l-\widehat{\Delta}-1}{2} \right)\Gamma \left(\frac{l+\widehat{\Delta}+1}{2}\right)}~,\\
\hat{f}_{22{\widehat{\mathcal{O}}^{(l)}}}& =\kappa_2(\widehat{\Delta},l) \hat{f}_{12{\widehat{\mathcal{O}}^{(l)}}},\quad \kappa_2(\widehat{\Delta},l)\equiv -\frac{2 b_1 \Gamma \left(\frac{l+\widehat{\Delta}}{2}\right) \Gamma \left(\frac{d+l-\widehat{\Delta}}{2} \right)}{b_2\Gamma\left(\frac{d+l-\widehat{\Delta}-1}{2} \right)\Gamma \left(\frac{l+\widehat{\Delta}-1}{2} \right)}~.
\end{split}
\end{align}
For certain special values of the parameters $(\widehat{\Delta},l)$ some of the blocks on the r.h.s of \eqref{phiphiOgen} are themselves regular. These special values correspond to the poles of the gamma functions in \eqref{ssspinconstr} and read (see also table \ref{tab:selprottower})
\begin{itemize}
\item $\widehat{\Delta}=d+l-2$. 
	We have that $\kappa_{1}(\widehat{\Delta},l)\rightarrow \infty$ while $\kappa_{2}(\widehat{\Delta},l)$ remains finite. This sets $\hat{f}_{12{\widehat{\mathcal{O}}^{(l)}}}=\hat{f}_{22{\widehat{\mathcal{O}}^{(l)}}}=0$, while leaving $\hat{f}_{11{\widehat{\mathcal{O}}^{(l)}}}$ unconstrained. We denote these operators as $[\widehat{O}_1 \widehat{O}_1]_{0,l}$.

	\item $\widehat{\Delta}=d+l+2n-2$ with $n$ a positive integer.
	We have $\kappa_{1,2}(\widehat{\Delta},l)\rightarrow \infty$ and so $\hat{f}_{12{\widehat{\mathcal{O}}^{(l)}}}=0$ while $\hat{f}_{11{\widehat{\mathcal{O}}^{(l)}}},\hat{f}_{22{\widehat{\mathcal{O}}^{(l)}}}$ remain unconstrained.  Given that generically they appear in both OPEs, we could denote these operators both as $[\widehat{O}_1 \widehat{O}_1]_{n,l}$ or $[\widehat{O}_2 \widehat{O}_2]_{n-1,l}$. For definiteness, we choose to denote them as $[\widehat{O}_1 \widehat{O}_1]_{n,l}$.
	
	\item $\widehat{\Delta}=d+l+2n-1$ and $n \in \mathbb{N}$. We have $\kappa_{1,2}(\widehat{\Delta},l)=0$, which sets $\hat{f}_{11{\widehat{\mathcal{O}}^{(l)}}}=\hat{f}_{22{\widehat{\mathcal{O}}^{(l)}}}=0$ while leaving $\hat{f}_{12{\widehat{\mathcal{O}}^{(l)}}}$ unconstrained. We denote these operators as $[\widehat{O}_1 \widehat{O}_2]_{n,l}$. Importantly, all odd-spin operators in $\widehat{O}_1 \times \widehat{O}_2$ are of this type, as can be seen by combining \eqref{ssspinconstr} with Bose symmetry.
\end{itemize}
\begin{table}[htp!]
	\begin{center}
		\begin{tabular}{l|l|c|c}
			$\widehat{\Delta}-l$&conditions&independent OPE coeff&operator\nonumber\\
			\hline
			\hline
			&&\nonumber\\
			$d-2$& $\frac{b_1 }{b_2} \kappa_1(\widehat{\Delta},l)=\infty$&$\hat{f}_{11{\widehat{\mathcal{O}}^{(l)}}}$&$[\widehat{O}_1 \widehat{O}_1]_{0,l}$\nonumber\\
			&&\nonumber\\
			$d+2n-2,\,\,\, n>0$& $\frac{b_1 }{b_2} \kappa_1(\widehat{\Delta},l), \frac{b_2 }{b_1} \kappa_2(\widehat{\Delta},l)=\infty$&$\hat{f}_{11{\widehat{\mathcal{O}}^{(l)}}},\hat{f}_{22{\widehat{\mathcal{O}}^{(l)}}}$&$[\widehat{O}_1 \widehat{O}_1]_{n,l}$\nonumber\\
			&&\nonumber\\
			$d+2n-1$& $\frac{b_1 }{b_2} \kappa_1(\widehat{\Delta},l), \frac{b_2 }{b_1} \kappa_2(\widehat{\Delta},l)=0$&$\hat{f}_{12{\widehat{\mathcal{O}}^{(l)}}}$&$[\widehat{O}_1 \widehat{O}_2]_{n,l}$\nonumber
 \end{tabular}
 	\caption{ \label{tab:selprottower} Table of special multiplets and their selection rules. Recall that $b_1^2=1+2^{d-2}a_{\phi^2}$ and $b_2^2=\left(d-2\right)(1-2^{d-2}a_{\phi^2})$.}
	\end{center}
\end{table} 
The special cases listed above are related to the higher-spin symmetry of the bulk theory, as we will now explain. We recall that the $\phi\times \phi$ OPE \eqref{phiphiOPE} contains infinitely many higher-spin conserved currents $J_\ell$, with even spin $\ell \geq 2$ and scaling dimensions $\Delta_\ell=d+\ell-2$. The conservation of these currents is generically violated by terms localized on the boundary, leading to the following Ward identities
\begin{align}\label{ward_currents}
\langle\partial_{\mu} J_\ell^{\mu \mu_1\dots \mu_{\ell-1}}(\vec{x},y)\dots\rangle&=\delta (y)\langle\widehat{\mathcal{O}}_\ell^{\mu_1\dots \mu_{\ell-1}}(\vec{x})\dots\rangle~.
\end{align}
In this formula any subset among the $\ell-1$  symmetric traceless indices $\{\mu_1\dots \mu_{\ell-1}\}$ can be taken to be parallel to the boundary, with the remaining indices being orthogonal, i.e. in the $y$ direction. Therefore, the BCFT generically contains boundary operators $\text{D}_\ell^{(l)}$ and $\text{V}_\ell^{(l+1)}$ of spin $l$ and $l+1$, respectively, and protected dimensions $\widehat{\Delta}=d+\ell-2$, where $l$ is an even integer ranging from $0$ to $\ell-2$. By `generically' we mean that some of these operators might actually be absent from the spectrum in special cases. The equations \eqref{ward_currents} can be equivalently rephrased in terms of the bOPE, namely the operators $\text{D}_\ell^{(l)}, \text{V}_\ell^{(l+1)}$ have the correct dimensions and spins to appear in the bOPE of the bulk higher-spin current $J_\ell$ in a way that is compatible with its conservation in the bulk. Furthermore, spin selection rules and bulk conservation imply that $\text{V}_\ell^{(l+1)}$ cannot appear in the bOPE of any $J_{\ell'}$ with $\ell'\neq \ell$, while the only other bulk current besides $J_\ell$ that can contain $\text{D}_\ell^{(l)}$ in its bOPE is $J_l$.

The relation to the special cases of \eqref{ssspinconstr} now stems from the observation that when $\ell-l=2n$, with $n$ non-negative integer, $\text{D}^{(l)}_\ell$ has the right dimension to be the special operator $[\widehat{O}_1 \widehat{O}_1]_{n,l}$ in table \ref{tab:selprottower}. Similarly, when $\ell-l=2n+1$ with $n\in \mathbb{N}$, $\text{V}^{(l+1)}_\ell$ has the right dimension to be the special operator $[\widehat{O}_1 \widehat{O}_2]_{n,l+1}$. We show in general in the section \ref{ss:mtchBulk} of appendix \ref{app:exactrelDer}, and for the special case $\ell=2$ in the next subsection, that in fact whenever the operator $\text{D}^{(l)}_\ell$ is present in the bOPE of $J_\ell$, then it must appear in at least one of either the OPE of $\widehat{O}_1$ with itself or the OPE of $\widehat{O}_2$ with itself. Similarly, whenever $\text{V}^{(l+1)}_\ell$ is present in the bOPE of $J_\ell$, it must also appear in the OPE of $\widehat{O}_1$ with $\widehat{O}_2$.

\subsubsection{Displacement operator, flux operator and bulk-to-boundary crossing}\label{ss:displ_flux}
The case $\ell = 2$ deserves special attention because it corresponds to the bulk stress tensor $T^{\mu \nu}$. Then the scalar operator $\text{D}^{(0)}_2\equiv \text{D}$ is the so-called displacement operator, and we will refer to the spin 1 operator $\text{V}^{(1)}_2\equiv \text{V}^{(1)}$ as the \emph{flux operator}. Their general importance stems from the conservation of momentum $P^\mu$ along a time coordinate chosen parallel to the boundary. If we split $x^\mu = (\tau, \vec {z},y)$ then
\be
\begin{split}
\frac{d}{d\tau}P^\mu(\tau) &= \int d^{d-2}\vec{z} \int_0^\infty dy \,\partial_t \,T^{t \mu}(\tau,\vec {z},y) \\
&=  \int d^{d-2}\vec{z} \, \,T^{y\mu}(\tau,\vec{z},y \to 0)~,
\end{split}	
\ee
where in the second equality we used the conservation to trade the time derivative with a spatial one, and then rewrote the integral of the spatial derivative as a boundary term. Choosing $\mu = y$ orthogonal to the boundary we find that the limit $y \to 0$ gives the displacement operator $\text{D}$, which therefore measures the breaking of translations orthogonal to the boundary and must be non-zero for any sensible boundary condition. Choosing $\mu$ parallel to the boundary, on the other hand, we find the vector operator $\text{V}^{(1)}$ and so we conclude that it measures the flux of energy into the boundary. Theories with a non-trivial flux operator $\text{V}^{(1)}\neq 0$ may still have a conserved boundary-translation charge, if there is an additional boundary contribution to the charge $P^a_{\text{tot}} = P^a + \widehat{P}^a$ satisfying
\begin{equation}\label{eq:bcontrP}
\frac{d}{d\tau}\widehat{P}^a(\tau) =  - \int d^{d-2}\vec{z} \, \, \text{V}^{(1) a}(\tau,\vec{z}\,)~.
\end{equation}
However the flux operator must be absent in any local unitary BCFT setup. To see why, note that the locality condition on the boundary is that $\widehat{P}^a$, if non-trivial, should be expressible as the integral
\begin{equation}
\widehat{P}^a(\tau) = \int d^{d-2}\vec{z} \,\, \widehat{t}^{\,\,t a}(\tau,\vec {z})~,
\end{equation}
of a local boundary operator with two indices $\widehat{t}^{\,\,b a}$. The condition \eqref{eq:bcontrP} locally takes the form 
\begin{equation}\label{eq:local}
\partial_b \widehat{t}^{\,\, b a} = -\text{V}^{(1) a}~.
\end{equation}
Moreover, by repeating the argument for the other generators of the conformal group on the boundary, one can easily show that the operator $\widehat{t}^{\,\,b a}$ has spin 2, i.e. it is symmetric and traceless. Recalling that $\text{V}^{(1) a}$ has scaling dimension $d$ and therefore $\widehat{t}^{\,\,b a}$ has scaling dimension $d-1$, we see that eq. \eqref{eq:local} with $\text{V}^{(1)}\neq 0$ is incompatible with the unitarity bound of a spin 2 operator in $d-1$ dimensions. We conclude that indeed in any unitary BCFT locality implies that  
\begin{equation} \label{eq:loccond}
\text{V}^{(1)} = 0
\end{equation} 
in which case $\widehat{P}^a$ is trivial.
In practice this means that if we couple a bulk CFT (not necessarily our free scalar theory) to some local boundary degrees of freedom, perhaps triggering an RG flow to a new conformal boundary condition, then the flux operator must never appear.\footnote{It is curious that the statement of locality, which in the bulk is encoded by the \emph{presence} of a stress tensor, corresponds to the \emph{absence} of a specific vector operator (in the bOPE of $T^{\mu \nu}$) in the BCFT setup.} This is because local boundary degrees of freedom should not be able to hold a macroscopic amount of energy.

It might be instructive to consider some non-local setups that do feature a flux operator. The first is a conformal \emph{interface}, where there is an entire new CFT living on the half space with $y < 0$. In that case the stress tensor for each side `sees' a flux operator, but if the interface setup is local then these two flux operators are in fact the same operator and the interface cannot act as a simultaneous energy sink for both sides. Such a setup can be generalized to the case where our $d-1$ dimensional boundary is at the same time a conformal defect in some $d'$-dimensional auxiliary space in which it is coupled to an arbitrary $d'$-dimensional bulk CFT, perhaps even triggering a boundary/defect RG flow to some new conformal configuration. According to the general structure of the operator expansion near a defect of dimension $d-1$ in a $d'$-dimensional CFT, the $d'$-dimensional stress tensor can always provide a vector operator of precisely the requisite dimension\footnote{This simply follows from requiring conservation of the $d'$-dimensional stress tensor in the allowed bulk-defect correlators with a vector. For a generic defect CFT these two-point functions were classified in \cite{Billo:2016cpy}.} (which is $d$) and unless the two sides decouple we will observe this as a flux operator in the $d$-dimensional BCFT.\footnote{Note that a similar statement applies for $\ell > 2$, namely if the operator $\text{V}_\ell^{(\ell-1)}$ in the bOPE of $J_\ell$ vanishes then the BCFT admits a conserved charge of spin $\ell$, which is given by a bulk spatial integral of the higher-spin conserved current.} Lastly one could also try to create a non-local setup by adding a GFF on the boundary and coupling it, perhaps with other degrees of freedom, to the bulk field. But this scenario is captured by the previous one, because GFFs are just regular local fields in an auxiliary higher-dimensional space (albeit with non-integral $d'$).

The previous discussion applied to any BCFT, but for the free scalar theory there are a few additional results that we can derive. To this end we return to the $\vev{\phi\phi\widehat O}$ three-point function of equation \eqref{phiphiOgen} and take the third operator to be either the flux operator $\text{V}^{(1)}$ or the displacement operator $\text{D}$. Let us thererefore first consider:
\begin{align}\label{scalarVV}
\langle \phi(\vec{x}_1,y_1)\phi(\vec{x}_2,y_2)\text{V}^{(1)}(\theta,\infty)\rangle~.
\end{align}
The relations \eqref{ssspinconstr} in this case give that $\hat{f}_{11{\text{V}^{(1)}}}=\hat{f}_{22{\text{V}^{(1)}}}=0$ and leave $\hat{f}_{12{\text{V}^{(1)}}}$ undetermined. On the other hand, the bulk channel decomposition of \eqref{scalarVV} receives contribution only from the bulk stress tensor $T^{\mu\nu}$, because -- as we explained in the previous subsection for the general case -- spin selection rules and bulk conservation imply that that $\text{V}^{(1)}$ cannot appear in the bOPE of any other operator in the OPE  \eqref{phiphiOPE}. We find
\begin{align}\label{vec_crossing}
\langle \phi(x_1)\phi(x_2)\text{V}^{(1)}(\theta,\infty)\rangle &=\frac{c_{\phi\phi T}}{C_T} \bigg[{x}_{12}^\mu {x}_{12}^\nu\langle T_{\mu\nu}(x_2)\text{V}^{(1)}(\theta,\infty)\rangle+\dots\bigg]~,
\end{align}
where the ellipses represent contributions of bulk descendants, and the OPE data of the stress tensor are given in \eqref{cphiphiT2}. The two-point function on the r.h.s. is completely fixed by the boundary conformal symmetry up to a single bOPE coefficient $b_{T\text{V}^{(1)}}$ \cite{McAvity:1995zd,Liendo:2012hy}. We can further relate $b_{T\text{V}^{(1)}}$ to the two-point function coefficient
\begin{align}\label{VV}
\langle \text{V}^{(1)}(\theta_1,\vec{x})\text{V}^{(1)}(\theta_2,0)\rangle= C_{\text{V}^{(1)}}\frac{(\theta_1\cdot I(\vec{x})\cdot \theta_2)}{|\vec{x}|^{2d}},\quad I^{ab}(\vec{x})\equiv \delta^{ab}-2 \frac{x^a x^b}{|\vec{x}|^2}~.
\end{align}
From the exact correlator \eqref{phiphiOgen} we find (details in appendix \ref{ss:mtchBulk})
\begin{align}\label{bTVlambda_rel}
\hat{f}_{12\text{V}^{(1)}}=-\frac{S_d (d-2)}{2b_1 b_2}\,b_{T\text{V}^{(1)}}=-\frac{ S_d (d-2)}{b_1 b_2}C_{\text{V}^{(1)}}
\end{align}
and we conclude that the flux operator appears in the $\widehat{O}_1 \times  \widehat{O}_2$ OPE if it also appears in the bOPE of the stress tensor.\footnote{These conclusions again have a natural generalization to all the odd-spin protected boundary operators $\text{V}^{(l)}_\ell$ defined in \eqref{ward_currents}, see appendix \ref{ss:mtchBulk} for more details.}

Next we consider the three-point function involving the displacement operator
\begin{align}\label{scalarDD}
\langle \phi(\vec{x}_1,y_1)\phi(\vec{x}_2,y_2)\text{D}(\infty)\rangle~.
\end{align}
In this case, the relations \eqref{ssspinconstr} give that $\hat{f}_{12{\text{D}}}=0$ while leaving $\hat{f}_{11{\text{D}}}$ and $\hat{f}_{22{\text{D}}}$ undetermined. Using again the general argument from the previous subsection about spin selection rules and conservation, we have that the only operators in the OPE  \eqref{phiphiOPE} that contribute to the bulk channel decomposition of \eqref{scalarDD} are $\phi^2$ and the bulk stress tensor. Therefore we have
\begin{align}\label{phiphiDblockexpalead}
\langle \phi(x_1)\phi(x_2)\text{D} (\infty)\rangle &=\bigg[\langle \phi^2(x_2)\text{D}(\infty)\rangle+\dots\bigg]+\frac{c_{\phi\phi T}}{C_T}\bigg[{x}_{12}^\mu {x}_{12}^\nu\langle T_{\mu\nu}(x_2)\text{D}(\infty)\rangle+\dots\bigg]~,
\end{align}
where again the ellipses denote contributions of bulk descendants.  Using the Ward identities for the displacement operator \cite{McAvity:1995zd}, the bulk-boundary two-point functions in the r.h.s. are determined in terms of the parameter $a_{\phi^2}$ in \eqref{aphi2_range}, as well as the coefficient $C_{\text{D}}$ in the two-point function of the displacement operator 
\begin{align}\label{defCD}
\langle \text{D}(\vec{x})\text{D}(0)\rangle = \frac{C_{\text{D}}}{|\vec{x}|^{2d}}~.
\end{align}
Comparing with the boundary channel correlator \eqref{phiphiOgen}, after some algebra which we relegate to appendix \ref{subsec:dispDet}, we find
\begin{align}\label{displRel}
{}\hat{f}_{11\text{D}}=\frac{(d-2) \left(a_{\phi^2} 2^d+2 C_{\text{D}} S_d^2\right)}{4 (d-1) S_d b_1^2}, \quad 
{}\hat{f}_{22\text{D}}=\frac{(d-2) \left(2 C_{\text{D}} S_d^2-a_{\phi^2} 2^d\right)}{2 S_d b_2^2}.
\end{align}
The unitarity requirement $C_\text{D}\geq 0$ implies that:
\begin{align}
\hat{f}_{11\text{D}}\geq \frac{(d-2) 2^d}{4 (d-1) S_d}\frac{a_{\phi^2}}{b_1^2}, \quad 
{}\hat{f}_{22\text{D}}\geq -\frac{(d-2) 2^d}{2 S_d}\frac{a_{\phi^2}}{b_2^2}.
\end{align}

\subsubsection{The three-point function with the boundary modes of $\phi$}
Another interesting special case of \eqref{phiphiOgen} arises when the boundary operator is one of the boundary modes of $\phi$, i.e.
\begin{align}\label{scalarO}
\langle \phi(\vec{x}_1,y_1)\phi(\vec{x}_2,y_2)\widehat{{O}}_1(\infty)\rangle, \quad \langle \phi(\vec{x}_1,y_1)\phi(\vec{x}_2,y_2)\widehat{{O}}_2(\infty)\rangle.
\end{align}
On general grounds, due to Bose symmetry, there are four independent boundary OPE coefficients that enter these correlators: $\hat{f}_{111},\hat{f}_{112},\hat{f}_{221},\hat{f}_{222}$. The latter are further related to each other, by means of the three independent constraints provided by regularity of the $\phi\times \phi$ OPE \eqref{ssspinconstr}:
\begin{align}\label{3pt123}
\hat{f}_{112} = -2& \frac{b_1}{b_2} \frac{\Gamma \left ( \frac{d}{4} \right )^2}{\Gamma \left ( \frac{d-2}{4} \right )^2} \hat{f}_{111}~,\quad \hat{f}_{221} = \frac{1}{4} \frac{b_1^2}{b_2^2} (d-2)(d-4) \hat{f}_{111}~. \nonumber\\
&\hat{f}_{222} = -\frac{1}{2} \frac{b_1^3}{b_2^3} (d-2)(d-4) \frac{\Gamma \left ( \frac{d}{4} \right )^2}{\Gamma \left ( \frac{d-2}{4} \right )^2} \hat{f}_{111}~.
\end{align}
Hence, the bulk three-point function $\langle \phi \phi \phi \rangle$  is completely controlled by a single boundary OPE coefficient, e.g. $\hat{f}_{111}$. The latter can be non-zero only if the boundary condition breaks the $\mathbb{Z}_2$ global symmetry $\phi\to-\phi\,$, under which both $\widehat{{O}}_i$ are odd.

\section{Examples}
\label{sec:examples}

In this section we explore some examples of conformal boundary conditions for a free scalar. We start by reviewing the free boundary conditions, i.e. Neumann and Dirichlet, and then construct examples of interacting boundary conditions using conformal perturbation theory around the free ones. As we will see, these constructions rely on some ad-hoc assumptions on the spectrum of an additional local 3d sector living on the boundary, which we couple to the bulk, and therefore they do not prove rigorously the existence of interacting boundary conditions. On the other hand, they will provide useful benchmarks to compare to our numerical results in section \ref{sec:numerics}.

\subsection{Free boundary conditions}\label{sec:freebc}
Suppose the theory is fully described by the free bulk action\footnote{Note that using this canonical normalization of the action the operator $\phi$ has a different normalization compared to the one in equation \eqref{phiphiOPE}, namely $\phi\vert_{\eqref{eq:freebulkaction}} = \sqrt{\frac{\Gamma\left(\frac{d}{2}-1\right)}{4\pi^{\frac{d}{2}}}} \phi\vert_{\eqref{phiphiOPE}}$. We will specify which normalization we are using whenever important. \label{footnotenormalization}}
\begin{equation}\label{eq:freebulkaction}
S = \int d^{d-1} \vec{x} \int_0^\infty dy\, \frac{1}{2}(\partial_\mu \phi)^2~,
\end{equation}
without any boundary-localized interaction. In order to have a stationary action, besides the bulk equation of motion $\square \phi =0$ we need to set to zero the boundary term
\begin{align}\label{freebc}
\delta S =- \int d^{d-1}\vec{x}\,\, \left.\delta\phi\,\,\partial_y \phi \right\vert_{y=0}=0~.
\end{align}
The two solutions to this condition that preserve boundary conformal invariance are
\begin{align}\label{freebcND}
\text{Neumann:}\quad & \partial_y\phi\vert_{y=0}=0~,\nonumber\\
&\text{or}\\
\text{Dirichlet:}\quad& \phi\vert_{y=0}=0\nonumber~.
\end{align}
We can rephrase these conditions in terms of the bOPE of the scalar field. Namely, in \eqref{BOPEphi} we have $b_2 = 0$ and $\phi\vert_{y=0} \propto \widehat{O}_1$ in the case of Neumann boundary condition, and $b_1 = 0$ and $\partial_y\phi\vert_{y=0} \propto \widehat{O}_2$ in the case of Dirichlet boundary condition. In either case, there is only one boundary operator in the bOPE of $\phi$, and the full set of boundary correlators can be simply characterized as the mean-field theory of this operator. This implies that all the correlation functions of these BCFTs are completely disconnected, i.e. they are computed by Wick contractions as products of two-point functions. For this reason we call these boundary conditions `free boundary conditions'.

We can also consider additional free boundary conditions that are not Neumann and Dirichlet. Such a boundary condition is obtained requiring both $\widehat{O}_1$ and $\widehat{O}_2$ to appear in the bOPE \eqref{BOPEphi}, i.e. $b_1 b_2 \neq 0$, and postulating that these operators are two decoupled generalized free fields.\footnote{This includes the case of `no boundary', or more precisely the `trivial interface', where the theory on the full $\mathbb R^d$ is re-interpreted as a BCFT. In that case $a_{\phi^2} = 0$ so according to \eqref{b1b2def} this corresponds to $b_1^2 = 1$ and $b_2^2 = (d-2)$.} However this implies that there is a spin 1 operator of dimension 4 in the spectrum of the boundary theory, namely the vector `double-trace' operator in the OPE of $\widehat{O}_1$ with $\widehat{O}_2$, schematically $~ \widehat{O}_1 \partial_a \widehat{O}_2 - \widehat{O}_2 \partial_a \widehat{O}_1 $. It is easy to check that this operator also appears in the bOPE of the bulk stress tensor, hence for these boundary conditions we have a non-vanishing flux operator $\text{V}^{(1)} \neq 0$. Therefore, following the discussion in the previous section, these are non-local boundary conditions. We conclude that the only local free boundary conditions are the familiar Neumann and Dirichlet boundary conditions reviewed above.

\subsection{Interacting boundary conditions in perturbation theory}\label{example4d}

In order to look for examples of interacting boundary conditions, a natural strategy is to couple the bulk scalar to a CFT$_{d-1}$ living on the boundary. We turn on some relevant interaction between the two sectors and then flow to the IR, hoping to reach a non-trivial BCFT fixed point. Concretely, we add to the free bulk action \eqref{eq:freebulkaction} a boundary action of the form
\begin{align}\label{4d3dexamplegen2}
S_{\partial}=S_{\text{CFT}_{d-1}} +\sum_I{g_I}{}\int_{y=0} d^{d-1} \vec{x}~ \widehat{\sigma}_I~,
\end{align}
where $\widehat{\sigma}_I$ are some scalar composites made of $\phi\vert_{y=0}$ or $\partial_y \phi\vert_{y=0}$, depending on whether we start with Neumann or Dirichlet boundary condition, as well as of local operators of the CFT$_{d-1}$. In order to have perturbative control over the resulting RG flow, we will assume that the operators $\widehat{\sigma}_I$ have scaling dimensions 
\begin{align}\label{eq:marginality}
\widehat{\Delta}_{I}=d-1-\epsilon_I,\quad 0< \epsilon_I \ll 1~,
\end{align}
i.e. the deformations are weakly relevant. Then one can systematically expand observables of the BCFT at the putative IR fixed point as a series in $\epsilon_I$. 

We will further assume that the boundary degrees of freedom are local. Technically, this means that in the absence of bulk-boundary couplings, i.e.  for $g_I = 0$, the spectrum of the CFT$_{d-1}$ contains a stress tensor $\widehat{t}_{ab}$, which is a conserved, spin 2 primary operator, of protected dimension $\widehat{\Delta}_{\widehat{t}}=3$. At the perturbative BCFT fixed point this operator gets a small anomalous dimension, which must be non-negative by unitarity, and actually strictly positive if the bulk and the boundary are not decoupled. We refer to this spin 2 operator at the interacting fixed point as `pseudo stress tensor'. In the next subsection we show how to compute the leading order contribution to this anomalous dimension for a rather generic interaction of the form \eqref{4d3dexamplegen2}, using multiplet recombination. We will then consider more specific examples for the perturbation, and compute the leading order corrections to the observables $a_{\phi^2}$ and $C_\text{D}$, defined in eqns. \eqref{one_point} and \eqref{defCD}, respectively. 

Typically, when computing (B)CFT observables in perturbation theory, one first computes the corrections as a function of the coupling constants, and then plugs the value of the coupling constants at the fixed point, obtained by solving for the zeroes of the beta functions. However, by restricting to the case with a single bulk-boundary coupling, we can also avoid the computation of the beta function and simply assume that a perturbative fixed point exists. This is sufficient because we can consider ratios of the leading order corrections to the observables mentioned above, in such a way that the coupling cancels out from the ratios. It would be interesting, but much more laborious, to actually compute the beta functions in terms of the data of the CFT$_{d-1}$. This would actually be necessary if one wants to verify the existence of the fixed point, or consider higher order corrections/multiple bulk-boundary couplings. The beta function needed in this setup starts at cubic order in the coupling, and the coefficient of the cubic term is given by a regularized integral of the four-point function of the deformation, see e.g. \cite{Komargodski:2016auf,Behan:2017emf} and also \cite{Behan:2017mwi} for the case of 1d CFTs.\footnote{The computation of the beta functions for bulk-boundary couplings in terms of the data of the CFT$_{d-1}$ was performed in \cite{Stanic} for some examples of perturbations around Dirichlet and Neumann. Some perturbative constructions of interacting boundary conditions for free theories can also be found in \cite{Herzog:2017xha,Prochazka:2019fah,Giombi:2019enr}.}

\subsubsection{Anomalous dimension of the pseudo stress tensor}\label{pseudoStress}

We now consider a slightly more specific bulk-to-boundary interaction, with a single coupling, of the form
\begin{align}\label{4d3dexamplegen}
S_{\partial}=S_{\text{CFT}_{d-1}} +g\int_{y=0} d^{d-1} \vec{x}~ \widehat{\Omega} \, \widehat{\chi}~.
\end{align}
In the expression above, $\widehat{\chi}$ denotes an operator in the CFT$_{d-1}$ and $\widehat{\Omega}$ is any local boundary operator built out of $\phi\vert_{y=0}$ or $\partial_y\phi\vert_{y=0}$, depending on whether we are perturbing a Neumann or Dirichlet free boundary condition. The assumption \eqref{eq:marginality} in this case takes the form
\begin{align}
\widehat{\Delta}_{\widehat{\Omega}}+\widehat{\Delta}_{\widehat{\chi}}=d-1-\epsilon,\quad 0< \epsilon \ll 1~.
\end{align}
In the presence of the interaction \eqref{4d3dexamplegen} the conservation and the tracelessness of the stress tensor $\widehat{t}_{ab}$ of the CFT$_{d-1}$ is violated as follows
\begin{align}\label{Wardid0}
\partial^a\widehat{t}_{ab}=g\,\widehat{\Omega}\, \partial_b \widehat{\chi}~,\quad
\widehat{t}_{a}^a=g\,\widehat{\Delta}_{\widehat{\chi}}\widehat{\chi}~.
\end{align}
Assuming a nearby fixed point with $g^2\propto \epsilon$,\footnote{This is the correct scaling with $\epsilon$ if the three-point function of the operator $\widehat{\Omega}$ vanishes for the free boundary conditions, as in the examples we will consider below. One can also consider cases in which the three-point function of $\widehat{\Omega}$ is non-vanishing, e.g. $\widehat{\Omega} = \phi^2\vert_{y=0}$ for a perturbation of Neumann, in which case $g\propto \epsilon$ at the fixed point. In any case the precise scaling does not affect any result in this section.} we have two seemingly problematic features in the above equations, namely the divergence is not expressed in terms of a primary operator of the undeformed theory, and the operator does not have only a spin 2 component because the trace is non-zero. Both these issues are solved by defining the `corrected' operator \begin{align}\label{improvedstress}
\widehat{\tau}_{ab}=\widehat{t}_{ab}-\frac{g\,\widehat{\Delta}_{\widehat{\chi}}}{(d-1)}\,\delta_{ab}\,\widehat{\Omega}\,\widehat{\chi}~.
\end{align}
Taking the divergence we then obtain
\begin{align}\label{multipletrec}
\partial^a\widehat{\tau}_{ab}=\frac{g}{ d-1}\left( (d-1-\widehat{\Delta}_{\widehat{\chi}}) \widehat{\Omega} \,\partial_b \widehat{\chi}-\widehat{\Delta}_{\widehat{\chi}}\,\widehat{\chi}\partial_b \widehat{\Omega}\right)~.
\end{align}
The new operator $\widehat{\tau}_{ab}$ is a symmetric traceless tensor, and its divergence \eqref{multipletrec} is a primary spin 1 operator of the undeformed theory, making the recombination of the multiplets manifest. Note that \eqref{multipletrec} is a manifestation in perturbation theory of the locality condition that we discussed in \ref{ss:displ_flux}. If the boundary degrees of freedom were non-local they would not have the operator $\widehat{t}_{ab}$ and then the right hand side of \eqref{multipletrec} would be a primary operator of spin 1 and protected dimension $d$ (it is easily checked that indeed this operator would appear in the bulk-to-boundary OPE of the bulk stress tensor).

We can exploit the recombination to compute the leading order anomalous dimension of $\widehat{\tau}_{ab}$ at the interacting fixed point. Let us consider computing the two-point function
\begin{align}\label{dtau2pt}
\langle \partial^a\widehat{\tau}_{ab}(\vec{x})\,\partial^c\widehat{\tau}_{cd}{(0)}\rangle~.
\end{align}
On the one hand, we can take derivatives of the two-point function of $\widehat{\tau}_{ab}$, which is fixed by boundary conformal invariance to be \cite{Osborn:1993cr}
\begin{align}\label{eq:tautau}
\langle \widehat{\tau}_{ab}(\vec{x})\widehat{\tau}_{cd}(0)\rangle & ={C_{\widehat\tau}(g)}{}\frac{I^{ab,cd}(\vec{x})}{|\vec{x}|^{2\widehat{\Delta}_{\widehat\tau}(g)}}~,\nonumber\\
I^{ab,cd}(\vec{x})\equiv \frac{1}{2}[I^{ac}(\vec{x})I^{bd}(\vec{x}) & +I^{ad}(\vec{x})I^{bc}(\vec{x})]-\frac{1}{d-1}\delta^{ab}\delta^{cd}~.
\end{align}
The definition of $I^{ab}$ was given in \eqref{parjpoly}, and we introduced
\begin{align}
\begin{split}
C_{\widehat\tau}(g)&=C_{\widehat\tau}^{(0)}+O(g^2)~,\nonumber\\
\widehat{\Delta}_{\widehat\tau}(g) = d-1+\widehat{\gamma}_{\widehat\tau}(g) & = d-1 +\widehat{\gamma}^{(1)}_{\widehat\tau} g^2 +O(g^4) ~.
\end{split}
\end{align}
The constant $C_{\widehat\tau}^{(0)}$ is the `central charge' of the CFT$_{d-1}$ that the bulk scalar couples to, i.e. the coefficient appearing the two-point function of the stress tensor $t_{ab}$ before we turn on the interaction. On the other hand we can compute \eqref{dtau2pt} at the leading order in $g$ by directly using the r.h.s. of \eqref{multipletrec}. By comparing the two results, we find
\begin{align}\label{anomstressdgen}
\widehat{\gamma}^{(1)}_{\widehat\tau} = 2\,\frac{\widehat{\Delta}_{\widehat{\chi}}(d-1-\widehat{\Delta}_{\widehat{\chi}})}{(d+1)(d-2)}\frac{C_{\widehat{\Omega}}^{(0)}C_{\widehat\chi}^{(0)}}{C_{\widehat\tau}^{(0)}}~,
\end{align}
where $C_{\widehat{O}}^{(0)}$ denotes the coefficient of the two-point function of the boundary operator $\widehat{O}$ computed at $g=0$. With the canonical normalization \eqref{eq:freebulkaction} of the bulk action we have
\begin{align}\label{canonicalNormal}
C_{\phi}^{(0)}=\frac{\Gamma{\left(\frac{d}{2}-1\right)}}{2\pi^{\frac{d}{2}}}~,\quad
C_{\partial_y\phi }^{(0)}=\frac{\Gamma{\left(\frac{d}{2}\right)}}{\pi^{\frac{d}{2}}}~.
\end{align}
We note that the leading order anomalous dimension is essentially controlled by the central charge $C_{\widehat\tau}^{(0)}$ of the CFT$_{d-1}$.

\subsubsection{Modified Dirichlet boundary conditions and perturbation theory}\label{sec:Dpert}
We now further specialize to the case in which the free boundary condition is Dirichlet, and the operator $\widehat{\Omega}$ is $\partial_y\phi\vert_{y=0}$, namely we take a deformation of the form
\begin{align}\label{4d3dexampleDirichlet}
S_{\partial}^{(D)}=S_{\text{CFT}_{d-1}} +g\int_{y=0} d^{d-1} \vec{x}~ \partial_y\phi \,\widehat{\chi}~.
\end{align}
The interaction term leads to the the following modified Dirichlet boundary condition\footnote{This can be obtained by varying the action \eqref{4d3dexampleDirichlet}, supplemented by the boundary term $\int  \phi \partial_y\phi$, with respect to $\partial_y \phi$.}
\begin{align}\label{modifD}
 \phi\vert_{y=0}=-g\,\widehat{\chi}~.
\end{align}
In this case the condition \eqref{eq:marginality} gives $\widehat{\Delta}_{\widehat{\chi}}=\frac{d}{2}-1-\epsilon$, with $0< \epsilon \ll 1$. As we discussed above, we assume the existence of a perturbative fixed point with $g^2\propto \epsilon$. Plugging in eq. \eqref{anomstressdgen} we obtain
\begin{align}\label{anomstressdD}
\widehat{\gamma}^{(1)}_{\widehat\tau} =\frac{\Gamma\left(\frac{d}{2}+1\right)}{\pi^{\frac{d}{2}}(d+1)}\frac{C_{\widehat\chi}^{(0)}}{C_{\widehat\tau}^{(0)}}~.
\end{align}

Let us now consider the leading order correction to the one-point function coefficient $a_{\phi^2}(g)$
\begin{align}
a_{\phi^2}(g)=-{2^{2-d}}+\delta a_{\phi^2}(g^2) = -{2^{2-d}}+\delta a^{(1)}_{\phi^2}g^2 + O(g^4)~.
\end{align}
The coefficient $\delta a^{(1)}_{\phi^2}$ must be non-negative as a consequence of the unitarity bound \eqref{aphi2_range}. 
To compute its value, note that the modified Dirichlet boundary condition \eqref{modifD} determines the
bOPE coefficient $b_1$ to be\footnote{This formula is simply obtained by appropriately normalizing the operators involved, namely $\widehat{O}_1$ needs to have unit-normalized two-point function and the bulk scalar field needs to have unit-normalized contribution of the identity in the bulk OPE, see also footnote \ref{footnotenormalization}.}
\begin{equation}
b_1 = - g \sqrt{\frac{4 \pi^{\frac{d}{2}}}{\Gamma\left(\frac{d}{2}-1\right)}C^{(0)}_{\widehat{\chi}}}\left(1+ O(g^2)\right) ~.
\end{equation} 
Plugging this result in the crossing relations \eqref{b1b2def}, we find
\begin{equation}\label{eq:aphicorrD}
\delta a^{(1)}_{\phi^2} = \frac{2^{4-d} \pi^{\frac{d}{2}}C^{(0)}_{\widehat{\chi}}}{\Gamma\left(\frac{d}{2}-1\right)}~.
\end{equation}
Having obtained two observables we can form a ratio that does not depend on the value of the coupling at the putative fixed point, namely
\begin{equation}\label{eq:gammatovaD}
\frac{\widehat{\gamma}_{\widehat{\tau}}(g)}{\delta a_{\phi^2}(g)} = \frac{\widehat{\gamma}^{(1)}_{\widehat{\tau}}}{\delta a^{(1)}_{\phi^2}} + O(g^2) = \frac{2^{d-4}\Gamma\left(\frac{d}{2}-1\right)\Gamma\left(\frac{d}{2}+1\right)}{ \pi^d(d+1)}\frac{1}{C_{\widehat\tau}^{(0)}} + O(g^2)~.
\end{equation}
This quantity depends on the central charge $C_{\widehat\tau}^{(0)}$ of the CFT$_{d-1}$ that the bulk scalar couples to.

Next, we consider the leading order correction to the coefficient $C_{\text{D}}$ in the two-point function of the displacement operator
\begin{equation}
C_{\text{D}} = C^{(D)}_{\text{D}} + \delta C_{\text{D}}(g) = C^{(D)}_{\text{D}} + \delta C^{(1)}_{\text{D}} g^2 +O(g^4)~, 
\end{equation}
where $C^{(D)}_{\text{D}}$ denotes the value at the free Dirichlet boundary condition. The displacement operator in this theory is \cite{McAvity:1993ue,McAvity:1995zd}
\begin{align}\label{GenericDisp}
\text{D}=\left.\left(\frac{1}{2}(\partial_y \phi)^2-\frac{1}{2}(\partial_a \phi)^2+\frac{1}{4}\frac{d-2}{d-1}\partial_a^2 \phi^2\right)\right\vert_{y=0}~.
\end{align}
Note that this formula makes sense even for an interacting boundary condition if we interpret the composite operators on the right hand side as products of $\phi\vert_{y=0}$ and $\partial_y\phi\vert_{y=0}$, made finite by subtracting all the singular terms in the OPE. This is because $\text{D}(\vec x) = \lim_{y\to0} T^{yy}(\vec x, y)$ and the bulk operator $T^{yy}$ is always equal to \eqref{GenericDisp}. (The scaling dimension of $\text{D}$ is guarantueed to come out correctly because $\phi\vert_{y=0}$ and $\partial_y\phi\vert_{y=0}$ have protected dimensions.) This observation allows us to easily compute the two-point function of $\text{D}$ in conformal perturbation theory for the modified Dirichlet condition \eqref{modifD}. We find that the contributions from $\phi\vert_{y=0}$ are $O(g^4)$ whereas the two-point function of $\partial_y\phi\vert_{y=0}$ is corrected already at $O(g^2)$ by the interaction term \eqref{4d3dexampleDirichlet} and is given by:
\begin{align}
\begin{split}
& \langle \partial_y\phi(0,\vec{x})\partial_y\phi(0,\vec{x}')\rangle  \\& = \frac{\Gamma{\left(\frac{d}{2}\right)}}{\pi^{\frac{d}{2}}}\frac{1}{|\vec{x}-\vec{x}'|^d }+ g^2 \left(\frac{\Gamma{\left(\frac{d}{2}\right)}}{\pi^{\frac{d}{2}}}\right)^2 \int \frac{d^{d-1} \vec{u}}{|\vec{x}-\vec{u}|^d}  \int \frac{d^{d-1} \vec{u}'}{|\vec{x}' - \vec{u}'|^d} \frac{C^{(0)}_{\widehat{\chi}}}{|\vec{u}-\vec{u}'|^{d-2}} + O(g^4)\\
& = \frac{\Gamma{\left(\frac{d}{2}\right)}}{\pi^{\frac{d}{2}}}\frac{1}{|\vec{x}-\vec{x}'|^d }\left(1-g^2 \frac{2 \pi^{\frac{d}{2}}C^{(0)}_{\widehat{\chi}}}{\Gamma\left(\frac{d}{2}-1\right)}\right) + O(g^4)~.\label{eq:correction2ptdphi}
\end{split}
\end{align}
Note that the integrals have a power-law UV divergence for $\vec{u}\sim \vec{x}$ and $\vec{u}'\sim \vec{x}'$ that we subtracted. As a check, the result \eqref{eq:correction2ptdphi} implies
\begin{equation}
b_2 = \sqrt{2(d-2)}\left(1-g^2 \frac{\pi^{\frac{d}{2}}C^{(0)}_{\widehat{\chi}}}{\Gamma\left(\frac{d}{2}-1\right)}\right)+O(g^4) ~, 
\end{equation}
which is in perfect agreement with the correction \eqref{eq:aphicorrD} that we computed for $a_{\phi^2}$ and the crossing relations \eqref{b1b2def}. Using \eqref{eq:correction2ptdphi} to compute the two-point function of $\left.\frac 12 (\partial_y\phi)^2\right\vert_{y=0}$ and therefore of D, we obtain
\begin{equation}
C^{(D)}_{\text{D}} = \frac{\Gamma\left(\frac{d}{2}\right)^2}{2\pi^d}~,~~\delta C^{(1)}_{\text{D}} = -\frac{(d-2)\Gamma\left(\frac{d}{2}\right)C^{(0)}_{\widehat{\chi}}}{\pi^{\frac{d}{2}}}~.\label{eq:CDD}
\end{equation}
We can then form another ratio of observables that is independent of the coupling at the putative perturbative fixed point
\begin{equation}\label{eq:CDovaD}
\frac{\delta C_{\text{D}}(g)}{\delta a_{\phi^2}(g)} = \frac{\delta C^{(1)}_{\text{D}}}{\delta a^{(1)}_{\phi^2}} + O(g^2) = -\frac{2^{d-3}\Gamma\left(\frac{d}{2}\right)^2}{\pi^d} + O(g^2)~.
\end{equation}
Note that this ratio does not depend on any data of the CFT$_{d-1}$ and therefore it is a universal result for deformations of the form \eqref{4d3dexampleDirichlet} of the Dirichlet boundary condition.

\subsubsection{Modified Neumann boundary conditions and perturbation theory}
As a final example, we consider deformations of the Neumann free boundary condition by the following interaction 
\begin{align}\label{4d3dexampleN}
S_{\partial}^{(N)}=S_{\text{CFT}_{d-1}} +g\int_{y=0} d^{d-1} \vec{x}~ \phi\,\widehat{\chi}~.
\end{align}
The interaction gives rise to the following modified Neumann boundary condition
\begin{align}\label{modifN}
\partial_y \phi\vert_{y=0}=g\,\widehat{\chi}~.
\end{align}
The condition \eqref{eq:marginality} now gives $\widehat{\Delta}_{\widehat{\chi}}=\frac{d}{2}-\epsilon$, with $0< \epsilon \ll 1$, and again we will assume the existence of a perturbative fixed point with $g^2 \propto \epsilon$. Plugging in eq. \eqref{anomstressdgen} we obtain
\begin{align}\label{anomstressdN}
\widehat{\gamma}^{(1)}_{\widehat\tau} =\frac{\Gamma\left(\frac{d}{2}-1\right)}{4\pi^{\frac{d}{2}}}\frac{d}{d+1}\frac{C_{\widehat\chi}^{(0)}}{C_{\widehat\tau}^{(0)}}~.
\end{align}

To compute the variation of the parameter $a_{\phi^2}$ we use the same strategy as in the previous example, namely it follows from the modified Neumann condition that
\begin{equation}
b_2 = g \sqrt{\frac{4 \pi^{\frac{d}{2}}}{\Gamma\left(\frac{d}{2}-1\right)}C^{(0)}_{\widehat{\chi}}}\left(1+ O(g^2)\right) ~,
\end{equation}
and using the crossing relations \eqref{b1b2def} this gives
\begin{equation}\label{eq:aphicorrN}
\delta a^{(1)}_{\phi^2} = -\frac{2^{3-d} \pi^{\frac{d}{2}}C^{(0)}_{\widehat{\chi}}}{\Gamma\left(\frac{d}{2}\right)}~.
\end{equation}
Note that this has an opposite sign compared to eq. \eqref{eq:aphicorrD}, in agreement with the unitarity bounds \eqref{aphi2_range}. The coupling-independent ratio then is
\begin{equation}
\frac{\widehat{\gamma}_{\widehat{\tau}}(g)}{\delta a_{\phi^2}(g)} = \frac{\widehat{\gamma}^{(1)}_{\widehat{\tau}}}{\delta a^{(1)}_{\phi^2}} + O(g^2) = -\frac{2^{d-4}\Gamma\left(\frac{d}{2}-1\right)\Gamma\left(\frac{d}{2}+1\right)}{ \pi^d(d+1)}\frac{1}{C_{\widehat\tau}^{(0)}} + O(g^2)~.
\end{equation}

Like in the previous example we now compute the correction to $C_{\text{D}}$, again using the definition \eqref{GenericDisp} as the starting point. The main difference is that in this case the leading-order correction comes from the second and third terms in eq. \eqref{GenericDisp}, namely those involving $\phi\vert_{y=0}$, while the first term involving $\partial_y\phi\vert_{y=0}$ only starts contributing at subleading order $O(g^4)$. We will then only need the two-point function of $\phi\vert_{y=0}$ up to $O(g^2)$ corrections, that is
\begin{align}
\begin{split}
& \langle \phi(0,\vec{x}) \phi(0,\vec{x}')\rangle  \\& \!\!= \frac{\Gamma{\left(\frac{d}{2}-1\right)}}{2\pi^{\frac{d}{2}}}\frac{1}{|\vec{x}-\vec{x}'|^{d-2} }+ g^2 \left(\frac{\Gamma{\left(\frac{d}{2}-1\right)}}{2\pi^{\frac{d}{2}}}\right)^2 \int \frac{d^{d-1} \vec{u}}{|\vec{x}-\vec{u}|^{d-2}}  \int \frac{d^{d-1} \vec{u}'}{|\vec{x}' - \vec{u}'|^{d-2}} \frac{C^{(0)}_{\widehat{\chi}}}{|\vec{u}-\vec{u}'|^d} + O(g^4)\\
& \!\!= \frac{\Gamma{\left(\frac{d}{2}-1\right)}}{2\pi^{\frac{d}{2}}}\frac{1}{|\vec{x}-\vec{x}'|^d }\left(1-g^2 \frac{ \pi^{\frac{d}{2}}C^{(0)}_{\widehat{\chi}}}{\Gamma\left(\frac{d}{2}\right)}\right) + O(g^4)~.\label{eq:correction2ptphi}
\end{split}
\end{align}
The integrals have a power-law UV divergence for $\vec{u}\sim \vec{u}'$ that we subtracted. As a check, from \eqref{eq:correction2ptdphi} we obtain
\begin{equation}
b_1 = \sqrt{2}\left(1-g^2 \frac{ \pi^{\frac{d}{2}}C^{(0)}_{\widehat{\chi}}}{2\Gamma\left(\frac{d}{2}\right)}\right) +O(g^4)~,
\end{equation}
which, upon substitution in the crossing equations \eqref{b1b2def}, gives a correction to $a_{\phi^2}$ in agreement with \eqref{eq:aphicorrN}. Using \eqref{eq:correction2ptphi} we obtain
\begin{equation}
C^{(N)}_{\text{D}} = \frac{\Gamma\left(\frac{d}{2}\right)^2}{2\pi^d}~,~~\delta C^{(1)}_{\text{D}} = -\frac{\Gamma\left(\frac{d}{2}\right)C^{(0)}_{\widehat{\chi}}}{\pi^{\frac{d}{2}}}~.
\end{equation}
Comparing with \eqref{eq:CDD} we see that the value at the free boundary condition is the same for Neumann and Dirichlet, while the leading correction differs by a factor of $d-2$. Taking the ratio with $\delta a_\phi^2$ we get
\begin{equation}
\frac{\delta C_{\text{D}}(g)}{\delta a_{\phi^2}(g)} = \frac{\delta C^{(1)}_{\text{D}}}{\delta a^{(1)}_{\phi^2}} + O(g^2) = -\frac{2^{d-3}\Gamma\left(\frac{d}{2}\right)^2}{\pi^d} + O(g^2)~, \label{eq:CDovaN}
\end{equation}
which notably is the same as the one obtained for the deformation of Dirichlet in eq. \eqref{eq:CDovaD}. Like in that example, this ratio is universal for deformations of the form \eqref{4d3dexampleN} of the Neumann boundary condition, because it does not depend on data of the CFT$_{d-1}$.

\section{Bootstrapping boundary conditions for a free scalar}
\label{sec:crossingeq}

\subsection{Crossing equations}\label{ssec:crossingeq}
In this section we present the crossing equation for the mixed system of four-point functions of the boundary modes of $\phi$, namely
\begin{align}
\langle \widehat{{O}}_i(\vec{x}_1)\widehat{{O}}_j(\vec{x}_2)\widehat{{O}}_m(\vec{x}_3) \widehat{{O}}_n(\vec{x}_4) \rangle.
\end{align}
The crossing equations for a generic mixed system of scalars, labelled by indices $i,j,m,n$, were derived in \cite{Kos:2014bka} and read
\begin{align}\label{crossEq}
\sum_{\widehat{\mathcal{O}}^{(l)}}[{}\hat{f}_{ij\widehat{\mathcal{O}}^{(l)}}{\hat{f}}_{mn}{}^{\widehat{\mathcal{O}}^{(l)}}F_{\mp,\widehat{\Delta},l}^{ij,mn}(u,v)\pm \hat{f}_{mj}{}^{\widehat{\mathcal{O}}^{(l)}}\hat{f}_{in\widehat{\mathcal{O}}^{(l)}}F_{\mp,\widehat{\Delta},s}^{mj,in}(u,v)]=0,
\end{align}
where $u=\frac{x_{12}^2x_{34}^2}{x_{13}^2x_{24}^2}$ and $v=\frac{x_{23}^2x_{14}^2}{x_{13}^2x_{24}^2}$. The functions $F^{ij,mn}_{\pm,\widehat{\Delta},l}$ are the following combinations of the ordinary $s$-channel conformal blocks $g_{\widehat{\Delta},l}^{\widehat{\Delta}_{ij},\widehat{\Delta}_{mn}}$
\begin{align}\label{FpmDef}
F^{ij,mn}_{\pm,\widehat{\Delta},l}(u,v)\equiv v^{\frac{1}{2}(\widehat{\Delta}_m+\widehat{\Delta}_j)}g_{\widehat{\Delta},l}^{\widehat{\Delta}_{ij},\widehat{\Delta}_{mn}}(u,v)\pm u^{\frac{1}{2}(\widehat{\Delta}_m+\widehat{\Delta}_j)}g_{\widehat{\Delta},l}^{\widehat{\Delta}_{ij},\widehat{\Delta}_{mn}}(v,u).
\end{align}
Note that not all equations in this system \eqref{crossEq} are independent, since\footnote{Recall that \cite{Dolan:2000ut,Dolan:2011dv}
	\begin{align}
	g_{\Delta,\ell}^{\Delta_{12},\Delta_{34}}\left({u}/{v},{1}/{v}\right)=(-1)^\ell v^{\frac{\Delta_{34}}{2}}g_{\Delta,\ell}^{-\Delta_{12},\Delta_{34}}(u,v)=(-1)^\ell v^{\frac{-\Delta_{12}}{2}}g_{\Delta,\ell}^{\Delta_{12},-\Delta_{34}}(u,v),\nonumber
	\end{align}}
\begin{align}
F_{\pm,\widehat{\Delta},l}^{ij,mn}(u,v)=F_{\pm,\widehat{\Delta},l}^{mn,ij}(u,v),\quad
F_{\pm,\widehat{\Delta},l}^{ij,ij}(u,v)=F_{\pm,\widehat{\Delta},l}^{ji,ji}(u,v),\quad
F_{\pm,\widehat{\Delta},l}^{ij,kk}(u,v)=F_{\pm,\widehat{\Delta},l}^{ji,kk}(u,v).
\end{align}
Due to Bose symmetry \eqref{Bosesymm} we have the OPE selection rule
\begin{align}\label{OhiOPEs}
\hat{f}_{ij}{}^{{\widehat{\mathcal{O}}^{(l)}}} = (-1)^l \, \hat{f}_{ji}{}^{{\widehat{\mathcal{O}}^{(l)}}}~.
\end{align}
If we specialize all these ingredients to our problem where $i,j,m,n \in \{1,2\}$ then we find 7 independent crossing equations:
\begin{align}
\begin{split}\label{cross_eq_red}
0=&\sum_{\widehat{\mathcal{O}}^{(l)}}{}\hat{f}_{11}{}^{{\widehat{\mathcal{O}}^{(l)}}}\hat{f}_{11{\widehat{\mathcal{O}}^{(l)}}} F_{-,\widehat{\Delta},l}^{11,11}(u,v),\\
0=&\sum_{\widehat{\mathcal{O}}^{(l)}}{}\hat{f}_{22}{}^{{\widehat{\mathcal{O}}^{(l)}}}\hat{f}_{22{\widehat{\mathcal{O}}^{(l)}}} F_{-,\widehat{\Delta},l}^{22,22}(u,v),\\
0=&\sum_{\widehat{\mathcal{O}}^{(l)}}{}\hat{f}_{12}{}^{{\widehat{\mathcal{O}}^{(l)}}}\hat{f}_{12{\widehat{\mathcal{O}}^{(l)}}} F_{-,\widehat{\Delta},l}^{12,12}(u,v),\\
0=&\sum_{\widehat{\mathcal{O}}^{(l)}}{}\hat{f}_{11}{}^{{\widehat{\mathcal{O}}^{(l)}}}  {}\hat{f}_{12{\widehat{\mathcal{O}}^{(l)}}}  F_{-,\widehat{\Delta},l}^{11,12}(u,v),\\
0=&\sum_{\widehat{\mathcal{O}}^{(l)}}{}\hat{f}_{12}{}^{{\widehat{\mathcal{O}}^{(l)}}}  {}\hat{f}_{22{\widehat{\mathcal{O}}^{(l)}}} F_{-,\widehat{\Delta},l}^{12,22}(u,v),\\
0=&\sum_{\widehat{\mathcal{O}}^{(l)}}(-1)^l {}\hat{f}_{12}{}^{{\widehat{\mathcal{O}}^{(l)}}}\hat{f}_{12{\widehat{\mathcal{O}}^{(l)}}} F_{\mp,\widehat{\Delta},l}^{12,21}(u,v)\pm{}\hat{f}_{11}{}^{{\widehat{\mathcal{O}}^{(l)}}}  {}\hat{f}_{22{\widehat{\mathcal{O}}^{(l)}}}  F_{\mp,\widehat{\Delta},l}^{11,22}(u,v)~.
\end{split}
\end{align}
In our case the operators must also obey the OPE relations \eqref{ssspinconstr}. Imposing those, the system of equations can be rewritten as follows (details can be found in appendix \ref{app:crossing})
\begin{align}
\begin{split}\label{crossEqVecGenFinal}
0=&\quad\vec{V}_{\id}+\sum_{l= \text{even}}{}\hat{f}_{12{\widehat{\mathcal{O}}^{(l)}}}\hat{f}_{12}{}^{\widehat{\mathcal{O}}^{(l)}} \vec{V}_{+,\widehat{\Delta},l}\\
+&\sum_{\underset{n=0,1,\dots}{\underset{\widehat{\Delta}=d+l-1+2n}{l=\text{odd},\dots}}}{\hat{f}_{12{\widehat{\mathcal{O}}^{(l)}}}\hat{f}_{12}{}^{{\widehat{\mathcal{O}}^{(l)}}}}{} \vec{V}_{-,\widehat{\Delta},l}\\
+&\sum_{{\ell \in 2\mathbb{N}}{} }\,\sum_{{}{\underset{\widehat{\Delta}=d+\ell-2}{\text{even}\,\,l<\ell}}}\begin{pmatrix}
{}\hat{f}_{11{\widehat{\mathcal{O}}^{(l)}}} &
{}\hat{f}_{22{\widehat{\mathcal{O}}^{(l)}}}
\end{pmatrix} \vec{V}_{\textbf{0},\widehat{\Delta},l}\begin{pmatrix}
{}\hat{f}_{11}{}^{{\widehat{\mathcal{O}}^{(l)}}} \\
{}\hat{f}_{22}{}^{{\widehat{\mathcal{O}}^{(l)}}} \\
\end{pmatrix}~.
\end{split}
\end{align}
The quantities $\vec{V}_{\pm,\widehat{\Delta},l},\vec{V}_{\id,\widehat{\Delta},l}$ are 7-component vectors defined in \eqref{VminusDefIgen}, \eqref{VplusDefgen} and $\vec{V}_{\textbf{0},\widehat{\Delta},l}$ are vectors of $2\times 2$ matrices defined in \eqref{Vodefgen}. Beside the identity, the first line accounts for `unprotected' primaries, i.e. operators of generic scaling dimension away from the poles of the gamma functions in \eqref{ssspinconstr}. The last two lines take into account the tower of `protected' operators listed in table \ref{tab:selprottower}, which can have both odd and even spin. The indices of the OPE coefficients are contracted according to the conventions of Appendix \ref{app:conventions}.

\subsection{Implementing the exact relations}
We will numerically `bootstrap' a set of crossing equation in the sense of \cite{Rattazzi:2008pe}. For most problems, the fastest program available for this task is the semidefinite program solver \texttt{SDPB} \cite{SimmonsDuffin:2015}. We have used the recent version \cite{Landry:2019} which supports the `hotstarting' algorithm suggested in \cite{Go:2019}. In order to use \texttt{SDPB} efficiently, the ingredients of the crossing equations must be approximated as rational functions of the scaling dimension such that all poles of odd order lie at or below the unitarity bound. This is always possible when the basis functions are conformal blocks and we refer the reader to \cite{Kos:2014bka} for details. The complication in this work is that the vector $\vec{V}_{+,\widehat{\Delta},l}$ in \eqref{crossEqVecGenFinal} has several occurences of $\widehat{\Delta}$ which are not in conformal blocks.

As shown in Appendix \ref{app:crossing}, we must consider linear combinations in which the coefficients are various products of $\kappa_1(\widehat{\Delta},l)$ and $\kappa_2(\widehat{\Delta},l)$ -- the functions from \eqref{ssspinconstr}. While these types of blocks were first introduced for studying the long-range Ising model, in the numerical analysis of \cite{Behan:2018hfx} only the last two components of \eqref{VplusDefgen} were used. In addition, $\kappa_1(\widehat{\Delta},l)\kappa_2(\widehat{\Delta},l)$ was treated as a vector with $> 1000$ discrete evaluations, thereby eschewing some of the benefits of semidefinite programming. In this work, we do not need to limit ourselves to those crossing equations that involve only the product $\kappa_1(\widehat{\Delta},l)\kappa_2(\widehat{\Delta},l)$ where $b_1$ and $b_2$ cancel out. Thanks to \eqref{b1b2def}, the other products only introduce one new parameter and it has a clear physical meaning in the BCFT context. To remedy the second problem, we need to discuss the approximation theory of gamma functions.

To start, it is easily verified that
\begin{align}
\begin{split}\label{approx1}
\kappa_1(\widehat{\Delta},l) &= -\frac{b_2}{2b_1} \kappa(\widehat{\Delta},l)  \\
\kappa_2(\widehat{\Delta},l) &= -\frac{b_1}{2b_2} (\widehat{\Delta} + l - 1) (d + l - \widehat{\Delta} - 2) \kappa(\widehat{\Delta},l) 
\end{split}
\end{align}
with
\begin{equation}
\kappa(\widehat{\Delta},l) = \frac{\Gamma \left(\frac{l+\widehat{\Delta}}{2}\right) \Gamma \left(\frac{d+l-\widehat{\Delta}-2}{2} \right)}{\Gamma\left(\frac{d+l-\widehat{\Delta}-1}{2} \right)\Gamma \left(\frac{l+\widehat{\Delta}+1}{2}\right)}. \label{approx2}
\end{equation}
As such, a rational approximation\footnote{With infinitely many poles above the unitarity bound it is clear that any rational approximation for $\kappa(\widehat{\Delta},l)^2$ is going to have significant errors in the semi-infinite range of allowed values for $\widehat{\Delta}$. Extremely large values of $\widehat{\Delta}$ should however be unimportant for the numerical results, and for a finite window of values a rational approximation is perfectly feasible. We have attempted to account for this in practice by using the simpler crossing equations \eqref{cross_eq_red} to cover the `tail' of the more constraining crossing equations. Each time we approximate $\kappa(\widehat{\Delta},l)^2$ with $\widehat{\Delta} \in [\Delta_0, \infty)$, we allow these extra conformal blocks, multiplying independent $f_{ij\widehat{\mathcal{O}}}$ coefficients, to have an exchanged scaling dimension in $[\Delta_0 + 20, \infty)$.} for $\kappa(\widehat{\Delta},l)^2$ will cover the cases of $\kappa_1(\widehat{\Delta},l)^2$, $\kappa_2(\widehat{\Delta},l)^2$ and $\kappa_1(\widehat{\Delta},l) \kappa_2(\widehat{\Delta},l)$. The most expensive step for our purposes will be the Weierstrass formula
\begin{equation}
\Gamma(z) = \frac{e^{-\gamma z}}{z} \prod_{k = 1}^{\infty} \left ( 1 + \frac{z}{k} \right )^{-1} e^{z / k} \label{asymptotic1}
\end{equation}
which introduces a new series of poles for each gamma function. Unlike \cite{Behan:2018hfx}, which suggested using \eqref{asymptotic1} on the full function, we will only apply it to the $-\widehat{\Delta}$ part of $\kappa(\widehat{\Delta},l)^2$.

The $+\widehat{\Delta}$ part, since it is regular, should be approximated with one of the many expressions for the Wallis ratio. This is a quantity which has attracted interest for hundreds of years due to the application of calculating $\pi$. In particular, we note the asymptotic formula
\begin{equation}
\frac{\Gamma(z + 1)}{\Gamma \left ( z + \frac{1}{2} \right )} \lesssim \sqrt{z + \frac{1}{4} + \frac{1}{32z + 8}} \; , \; z \rightarrow \infty. \label{asymptotic2}
\end{equation}
It was found in \cite{Mortici:2010} that \eqref{asymptotic2} is the $n = 1$ case in a sequence of approximants that have the schematic form $\left ( z^n + \dots \right )^{\frac{1}{2n}}$. We cannot use these higher radicals due to the requirement that $\kappa(\widehat{\Delta},l)^2$ be a rational function but it is still possible to make \eqref{asymptotic2} arbitrarily accurate. One simply applies the functional equation $n$ times to arrive at
\begin{equation}
\frac{\Gamma(z + 1)}{\Gamma \left ( z + \frac{1}{2} \right )} \lesssim \frac{\left ( z + \frac{1}{2} \right )_n}{(z + 1)_n} \sqrt{z + n + \frac{1}{4} + \frac{1}{32z + 32n + 8}} \; , \; z \rightarrow \infty. \label{asymptotic3}
\end{equation}
We have not found it necessary to choose a large value of $n$. For example, even when $z = \frac{1}{4}$, Weierstrass does not become better than \eqref{asymptotic2} until $k = 36$.

We will now quote expressions for the two factors of $\kappa(\widehat{\Delta},l)^2$. Using \eqref{asymptotic3} with $n = 1$,
\begin{align}
\begin{split}\label{apReg}
\frac{\Gamma \left ( \frac{\widehat{\Delta} + l}{2} \right )^2}{\Gamma \left ( \frac{\widehat{\Delta} + l + 1}{2} \right )^2} &= \left [ \frac{2(\widehat{\Delta} + l + 1)}{(\widehat{\Delta} + l)(\widehat{\Delta} + l + 2)} \frac{\Gamma \left ( \frac{\widehat{\Delta} + l + 4}{2} \right )}{\Gamma \left ( \frac{\widehat{\Delta} + l + 3}{2} \right )} \right ]^2  \\
&\approx \left [ \frac{2(\widehat{\Delta} + l + 1)}{(\widehat{\Delta} + l)(\widehat{\Delta} + l + 2)} \right ]^2 \frac{8\widehat{\Delta}^2 + 8(2l + 5)\widehat{\Delta} + 8l^2 + 40l + 51}{8(2\widehat{\Delta} + 2l + 5)}~.
\end{split} 
\end{align}
The singular part requires a cutoff which we call $k_{\mathrm{max}}$.
\begin{align}
\begin{split}\label{apSing}
\frac{\Gamma \left ( \frac{d + l - \widehat{\Delta} - 2}{2} \right )^2}{\Gamma \left ( \frac{d + l - \widehat{\Delta} - 1}{2} \right )^2} &\approx e^{\gamma} \prod_{k = 1}^{k_{\mathrm{max}}} e^{\frac{1}{k}} \left [ \prod_{k = 0}^{k_{\mathrm{max}}} \frac{\widehat{\Delta} - d - l - 2k + 1}{\widehat{\Delta} - d - l - 2k + 2} \right ]^2 \\
&\approx \frac{1}{k_{\mathrm{max}}} \prod_{k = 0}^{k_{\mathrm{max}}} \frac{(\widehat{\Delta} - d - l - 2k + 1)^2}{(\widehat{\Delta} - d - l - 2k + 2)^2}~.
\end{split}
\end{align}
While it is optional to resum the exponent in the second step, the logarithmic behaviour of the harmonic series makes it convenient.

The expressions \eqref{apReg} and \eqref{apSing} have been implemented as a patch for the helper program \texttt{PyCFTBoot} \cite{Behan:2016}. For the poles exhibited here, which are two units apart, we have taken $k_{\mathrm{max}} = 20$. The poles coming from conformal blocks \cite{Kos:2014bka} are only one unit apart so we take $k_{\mathrm{max}} = 40$ for those. The standard way to account for poles in $\widehat{\Delta}$ is to absorb them into the OPE coefficients of \eqref{crossEqVecGenFinal} so that crossing symmetry becomes a statement about polynomials. The most desirable type of problem for \texttt{SDPB} is one in which these polynomials can be expressed in terms of an orthonormal basis \cite{SimmonsDuffin:2015}. Recently, \cite{Go:2020} gave an example of a problem which cannot be optimized in this way. In our case, this privileged basis of polynomials is again unavailable due to the 20 double poles of \eqref{apSing} that are above the unitarity bound. For this reason, we have opted to still use the simple crossing equations \eqref{cross_eq_red} for spins above a certain cutoff $l_0$. For most of the bounds in the next section, this is $l_0 = 4$ while some of them have been redone with $l_0 = 6$. Seeing almost no difference, we conclude that the exact relations for $l = 0$ and $l = 2$ are doing most of the work.

The other limitation of our approach is that the square root in \eqref{asymptotic3} can only be eliminated when the $\kappa_i(\widehat{\Delta},l)$ appear quadratically. This forces us to drop $\langle \widehat{O}_1\widehat{O}_1\widehat{O}_1\widehat{O}_2 \rangle$ which is linear in $\kappa_1(\widehat{\Delta},l)$ and $\langle \widehat{O}_2\widehat{O}_2\widehat{O}_2\widehat{O}_1 \rangle$ which is linear in $\kappa_2(\widehat{\Delta},l)$. According to standard lore, the bounds should be unaffected as these two correlators exchange the same operator families as the other three.

\section{Numerical results for $4d/3d$ systems}
\label{sec:numerics}

Let us collect the constraints used for the numerical bootstrap analysis. First, we take the crossing equations for $\vev{\widehat{O}_1\widehat{O}_1\widehat{O}_1\widehat{O}_1}$, for $\vev{\widehat{O}_1\widehat{O}_1\widehat{O}_2\widehat{O}_2}$ and $\vev{\widehat{O}_2\widehat{O}_2\widehat{O}_2\widehat{O}_2}$ given in \eqref{cross_eq_red}. These apply to any (possibly non-local) CFT containing the operators $\widehat{O}_1$ and $\widehat{O}_2$. Second, we have the exact OPE relations  \eqref{apReg} and \eqref{apSing} which are necessary for a solution of these crossing equations to be an admissible boundary condition for a massless free scalar in $d$ dimensions. These conditions reduce the crossing equations to equation \eqref{crossEqVecGenFinal} at the cost of introducing a new parameter, $a_{\phi^2}$, that our bounds will depend on. Notice that this in particular implies that the odd-spin operators can only have the scaling dimensions of the generalized free theory. Third, the boundary spectrum cannot have a stress tensor, so it is natural to demand that the first spin 2 operator has a dimension $\widehat{\Delta}_{\widehat{\tau}}$ strictly larger than $d-1$. Fourth, we should demand that the flux operator $\text{V}^{(1)}$ of dimension $d$ is absent to avoid interfaces and other possible sources of non-locality on the boundary, see the discussion in \ref{ss:displ_flux}. Fifth, we have the Ward identities for the displacement operator \eqref{displRel} which restrict $\hat{f}_{11\text{D}}$ and $\hat{f}_{22\text{D}}$ to a curve parametrized by $C_{\text{D}}$.

\begin{figure}[h!]
\centering
\includegraphics[width=0.8\textwidth]{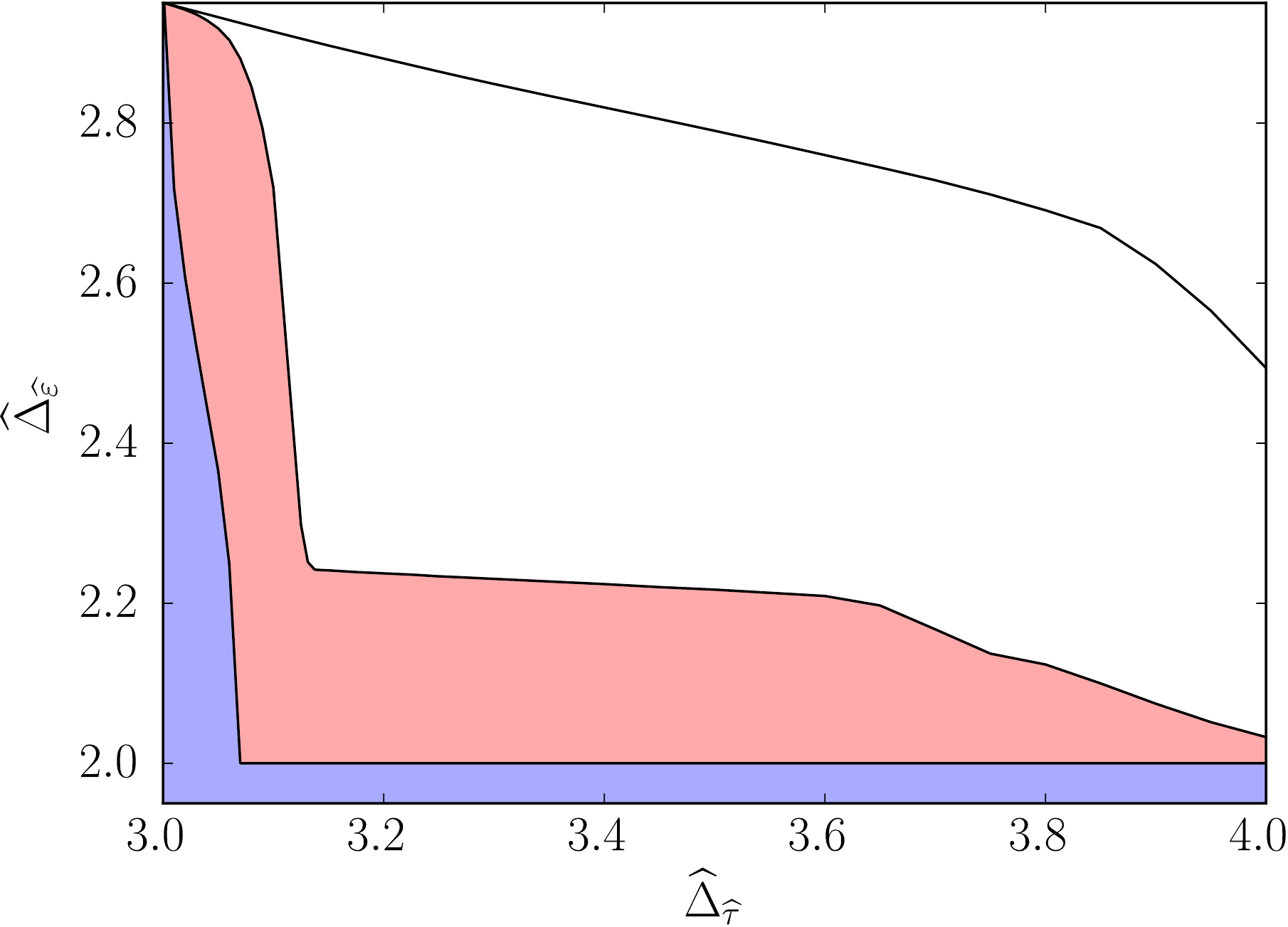}
\caption{A plot showing the upper bound on the dimension of ${\widehat{\varepsilon}}$, the first scalar, other than the identity, seen by any of the OPEs in our correlator system. The unshaded region is the one that follows from a single correlator $\left < \widehat{O}_1\widehat{O}_1\widehat{O}_1\widehat{O}_1 \right >$. The pink region, which is more restrictive, uses the multi-correlator system but the only inputs it uses from the exact relations are the odd-spin operator dimensions given in table \ref{tab:selprottower}. The blue region, more restrictive again, follows from a genuine use of the exact relations. Since these depend on $a_{\phi^2}$, we have extremized $\widehat{\Delta}_{\widehat{\varepsilon}}$ over a third axis which is not shown.}
\label{EvsT_4d}
\end{figure}

We will set $d = 4$ throughout in order to work with a correlator system that involves $3d$ conformal blocks. As discussed in \cite{Behan:2018,Fuente:2019}, similar problems with $2d$ blocks often require more experimentation with the gaps being imposed. These works are concerned with maximizing the gap in the scalar sector, and indeed, we can provide a nice preview of our results by doing the same. Figure \ref{EvsT_4d} bounds the dimension of the lightest exchanged scalar, which we call $\widehat{\Delta}_{\widehat{\varepsilon}}$, as a function of the pseudo stress tensor dimension $\widehat{\Delta}_{\widehat{\tau}}$. To obtain this plot we scanned over all the allowed values of $a_{\phi^2}$. The so obtained blue region is clearly smaller than the pink region, obtained without imposing the OPE relations, or the single correlator region delineated by the upper black line. Three further comments are worthwhile.

First, we observed that much of the constraining power came from our fourth constraint, i.e. the exclusion of the dimension $d$ vector $\text{V}^{(1)} = [\widehat{O}_1 \widehat{O}_2]_{0,1}$ from the spectrum. In fact, if we were to reinstate just this vector then the blue region would expand to almost the same size as the pink region. We emphasize that the OPE relations are essential to meaningfully impose this constraint: they prevent the appearance of vector operators of dimensions very close to $d$ that would numerically be indistinguishable from $\text{V}^{(1)}$. Furthermore, because of the fake primary effect \cite{Esterlis:2016,Karateev:2019} the block for $\text{V}^{(1)}$ can be mimicked in our numerical analysis by a spin 2 operator of dimension 3, and therefore the constraint that $\widehat{\Delta}_{\widehat{\tau}} > 3$ (strictly) is also essential to ensure that it is really absent. This latter argument relies on the observation that, for a spin 2 operator whose dimension $\widehat \Delta \to 3$, the corresponding combination of blocks that enters in the crossing equation \eqref{crossEqVecGenFinal} is:
\begin{equation}
\kappa_1(\widehat{\Delta}, 2)\kappa_2(\widehat{\Delta}, 2)g^{0,0}_{\widehat{\Delta}, 2}(u, v) - g^{-1,1}_{\widehat{\Delta}, 2}(u, v) = \frac{4}{5(\widehat{\Delta} - 3)} g^{-1,1}_{4,1}(u, v) + \dots \label{fake-primary}
\end{equation}
by virtue of the OPE relations discussed in section \ref{sec:crossingeq}, whose notation we follow here. Therefore we can recover a vector operator if we assume that its overall coefficient $\hat{f}_{12\widehat{\mathcal{O}}^{(l)}}\hat{f}_{12}{}^{\widehat{\mathcal{O}}^{(l)}} \propto (\widehat{\Delta} - 3)$.

Second, for the Dirichlet or Neumann boundary conditions either $\widehat{O}_1$ or $\widehat{O}_2$ vanishes so in some sense they are not within the reach of our numerical analysis. 
On the other hand we can find a more general solution including both $\widehat{O}_1$ and $\widehat{O}_2$ as independent GFFs, with any value of $a_{\phi^2}$. This solution does contain the vector $\text{V}^{(1)}$, and because of the fake primary effect we just discussed it corresponds to the point with $\widehat{\Delta}_{\widehat{\varepsilon}} = 2$ and $\widehat{\Delta}_{\widehat{\tau}} = 3$, which is well within the allowed region.

Third, one may check that these bounds are saturated by the following two extremal solutions. The point $(\widehat{\Delta}_{\widehat{\tau}}, \widehat{\Delta}_{\widehat{\varepsilon}}) = (4, 2)$ represents the `single GFF' solution where $\widehat{O}_1$ is a GFF and ${\widehat{\varepsilon}} = \widehat{O}_2 = \widehat{O}_1^2$. This satisfies our crossing equations because it consists entirely of protected operators in \eqref{crossEqVecGenFinal}. Also, the aforementioned vector is indeed absent from the spectrum of primaries because $[\widehat{O}_1 \widehat{O}_2]_{0,1}$ is a descendant of $\widehat{O}_1^3$ in this theory. Since the bound in figure \ref{EvsT_4d} can only decrease as a function of $\widehat{\Delta}_{\widehat{\tau}}$, we can be confident that it will stop changing once it hits $\widehat{\Delta}_{\widehat{\varepsilon}} = 2$. In the blue plot, this turns out to happen well before $\widehat{\Delta}_{\widehat{\tau}} = 4$. We can also understand the point $(\widehat{\Delta}_{\widehat{\tau}}, \widehat{\Delta}_{\widehat{\varepsilon}}) \approx (3, 2.95)$: here the four-point function of $\widehat{O}_1$ can be the extremal solution for a local three-dimensional CFT with $\widehat \Delta = 1$, which according to \cite{Atanasov:2018} has $\widehat{\Delta}_{\widehat{\varepsilon}} \approx 2.95$, and then $\widehat{O}_2$ can be a disconnected GFF. This setup satisfies all of the constraints we have imposed (we are of course not imposing $\widehat{\Delta}_{\widehat{\tau}} > 3$ here) except for the absence of $\text{V}^{(1)}$, which again manifests itself as a spin 2 operator of dimension 3.

We find it plausible that, with infinite computational power, the drop from $\widehat{\Delta}_{\widehat{\varepsilon}} \approx 2.95$ becomes infinitely sharp leading to a value of $\widehat{\Delta}_{\widehat{\varepsilon}} = 2$ almost everywhere. The remainder of this section is about what lies below $\widehat{\Delta}_{\widehat{\varepsilon}} = 2$.\footnote{The bulk has a global reflection symmetry $\phi \to - \phi$ under which $\widehat{O}_1$ and $\widehat{O}_2$ are odd but ${\widehat{\varepsilon}}$ is even. Since $\widehat{\Delta}_{\widehat{\varepsilon}} < 3$ always, and $\widehat{\Delta}_{\widehat{\varepsilon}} \leq 2$ seems likely, any non-trivial boundary condition must be (strongly) unstable even for RG flows that preserve the $\mathbb Z_2$ symmetry. The cases of Dirichlet and Neumann are not included in this discussion because as we explained in these cases one should remove many operators from the spectrum. The Neumann condition is also (strongly) unstable due to the operator $\phi^2$, while in the Dirichlet case the leading $\mathbb{Z}_2$ even deformation is $(\partial_y\phi)^2$ so this boundary condition is stable.
}

\subsection{A universal bound}
\begin{figure}[h!]
\centering
\includegraphics[width=0.8\textwidth]{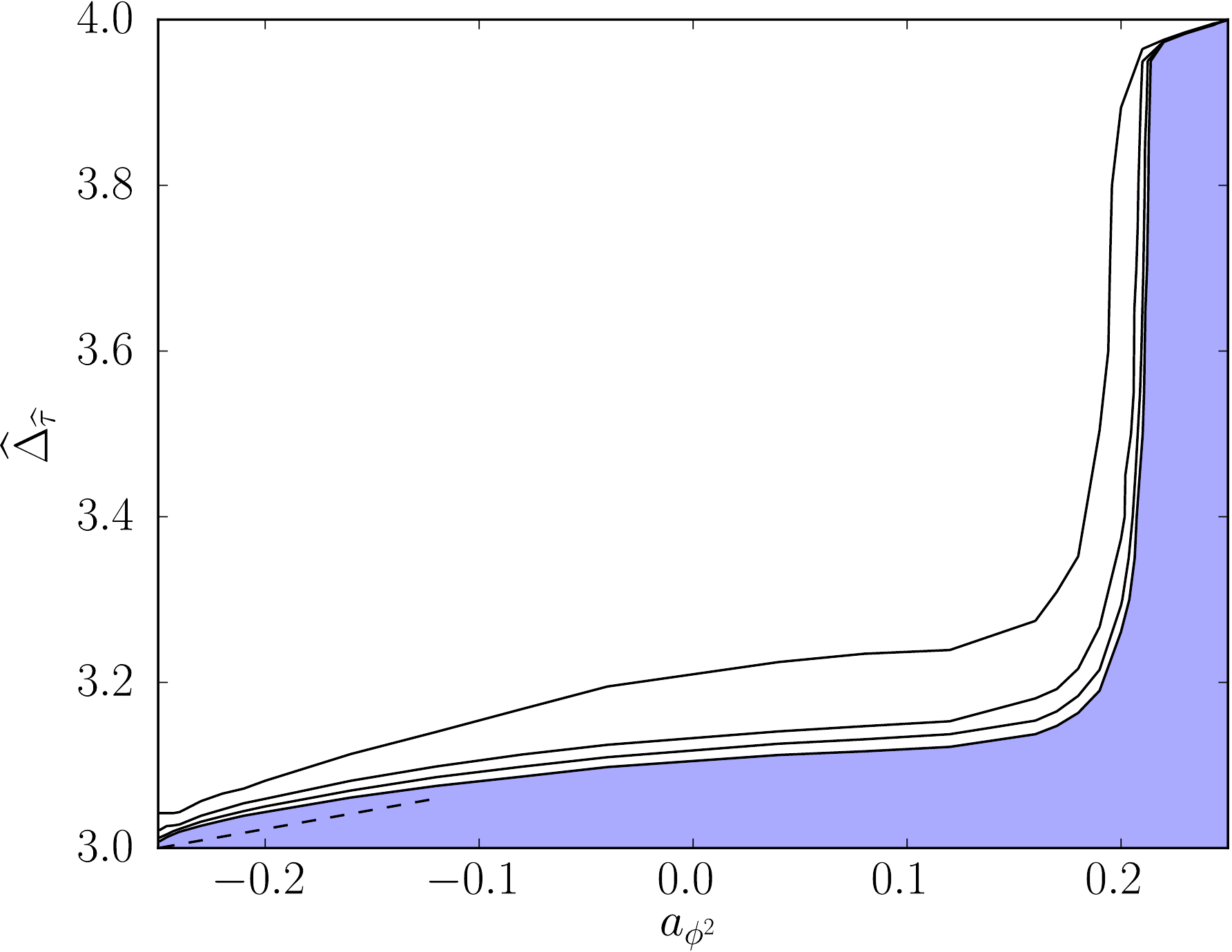}
\caption{Bounds on the dimension of the leading spin 2 operator $\widehat{\tau}$ over the range $-\frac{1}{4} < a_{\phi^2} < \frac{1}{4}$ with our best estimate for the allowed region shaded in blue. The curves have $n_{\mathrm{max}} = 5, 6, 7, 8$ in the notation of \cite{Nakayama:2016,Cappelli:2019}. As for the number of derivative components being kept in each crossing equation, these correspond to $21, 28, 36, 45$ respectively. The dotted line shows the maximum possible value for $\widehat{\Delta}_{\widehat{\tau}}$ from leading order conformal perturbation theory under the assumption that the Ising model is the 3d CFT with the lowest central charge.}
\label{TvsA_4d}
\end{figure}
The next parameter to introduce is $a_{\phi^2}$, which through \eqref{b1b2def} determines $b_1$ and $b_2$, in order to more fully exploit the exact relations. When scanning over $a_{\phi^2}$, it is instructive to first determine its value for the two extremal solutions at $(\widehat{\Delta}_{\widehat{\tau}}, \widehat{\Delta}_{\widehat{\varepsilon}}) = (4, 2)$ and at $(\widehat{\Delta}_{\widehat{\tau}}, \widehat{\Delta}_{\widehat{\varepsilon}}) \approx (3, 2.95)$ discussed above. In the first we found a spin 2 boundary operator with $\widehat{\Delta}_{\widehat{\tau}} = 4$ corresponding to an unprotected block in the first line of \eqref{crossEqVecGenFinal}. Since the overall coefficient of this combination is $\hat{f}_{12{\widehat{\mathcal{O}}^{(l)}}} \hat{f}_{12}{}^{\widehat{\mathcal{O}}^{(l)}}$, \eqref{ssspinconstr} tells us that we must be in a situation where $\kappa_1(4, 2)$ does not blow up. As shown in the middle row of table \ref{tab:selprottower}, this can only happen for an $\widehat{O}_1$ double-trace if $b_2 = 0$. Consequently, this solution sits at $a_{\phi^2} = \frac{1}{4}$. 
In the other extremal solution, the spin 2 operator with $\widehat{\Delta}_{\widehat{\tau}} = 3$ needs to be absent from $\widehat{O}_2 \times \widehat{O}_2$ since $\widehat{O}_2$ is a GFF. The only way to make this compatible with \eqref{ssspinconstr} is to have $a_{\phi^2} = -\frac{1}{4}$.
Proceeding to intermediate values of $a_{\phi^2}$, it is useful to maximize $\widehat{\Delta}_{\widehat{\tau}}$ since this can be interpreted as a measure of how non-local a CFT is. Figure \ref{TvsA_4d} presents this result. The four different lines correspond to four different search spaces, giving a sense of how close we are to having an optimal bound. Every other plot in this section uses the number of components corresponding to the second most restrictive region in figure \ref{TvsA_4d}.

Figure \ref{TvsA_4d} enables a comparison with the results of conformal perturbation theory, particularly around $a_{\phi^2} = -\frac{1}{4}$ where the bound is very strong and we can make a meaningful comparison with the deformation of the Dirichlet boundary conditions by a putative 3d CFT with a scalar operator with $\widehat{\Delta}_{\widehat\chi} = 1 - \epsilon$, that we studied in section \ref{sec:Dpert}. Recall that, according to equation \eqref{anomstressdD}, the anomalous dimension of the spin 2 operator depends on the unperturbed central charge $C_{\widehat\tau}^{(0)}$. Even though we do not know a theory with an operator that can play the role of $\widehat\chi$, it is clear that the central charge of a unitary 3d CFT cannot be arbitrarily small. In fact, we believe that it is not unreasonable to assume that the 3d Ising CFT with\footnote{We are using conventions such that $C_{\widehat\tau}^{\mathrm{free}} = \frac{3}{32\pi^2}$.}
\be
C_{\widehat\tau}^{(0)} \approx 0.95 C_{\widehat\tau}^{\mathrm{free}}
\ee
is the theory with the lowest possible central charge. An early indication for this conjecture was the local minimum corresponding to the Ising CFT found in \cite{ElShowk:2012,ElShowk:2014}, and recently in \cite{Dymarsky:2018} a rigorous lower bound was found that, with sufficient numerical precision, is likely to lie between about $0.6 C_{\widehat\tau}^{\mathrm{free}}$ and $0.95 C_{\widehat\tau}^{\mathrm{free}}$. For us this implies that
\begin{align}
C_{\widehat\tau}^{(0)} > 0.95 C_{\widehat\tau}^{\mathrm{free}} \;\;\; \underset{\eqref{eq:gammatovaD}}{\Rightarrow} \;\;\; \widehat{\gamma}_{\widehat\tau} < 0.46 \, \delta a_{\phi^2} \label{maxslope}
\end{align}
as a bound on the anomalous dimension of the first spin 2 operator. In figure \ref{TvsA_4d} it follows that every such example must lie below the dotted line. The possibilities obeying \eqref{maxslope} are all within the allowed region for now, but we will see that many of them are ruled out when we add more constraints.

\subsubsection{The kink and the extremal spectrum}

\begin{table}[h]
\centering
\begin{tabular}{|c|c|c|c|c|}
\hline
\multicolumn{5}{|c|}{$(a_{\phi^2}, \widehat{\Delta}_{\widehat{\tau}}, C_{\text{D}}) = (0.215, 3.966, 0.0050)$} \\
\hline
$l$ & $\widehat{\Delta}$ & $\hat{f}_{11\mathcal{O}}$ & $\hat{f}_{12\mathcal{O}}$ & $\hat{f}_{22\mathcal{O}}$ \\
\hline
$0$ & $0.561$ & $\textcolor{red}{-0.2549}$ & $0.6690$ & $\textcolor{red}{-2.4365}$ \\
$0$ & $1.815$ & $\textcolor{red}{-2.0305}$ & $1.0807$ & $\textcolor{red}{-2.4953}$ \\
$0$ & $3.381$ & $\textcolor{red}{-0.1749}$ & $0.9209$ & $\textcolor{red}{1.6047}$ \\
\hline
$0$ & $4$ & \multicolumn{3}{|c|}{See \eqref{row4-left}} \\
\hline
$0$ & $4.466$ & $\textcolor{red}{0.1032}$ & $0.4145$ & $\textcolor{red}{-1.6912}$ \\
$0$ & $6$ & $0.1065$ & $0$ & $0.3046$ \\
$1$ & $6$ & $0$ & $0.3702$ & $0$ \\
$2$ & $3.966$ & $\textcolor{red}{-0.7775}$ & $0.2489$ & $\textcolor{red}{-1.0387}$ \\
$2$ & $5.188$ & $\textcolor{red}{-0.0176}$ & $0.7099$ & $\textcolor{red}{1.2956}$ \\
$2$ & $6$ & $0.2549$ & $0$ & $0.8199$ \\
$3$ & $6$ & $0$ & $0.3666$ & $0$ \\
$4$ & $5.791$ & $\textcolor{red}{-0.0044}$ & $0.0755$ & $\textcolor{red}{-1.9929}$ \\
\hline
\end{tabular}
\quad
\begin{tabular}{|c|c|c|c|c|}
\hline
\multicolumn{5}{|c|}{$(a_{\phi^2}, \widehat{\Delta}_{\widehat{\tau}}, C_{\text{D}}) = (0.218, 3.970, 0.0050)$} \\
\hline
$l$ & $\widehat{\Delta}$ & $\hat{f}_{11\mathcal{O}}$ & $\hat{f}_{12\mathcal{O}}$ & $\hat{f}_{22\mathcal{O}}$ \\
\hline
$0$ & $0.555$ & $\textcolor{red}{-0.2399}$ & $0.6625$ & $\textcolor{red}{-2.5352}$ \\
$0$ & $1.826$ & $\textcolor{red}{-2.0122}$ & $1.0635$ & $\textcolor{red}{-2.5603}$ \\
$0$ & $3.227$ & $\textcolor{red}{-0.0973}$ & $0.9453$ & $\textcolor{red}{0.8731}$ \\
\hline
$0$ & $4$ & \multicolumn{3}{|c|}{See \eqref{row4-right}-\eqref{row4-right2}} \\
\hline
$0$ & $4.859$ & $\textcolor{red}{0.0181}$ & $0.3998$ & $\textcolor{red}{-0.3778}$ \\
$0$ & $6$ & $0.1057$ & $0$ & $0.4466$ \\
$1$ & $6$ & $0$ & $0.3707$ & $0$ \\
$2$ & $3.970$ & $\textcolor{red}{-0.7799}$ & $0.2319$ & $\textcolor{red}{-1.0137}$ \\
$2$ & $5.311$ & $\textcolor{red}{-0.0308}$ & $0.7837$ & $\textcolor{red}{2.8659}$ \\
$2$ & $6$ & $0.2533$ & $0$ & $0.8633$ \\
$3$ & $6$ & $0$ & $0.3652$ & $0$ \\
$4$ & $5.960$ & $\textcolor{red}{-0.0085}$ & $0.0334$ & $\textcolor{red}{-0.8782}$ \\
\hline
\end{tabular}
\caption{The low-lying spectrum at two points in $(a_{\phi^2}, \widehat{\Delta}_{\widehat{\tau}}, C_{\text{D}})$ space. The point associated with the left table is still visible after projecting down to just $(a_{\phi^2}, \widehat{\Delta}_{\widehat{\tau}})$ -- it is the kink in figure \ref{TvsA_4d}. Due to our maximization choice, we see every possible operator with odd spin. Protected operators (the ones with integer scaling dimensions) of even spin have vanishing mixed OPE coefficients and they start above the leading twist. Unprotected operators have their OPE coefficients related by \eqref{ssspinconstr} and thus we have shown calculated values in red.
}
\label{CDTab}
\end{table}

We now come to the most striking feature of figure \ref{TvsA_4d} which is the jump near the right hand side. Since the convergence appears to be rapid in this vicinity, we can be confident that the coordinate at which the curve flattens again is not tending towards $a_{\phi^2} = \frac{1}{4}$. In other words, there is a kink at $(a_{\phi^2}, \widehat{\Delta}_{\widehat{\tau}}) \approx (0.215, 3.966)$ which obeys the exact relations and cannot be one of the free boundary conditions. If it is truly a new boundary condition then it must obey a further constraint that we have not yet imposed: the Ward identity \eqref{displRel} for the displacement operator. This is one of the reasons we would like to investigate the spectrum at the kink in more detail.

\paragraph{The extremal spectrum}
The extremal functional method \cite{Poland:2011,ElShowk:2013} allows for the extraction of an approximate spectrum and OPE coefficients for any point on the boundary of an allowed region. We have done so at two points: the first is the kink in figure \ref{TvsA_4d} and the second involves a tuning of the displacement central charge $C_{\text{D}}$
using the procedure that will be explained in the next subsection. The CFT data for operators with $\widehat{\Delta} < 6.5$ are listed in table \ref{CDTab}. The black numbers were obtained from the output of the script in \cite{SimmonsDuffin:2017} and the red numbers were computed using the OPE relations. Our OPE coefficients are defined such that
\begin{equation}
g_{\widehat{\Delta}, l}(z, z) \sim (-1)^l \left [ \frac{4z}{(1 + \sqrt{1 - z})^2} \right ]^{\widehat{\Delta}} + \dots \label{blocknorm}
\end{equation}
for the standard cross-ratios $z$ and $\bar{z}$ approaching zero along the diagonal. 

Notice that the most stable result is obtained after maximizing an OPE coefficient near the boundary of the plot to make the functional as close to extremal as possible. We have chosen to optimize the coefficient of the $\text{V}^{(3)}_4$ operator, which is a spin 3 operator of dimension 6. Our reason for doing so is to avoid another interference from the fake primary effect: much like $\text{V}^{(1)}_2$ can be mimicked by a spin 2 operator of dimension almost 3, the $\text{V}^{(3)}_4$ operator can be mimicked by a spin 4 operator of dimension almost 5. However this scenario is unnatural, not only because of the existence of an operator very close to the unitarity bound but also because the absence of $\text{V}^{(3)}_4$ would imply that the higher-spin charge corresponding to the bulk spin 4 current is preserved by the boundary. We do not expect such an `integrable' boundary, and our optimization minimizes the chances of an unwanted spin 4 operator taking the place of $\text{V}^{(3)}_4$.

After going through this maximization, a reassuring feature we observe is that there is no even spin $l$ operator with $\widehat{\Delta} = d + l - 2$, as anticipated in the range of the sum in \eqref{crossEqVecGenFinal}. Such a block, if present, would have to be treated with $\vec{V}_{\textbf{0}, d + l - 2, l}$ because the exact relations degenerate at this point. But indeed, bulk spin $\ell$ currents only have boundary modes up to $l = \ell - 1$ in the bOPE and there is no reason for $l = \ell$ to be present as a protected operator.

\paragraph{The displacement Ward identity}
With the spectrum in hand we can investigate whether the Ward identity \eqref{displRel} for the displacement operator is satisfied. However this is again a rather subtle business, this time because there might be other scalar operators of dimension 4. The correct procedure is as follows.

Consider all the scalar operators of dimension 4 in the putative extremal solution at the kink. One of these operators is the displacement operator, and its coefficients $\hat{f}_{11\text{D}}$ and $\hat{f}_{22\text{D}}$ must obey \eqref{displRel}, i.e.~they are constrained to lie on a curve parameterized by $C_{\text{D}}$. Every other operator is then not a displacement operator, meaning that it must be absent from the bulk-to-boundary OPE of the stress tensor: $b_{T\text{D}^\prime} = 0$ for any `non-displacement' $\text{D}^\prime$. Repeating the arguments in appendix \ref{subsec:dispDet} this leads to the condition that:
\begin{equation}\label{nonDisplRel}
\hat{f}_{11\text{D}^\prime} = \frac{b_{\phi^2\text{D}^\prime}}{12 b_1^2}, \quad \hat{f}_{22\text{D}^\prime} = -\frac{b_{\phi^2\text{D}^\prime}}{2b_2^2}\,.
\end{equation}
with $b_{\phi^2 \text{D}^\prime}$ arbitrary. So from a physical perspective \eqref{displRel} and \eqref{nonDisplRel} are the equations to be checked.

On the numerical side of things we do not get these coefficients so cleanly; instead we are given the elements of the matrix corresponding to
\be
\sum_{\mathcal{O} = \text{D}, \text{D}^\prime} \begin{pmatrix} \hat{\lambda}_{11\mathcal{O}}^2 & \hat{\lambda}_{11\mathcal{O}}\hat{\lambda}_{22\mathcal{O}} \\ \hat{\lambda}_{11\mathcal{O}}\hat{\lambda}_{22\mathcal{O}} & \hat{\lambda}_{22\mathcal{O}}^2  \end{pmatrix}~,
\ee
where $\hat{\lambda}_{ij \mathcal{O}}$ denotes the OPE coefficient with unit-normalized two-point function of $\mathcal{O}$, and it is up to us to cook up a series of OPE coefficients $\hat{\lambda}_{ij \mathcal{O}}$ for operators D and D$'$ in order to fit this data. (The switch from $\hat{f}_{ij \mathcal{O}}$ to $\hat{\lambda}_{ij \mathcal{O}}$ is deliberate: numerically we obtain OPE coefficients for unit-normalized operators, so the coefficients in \eqref{displRel} should really be scaled by $\sqrt{C_{\text{D}}}$.)

For the spectrum on the left hand side of table \ref{CDTab} we numerically obtain a matrix of rank 1 whose factorization yields
\be
\begin{pmatrix}
\hat{\lambda}_{11\text{D}} \\ \hat{\lambda}_{22\text{D}}
\end{pmatrix}
=
\begin{pmatrix}
0.4695 \\ 1.1936
\end{pmatrix}~.
\ee
Since we have only one operator this must be the displacement, so we can check compatibility with the Ward identity \eqref{displRel}. Remarkably we find that it is well obeyed with $C_{\text{D}} = 0.0050$ -- a strong indication that the extremal solution is actually physical! We also obtain that
\begin{align}
\begin{pmatrix}
\hat{f}_{11\text{D}} \\ \hat{f}_{22\text{D}}
\end{pmatrix}
=
\begin{pmatrix}
0.0332 \\ 0.0844
\end{pmatrix}
\label{row4-left}
\end{align}
for the OPE coefficients.

For the spectrum on the right hand side of table \ref{CDTab} we find that
\be
\sum_{\mathcal{O} = \text{D}, \text{D}^\prime} \begin{pmatrix} \hat{\lambda}_{11\mathcal{O}}^2 & \hat{\lambda}_{11\mathcal{O}}\hat{\lambda}_{22\mathcal{O}} \\ \hat{\lambda}_{11\mathcal{O}}\hat{\lambda}_{22\mathcal{O}} & \hat{\lambda}_{22\mathcal{O}}^2  \end{pmatrix} = \begin{pmatrix} 0.2291 & 0.2397 \\ 0.2397 & 13.199 \end{pmatrix}.
\ee
This is a matrix of rank two and we need more than one operator. It is natural to try to see if we can fit it with one displacement and one non-displacement operator. Notice that the matrix has three independent entries but for the two operators we only have the two parameters $C_{\text{D}}$ and $b_{\phi^2 \text{D}^\prime}$. 
Two simple approaches can be taken at this point. In the first approach, we demand that a non-displacement is exactly present and extract $b_{\phi^2 \text{D}'}$:
\begin{equation}
\begin{pmatrix} \hat{\lambda}_{11\text{D}^\prime} \\ \hat{\lambda}_{22\text{D}^\prime} \end{pmatrix} = \begin{pmatrix} \frac{b_{\phi^2 \text{D}^\prime}}{12b_1^2} \\ -\frac{b_{\phi^2 \text{D}^\prime}}{2b_2^2} \end{pmatrix}
\Rightarrow b_{\phi^2 \text{D}^\prime} = 1.775, \; \begin{pmatrix} \hat{\lambda}_{11\text{D}} \\ \hat{\lambda}_{22\text{D}} \end{pmatrix} = \pm \begin{pmatrix} 0.472 \\ 1.088 \end{pmatrix}~. \label{row4-right} 
\end{equation}
To fit the rest of the matrix we need the given displacement OPE coefficients, which lead to $C_{\text{D}} = 0.0053$ from $\hat{\lambda}_{11\text{D}}$ and $C_{\text{D}} = 0.0049$ from $\hat{\lambda}_{22\text{D}}$. Alternatively, in the second approach, we demand that one of the outer products is an exact displacement and extract $C_\text{D}$ and OPE coefficients for the non-displacement:
\begin{equation}
\begin{pmatrix} \hat{\lambda}_{11\text{D}} \\ \hat{\lambda}_{22\text{D}} \end{pmatrix} =  \begin{pmatrix} \frac{2S_4^2C_{\text{D}} + 16a_{\phi^2}}{6S_4b_1^2 \sqrt{C_{\text{D}}}} \\ \frac{2S_4^2C_{\text{D}} - 16a_{\phi^2}}{S_4b_2^2 \sqrt{C_{\text{D}}}} \end{pmatrix}
\Rightarrow C_{\text{D}} = 0.0050, \; \begin{pmatrix} \hat{\lambda}_{11\text{D}^\prime} \\ \hat{\lambda}_{22\text{D}^\prime} \end{pmatrix} = \pm \begin{pmatrix} 0.085 \\ -3.452 \end{pmatrix}~. \label{row4-right2}
\end{equation}
This leads to $b_{\phi^2\text{D}^\prime} = 1.914$ from $\hat{\lambda}_{11\text{D}^\prime}$ and $b_{\phi^2\text{D}^\prime} = 1.767$ from $\hat{\lambda}_{22\text{D}^\prime}$. Although the small mismatches in both approaches might be due to numerical errors, it seems reasonable to conclude that this solution is not as physical as the solution on the left hand side of table \ref{CDTab}.

\subsection{Local boundary conditions}
\begin{figure}[h!]
\centering
\subfloat[][$a_{\phi^2} = -0.249$]{\includegraphics[scale=0.3]{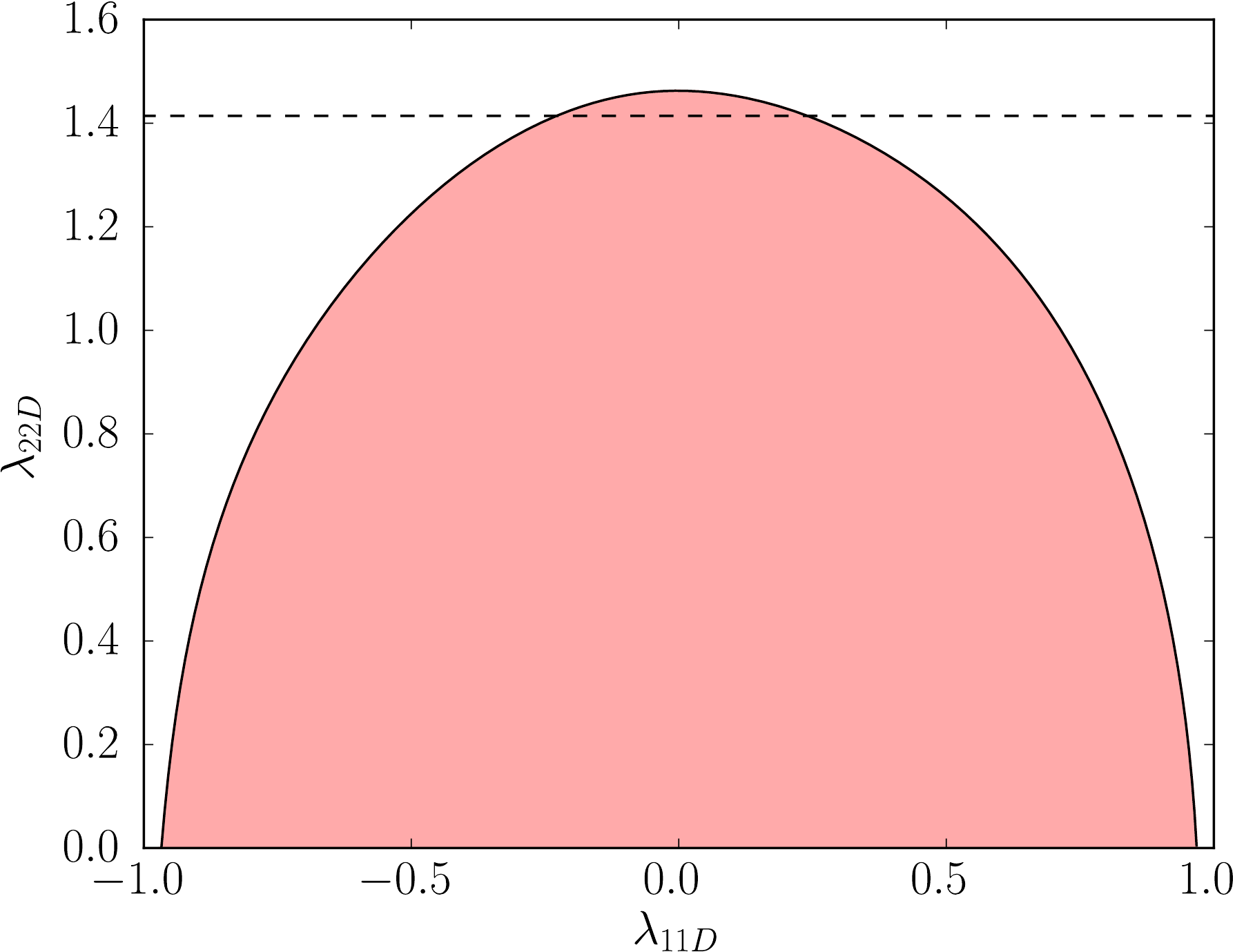}}
\subfloat[][$a_{\phi^2} = -0.24$]{\includegraphics[scale=0.3]{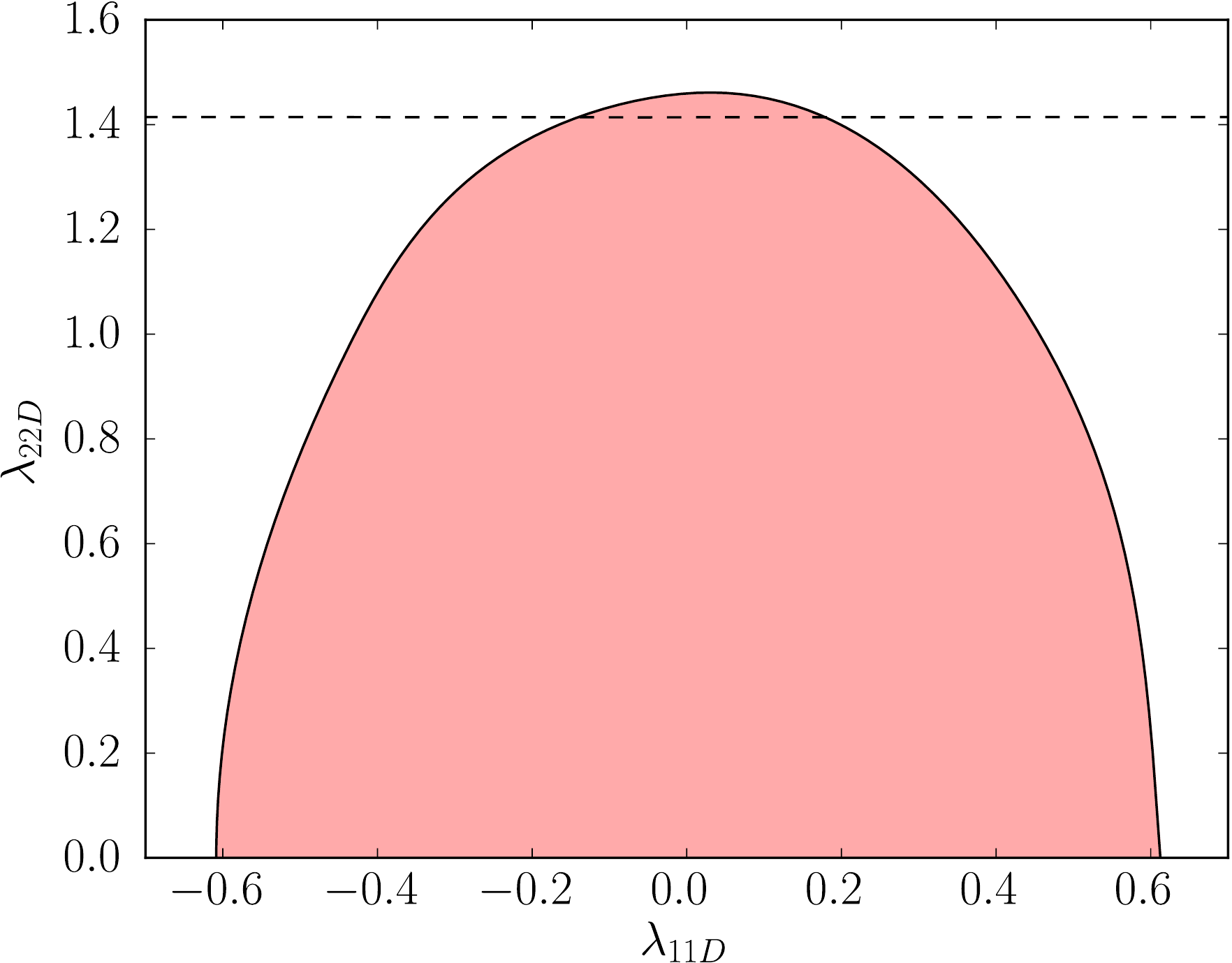}}
\subfloat[][$a_{\phi^2} = -0.04$]{\includegraphics[scale=0.3]{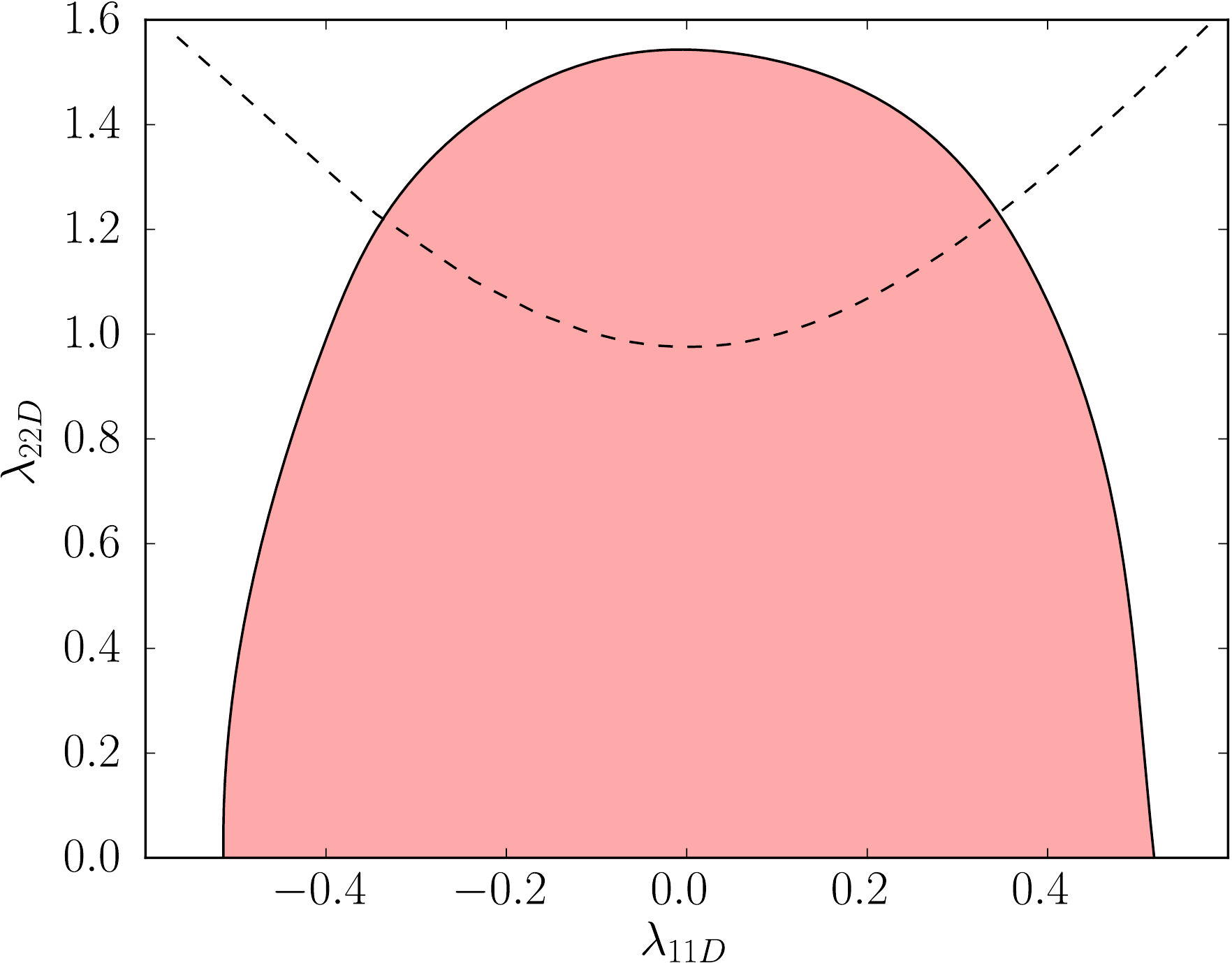}} \\
\subfloat[][$a_{\phi^2} = 0.04$]{\includegraphics[scale=0.3]{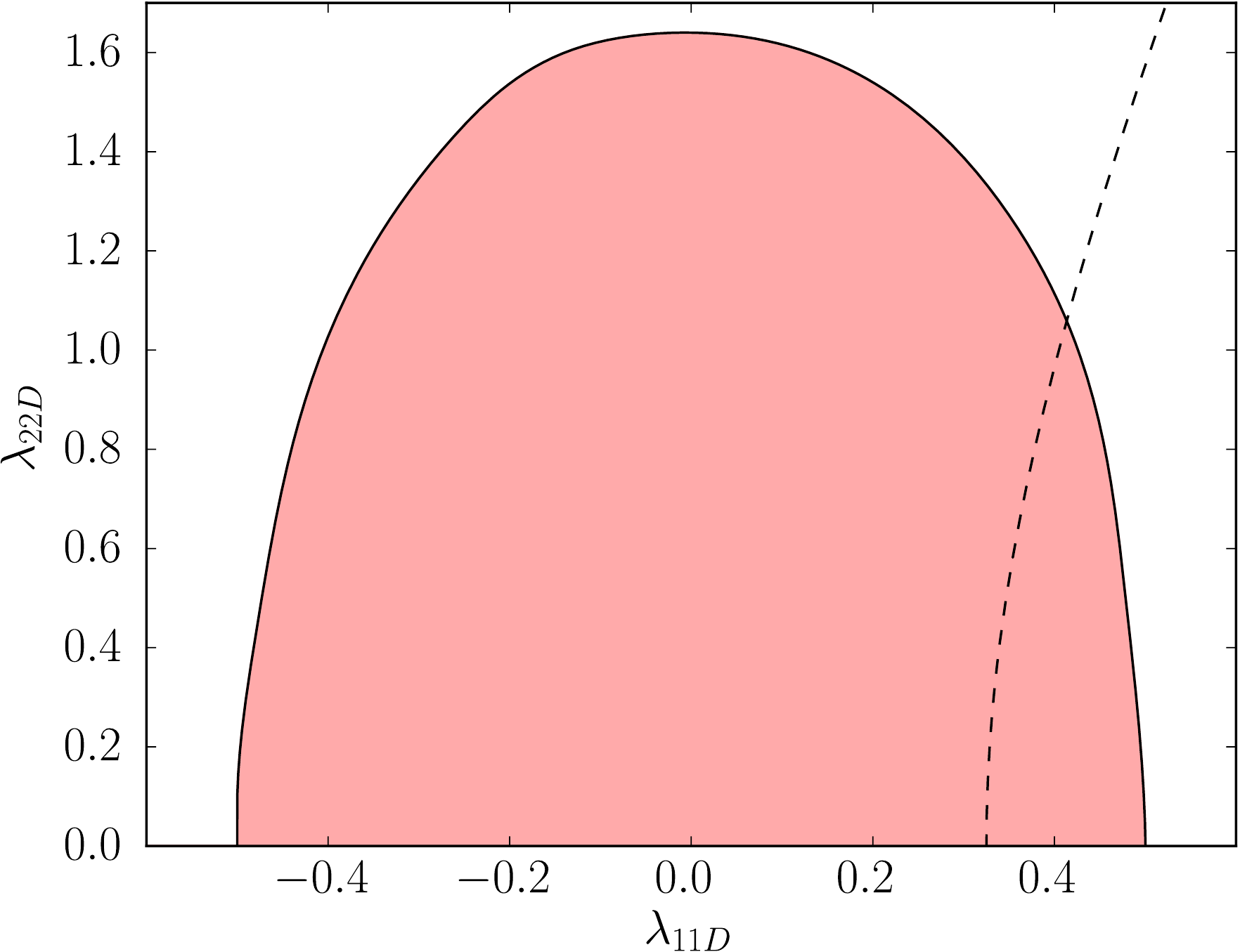}}
\subfloat[][$a_{\phi^2} = 0.24$]{\includegraphics[scale=0.3]{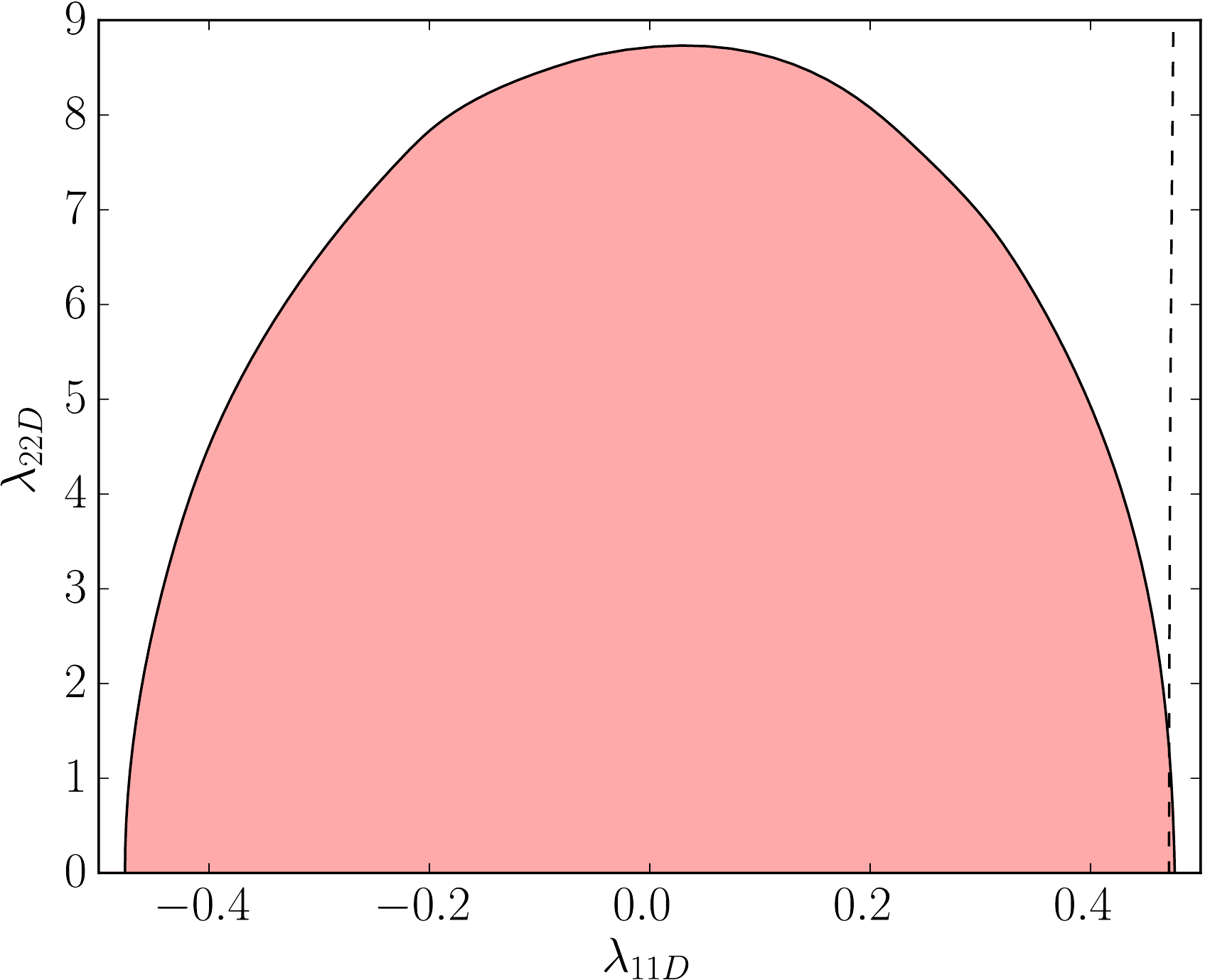}}
\subfloat[][$a_{\phi^2} = 0.249$]{\includegraphics[scale=0.3]{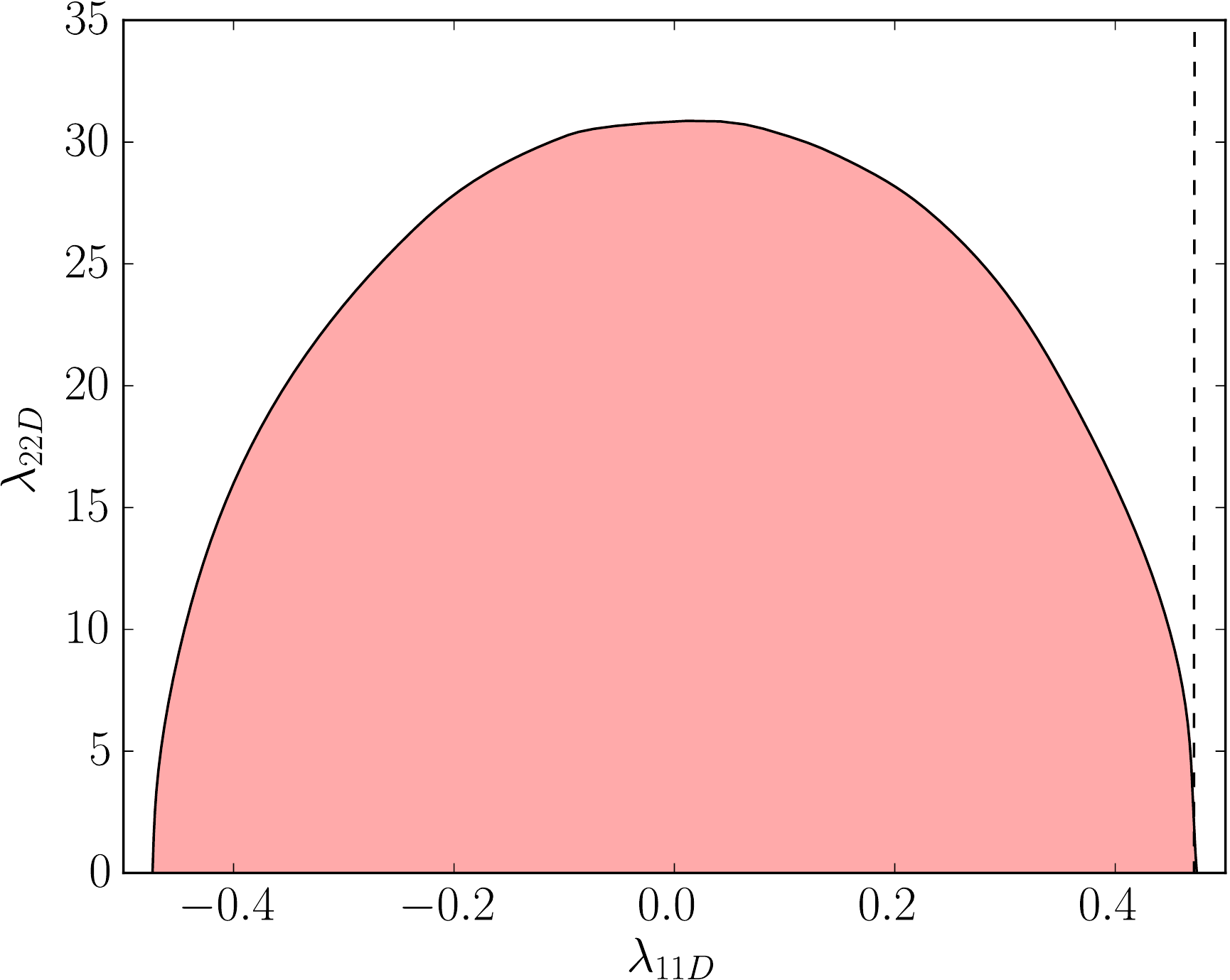}}
\caption{Six allowed regions for the OPE-space vector of the unit-normalized displacement. The dotted line shows the physical locus for $\hat{\lambda}_{11\text{D}}$ and $\hat{\lambda}_{22\text{D}}$, i.e. \eqref{displRel} divided by $\sqrt{C_{\text{D}}}$. When this line becomes vertical (defining a unique $\hat{\lambda}_{11\text{D}}$ in order for $\hat{\lambda}_{22\text{D}}$ to be finite), it saturates our bound. This does not quite happen in the opposite limit of the line becoming horizontal. Note that in the GFF example there are two candidates for the displacement. Both $[\widehat{O}_1 \widehat{O}_1]_{1, 0}$ and $[\widehat{O}_2 \widehat{O}_2]_{0, 0}$ are compatible with these bounds if we treat them as different operators that satisfy $\hat{\lambda}_{11\text{D}}\hat{\lambda}_{22\text{D}} = 0$.}
\label{OPEvsA_4d}
\end{figure}

As with numerical bounds on the gap, the exact relations also lead to significant improvements for bounds on OPE coefficients. Consider again the (unit-normalized) displacement operator which appears with the coefficients $\hat{\lambda}_{11\text{D}}$ and $\hat{\lambda}_{22\text{D}}$. To constrain them, we set
\begin{equation}
\begin{pmatrix}
\hat{\lambda}_{11\text{D}} \\ \hat{\lambda}_{22\text{D}}
\end{pmatrix}
\mapsto \hat{\lambda}_D
\begin{pmatrix}
\cos \theta \\ \sin \theta
\end{pmatrix}
\label{sincos}
\end{equation}
as in \cite{Kos:2016}, then apply standard methods for bounding the magnitude of an OPE-space vector \cite{Caracciolo:2010}. Figure \ref{OPEvsA_4d} shows the results of this exercise for different values of $a_{\phi^2}$.

A first thing to note is once more the importance of the exact OPE relations in \eqref{ssspinconstr}. Without them the allowed region would certainly be the union of all the regions in figure \ref{OPEvsA_4d}. However for $a_{\phi^2} \rightarrow \frac{1}{4}$ we observe an unbounded growth in the vertical direction (note the different vertical scales), and therefore $\hat{\lambda}_{22\text{D}}$ is really only bounded by virtue of the OPE relations.

\begin{figure}[h!]
\centering
\includegraphics[width=0.8\textwidth]{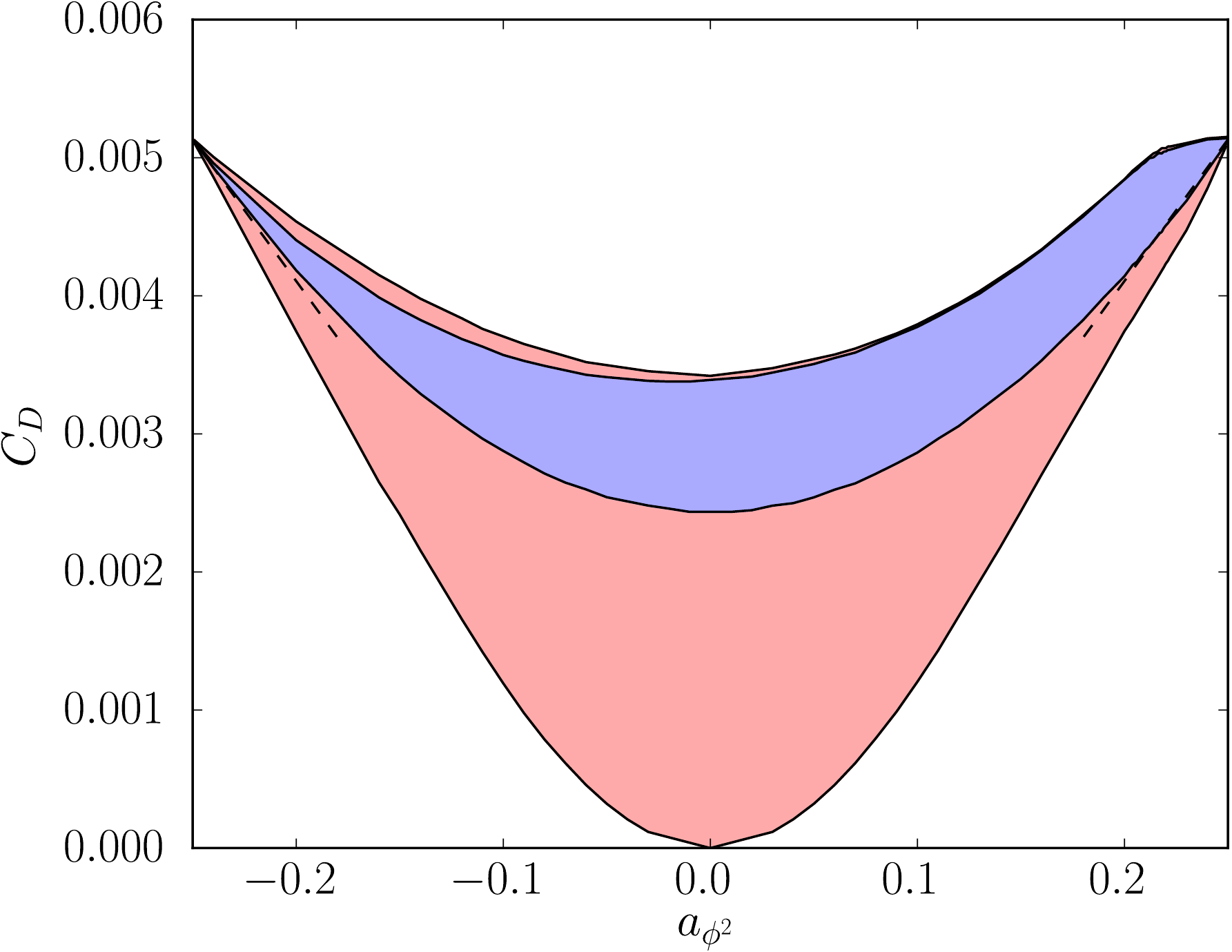}
\caption{An asymmetric plot showing the minimum and maximum $C_{\text{D}}$ as a function of $a_{\phi^2}$. In the blue region, all dimension 4 scalars not singled out by \eqref{addd} are constrained to satisfy $b_{T\text{D}^\prime} = 0$. No such constraint is made in the pink region which leads to weaker bounds. The dotted lines give the predictions of conformal perturbation theory which are model-independent at leading order. A slight kink in the upper right corner looks well positioned to be identified with the kink in figure \ref{TvsA_4d}.}
\label{CvsA_4d}
\end{figure}

The dotted line represents the combinations of OPE coefficients that obey the Ward identity \eqref{displRel}, parameterized by the displacement central charge $C_{\text{D}}$. As discussed above, the non-trivial fact about the kink in figure \ref{TvsA_4d} was that it happened to sit on this line.
The intersection of the dotted line with the allowed region also translates into a lower and upper bound for $C_{\text{D}}$ for each value of $a_{\phi^2}$. This is shown as the pink region in figure \ref{CvsA_4d}. 
This bound is certainly valid but rather crude: it does not take into account the restriction \eqref{nonDisplRel} on additional scalar operators of dimension 4 that are not the displacement. 
To do better we can assume a fixed displacement operator with a certain $C_{\text{D}}$ in the crossing equations by replacing
\begin{equation}
\vec{V}_{\mathds{1}} \mapsto \vec{V}_{\mathds{1}} + \frac{1}{S_4^2 C_{\text{D}}}
\begin{pmatrix}
\frac{2S_4^2 C_{\text{D}} + 16a_{\phi^2}}{6b_1^2} & \frac{2S_4^2 C_{\text{D}} - 16a_{\phi^2}}{b_2^2}
\end{pmatrix}
\vec{V}_{\textbf{0}, 4, 0}
\begin{pmatrix}
\frac{2S_4^2 C_{\text{D}} + 16a_{\phi^2}}{6b_1^2} \\ \frac{2S_4^2 C_{\text{D}} - 16a_{\phi^2}}{b_2^2}
\end{pmatrix}
\label{addd}
\end{equation}
and removing the scalar of dimension 4 from the special operators in the crossing equation \eqref{crossEqVecGenFinal}. We then bisect in $C_{\text{D}}$ to find the allowed region, and  this leads to the much improved blue region in figure \ref{CvsA_4d}.

One may wonder if the blue region allows for other scalar operators of dimension 4 that are not the displacement operator. The answer is that it does, because such operators lie in the allowed continuum of operators. Furthermore,
\be
\frac{\hat{f}_{22\text{D}^\prime}}{\hat{f}_{11\text{D}^\prime}} = \frac{\kappa_2(4, 0)}{\kappa_1(4, 0)} = -6 \frac{b_1^2}{b_2^2}, \label{ratioAgreement}
\ee
which implies that the limit of a scalar operator as $\hat \Delta \to 4$ in the continuum is actually exactly a D$'$ operator whose OPE coefficients automatically obey \eqref{nonDisplRel}. So fixing $C_\text{D}$ not only allows one to single out a displacement operator for which the Ward identities \eqref{displRel} are obeyed, it also ensures that \eqref{nonDisplRel} holds for every other scalar of dimension 4.

We can compare figure \ref{CvsA_4d} to the conformal perturbation theory results \eqref{eq:CDovaD} and \eqref{eq:CDovaN}. The lines corresponding to potential perturbative fixed points saturate the lower bound on $C_{\text{D}}$ once the proper constraints on dimension 4 scalars are imposed. These emanate from the points $(a_{\phi^2}, C_{\text{D}}) = \left ( \pm \frac{1}{4}, \frac{1}{2\pi^4} \right )$ at which the upper and lower bounds are forced to meet by the Ward identity. The other point that can be explained analytically is the origin which is associated with no boundary at all. This point has to be allowed by the pink region since it corresponds to adding zero in \eqref{addd}. Once we classify dimension 4 scalars into displacements and non-displacements, it appears that there are no longer any nearby solutions that would allow us to see this point in the blue region. To see that they cannot arise from a GFF construction, consider the explicit displacement operator \eqref{GenericDisp}. We may rewrite it as
\begin{equation}
\text{D} = \sqrt{\frac{C_{\text{D}}}{(d - 2)^2 b_1^4 + b_2^4}} \left ( (d - 2)b_1^2 [\widehat{O}_1 \widehat{O}_1]_{1, 0} + b_2^2 [\widehat{O}_2 \widehat{O}_2]_{0, 0} \right ) \label{DoubleTraceDisp}
\end{equation}
by using the bulk two-point function \eqref{2ptFunctionsDEF} to relate the double-traces of $\phi$ and $\partial_y \phi$ to those involving $\widehat{O}_1$ and $\widehat{O}_2$. The rules of GFF then allow us to go from \eqref{DoubleTraceDisp} to
\begin{equation}
\hat{f}_{11\text{D}} = \frac{b_1^2 (d - 2)^2}{\sqrt{2} (d - 1)} \sqrt{\frac{C_{\text{D}}}{(d - 2)^2 b_1^4 + b_2^4}} \; , \; \hat{f}_{22\text{D}} = b_2^2 \sqrt{\frac{2C_{\text{D}}}{(d - 2)^2 b_1^4 + b_2^4}}. \label{GFFExtension}
\end{equation}
For a generic $a_{\phi^2}$, there is no $C_{\text{D}}$ which can make both of these coefficients satisfy the Ward identity.

\begin{figure}[h!]
\centering
\includegraphics[width=0.8\textwidth]{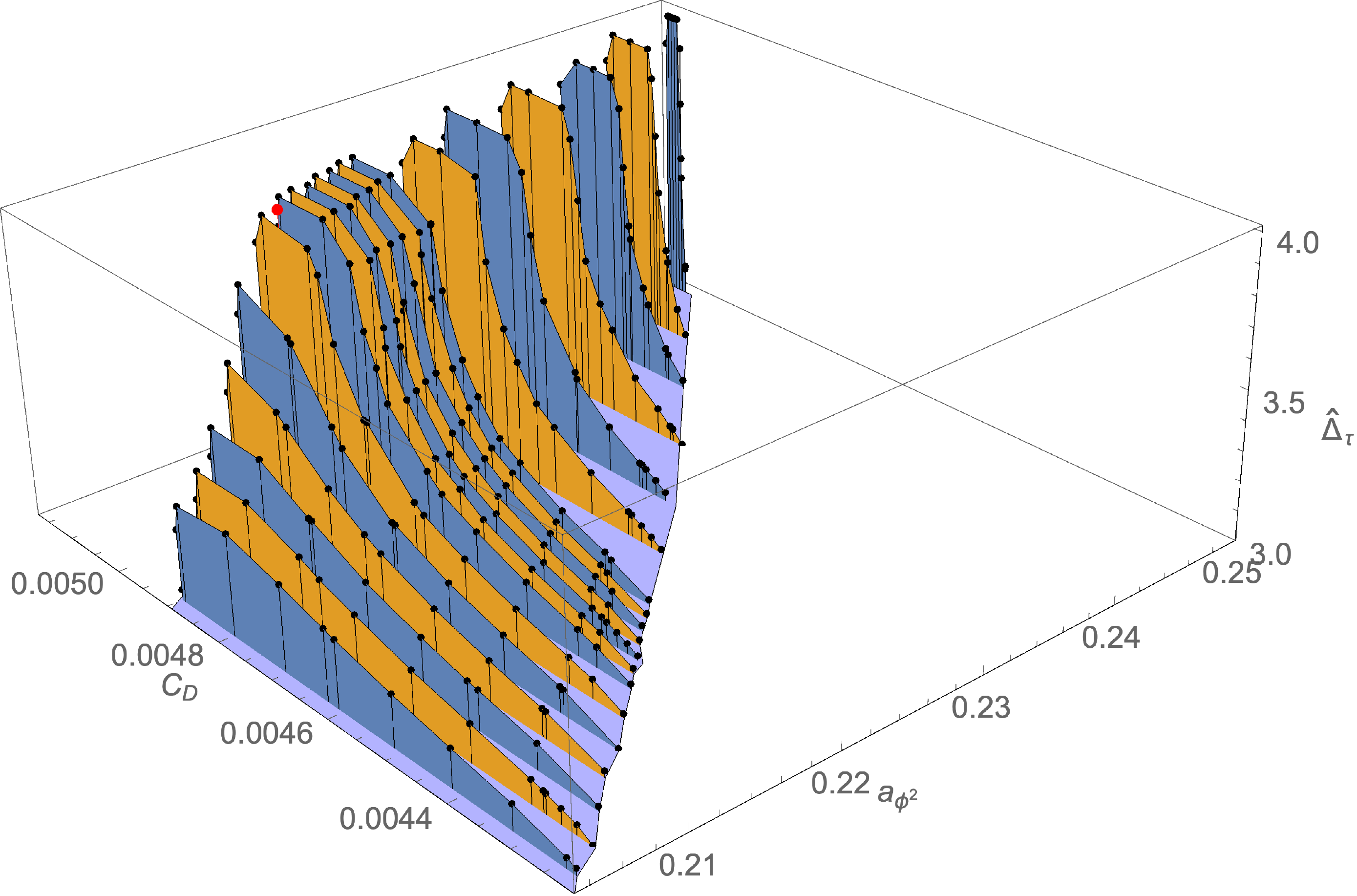}
\caption{The maximum possible $\widehat{\Delta}_{\widehat{\tau}}$ for several points in the most interesting region of figure \ref{CvsA_4d}. Planes are inserted below points with the same value of $a_{\phi^2}$ for visibility. The red point has its spectrum shown in the left columns in table \ref{CDTab}.}
\label{CvsAvsT_4d}
\end{figure}

After producing universal bounds in the $(a_{\phi^2}, \widehat{\Delta}_{\widehat{\tau}})$ and $(a_{\phi^2}, C_{\text{D}})$ planes, it is natural to try scanning in all three parameters. This means choosing a grid of points in the allowed blue region of figure \ref{CvsA_4d} and maximing the spin 2 gap at each one. The best feature of this plot is that every point with $\widehat{\Delta}_{\widehat{\tau}} > 3$ is guaranteed to obey all the constraints given above: exact OPE relations, no stress tensor, no flux operator, and the Ward identity for the displacement operator. For $a_{\phi^2} \geq 0.20$, which is the vicinity of the kink, the results are shown in figure \ref{CvsAvsT_4d}.

Before we discuss this figure, let us comment first on the analysis for more general $a_{\phi^2}$ and for which the data is not shown. This analysis indicated that the maximum spin 2 gap, so the points on the boundary of figure \ref{TvsA_4d}, correspond to the largest possible values of $C_\text{D}$, so the points near the upper boundary of figure \ref{CvsA_4d}. On the other hand, near the lower boundary of figure \ref{CvsA_4d} the spin 2 gap remains very close to $3$. Since the perturbative line in figure \ref{CvsA_4d} is near this lower boundary, it indicates that the corresponding line in figure \ref{TvsA_4d} must actually be quite a bit flatter than the slope determined by the Ising model central charge \eqref{maxslope}. In short, the (non-rigorous!) extrapolation of the one-loop analysis to small but finite values of $\delta a_{\phi^2}$ indicates that the Dirichlet boundary condition can only be driven to a weakly coupled fixed point if the 3d CFT that triggers the RG flow has a large central charge.

Let us return to figure \ref{CvsAvsT_4d}. Our best candidate for a non-trivial boundary condition, the left of table \ref{CDTab}, may be found by hugging the upper edge of figure \ref{CvsA_4d} and looking for where $\widehat{\Delta}_{\widehat{\tau}}$ jumps. As suggested by the extremal spectrum, this happens at $(a_{\phi^2}, \widehat{\Delta}_{\widehat{\tau}}, C_{\text{D}}) = (0.215, 3.966, 0.0050)$ and corresponds to the red point in the figure. We see that the two-dimensional kinks in previous figures have become a three-dimensional feature: a cliff appears to develop around this point.

The spectrum on the right of table \ref{CDTab} corresponds to $(a_{\phi^2}, \widehat{\Delta}_{\widehat{\tau}}, C_{\text{D}}) = (0.218, 3.970, 0.0050)$ which appears to be a more generic point in this three parameter space. We originally chose this point by hugging the lower edge of pink region in figure \ref{CvsA_4d}, i.e.~by bisecting in $\widehat{\Delta}_{\widehat{\tau}}$ without the constraint \eqref{nonDisplRel} for non-displacement scalars of dimension 4. This produces a jump at $(a_{\phi^2}, C_{\text{D}}) = (0.218, 0.0044)$ in that plot. However, re-interpreting the extra dimension 4 scalars found in that solution as discussed above shifted $C_{\text{D}}$ from $0.0044$ to $0.0050$. (As discussed above, this is under the assumptions that the Ward identities hold for this point.)

Notice also that $\widehat{\Delta}_{\widehat{\tau}} \to 3$ rather smoothly as $C_\text{D}$ approaches its lowest possible value. According to the dashed line in figure \ref{CvsA_4d} this is where we could find potential weakly coupled fixed points from the Neumann end. As for the Dirichlet end discussed above, one might take this as an indication that the anomalous dimension of the three-dimensional stress tensor cannot be too big.

\section{Outlook}
\label{sec:conclusions}
We set out to investigate whether a free real scalar field could have conformal boundary conditions other than Dirichlet or Neumann. The bulk equation of motion restricted the two- and three-point functions of $\phi$ so strongly that we found that all non-trivial boundary conditions must support a shadow pair of boundary operators of dimensions $\Delta_\phi$ and $\Delta_\phi + 1$. The numerical analysis in four bulk dimensions (so three boundary dimensions) of correlation functions of this shadow pair yielded interesting results. On the one hand, for a large range of values of $a_{\phi^2}$ (the one-point function of the bulk $\phi^2$ operator) there must be a spin 2 operator relatively close to the unitarity bound, providing some evidence for the absence of interesting boundary conditions. On the other hand, for $a_{\phi^2}$ near its upper bound of $1/4$ this maximal value shoots up and we observed an interesting kink in the data at about $a_{\phi^2} = 0.215$ with a spin 2 operator of dimension $3.966$ and $C_D$ approximately equal to $0.0050$. More numerical data is provided in table \ref{CDTab}. This could be a new conformal boundary condition for the free scalar field.

It the future it would be interesting to see whether the shadow relations can be explored analytically rather than numerically. Indeed, one could ask whether the shadow transform
\be
\tilde O(x) = \int d^d y \frac{1}{(x-y)^{2(d-\Delta)}} O(y)
\ee
can be applied to four-point functions and conformal blocks? As we have seen, the shadow transformation is singular for three-point functions when the scaling dimension of the third operator is of double-twist type, so it is not entirely obvious that shadow transforming one or more operators in a consistent four-point function automatically leads to another consistent four-point function. Our expectation is instead that contact terms will become important because they get magnified to non-trivial functions by the shadow transformation.

It would of course also be interesting to understand the possible new conformal boundary condition that corresponds to the kink in our numerical plots. One approach would be to try to `move' the kink by changing the problem. For example, we could try different spacetime dimensions $d$ or generalize the problem to $N > 1$ free scalar fields.\footnote{$O(N)$ invariant conformal boundary conditions for multiple free scalars were studied recently in \cite{Prochazka:2019fah,Giombi:2019enr}. } These would of course be interesting studies in their own right, but if we can dial a parameter like $d$ or $N$ to a value where the kink merges with a free boundary condition then that would provide strong indications for a possible perturbative approach to the problem. Another approach would be to put the free bulk theory in AdS: then we can add a mass term to the bulk field which would continuously change the scaling dimensions but which is not expected to spoil the conformality of the boundary and a conformal bootstrap analysis should always be possible \cite{QFTinAdS}.

An obvious direction for the future is to try to extend the analysis of this paper to other examples of free theories in the bulk, such as the free scalar in other spacetime dimensions, the free fermion in any $d$ or the free vector in $d=4$. In the latter case it would be extremely interesting to see if there is any signature of the continuous family of boundary conditions \cite{DiPietro:2019hqe} in the bootstrap approach, perhaps along the lines of the previous bootstrap study of conformal manifolds in \cite{Baggio:2017mas}.

More generally, the `landscape' of boundary conditions for a given CFT$_d$ is a wide open problem. It therefore remains an interesting target for further explorations. The subject is even richer because, as this paper exemplifies, we need to modify the usual crossing symmetry equations in surprising ways when defects, boundaries, or interfaces are taken into account.

\section*{Acknowledgements }

We thank the organizers of the `Bootstrap 2019' and the `Boundaries and Defects in Quantum Field Theory' workshops at Perimeter Institute where part of this work was carried out. We further thank P.~Liendo and X.~Zhao for collaboration on related topics and S.~Cremonesi, M.~Del Zotto, D.~Dorigoni, D.~Gaiotto, C.~Herzog, M.~Lemos, M.~Meineri, P.~Niro  and E.~Trevisani for interesting discussions. Research at Perimeter Institute is supported by the Government of Canada through Industry Canada and by the Province of Ontario through the Ministry of Research \& Innovation. EL and BvR are supported by the Simons Foundation grant $\#$488659 (Simons Collaboration on the non-perturbative bootstrap). LD is partially supported by INFN Iniziativa Specifica ST\&FI. LD also acknowledges support by the program ``Rita Levi Montalcini'' for young researchers. CB has received funding from the European Research Council (ERC) under the European Union's Horizon 2020 research and innovation programme (grant agreement $\#$787185). The numerical calculations were done on the University of Oxford Advanced Research Computing (ARC) facility \cite{Richards:2015} and the SeaWulf high-performance computing system maintained by Stony Brook Research Computing and Cyberinfrastructure and the Institute for Advanced Computational Science at Stony Brook.

\appendix

\section{Conventions}\label{app:conventions}

\subsection{bOPE}
Consider a scalar bulk operator $\mathcal{O}$, not necessarily free. 
The bOPE of $\mathcal{O}$ is completely determined by $SO(d,1)$ symmetry, up to a certain collection of CFT data \cite{McAvity:1993ue,McAvity:1995zd}
\begin{align}\label{bOPEphi2}
\mathcal{O}(\vec{x},y)=\sum_{\widehat{\mathcal{O}}}\sum_n \frac{b_{\mathcal{O}}{}^{\widehat{\mathcal{O}}}}{y^{\Delta_{\mathcal{O}} -\widehat{\Delta}_{\widehat{\mathcal{O}}}}}\frac{\left(-\frac{1}{4}y^2   \vec \nabla^2\right)^n }{n! \left(\widehat{\Delta}_{\widehat{\mathcal{O}}}+\frac{3-d}{2}\right)_n}\widehat{\mathcal{O}}(\vec{x})~.
\end{align}
One can check that the expression above reproduces the bulk-to-boundary correlators \eqref{BD2pt}, once applied to the boundary two-point functions
\begin{align}\label{Btwopt}
\langle \widehat{\mathcal{O}}(\vec{x}) \widehat{\mathcal{O}}'(0)\rangle =\frac{\widehat{C}_{\widehat{\mathcal{O}}\widehat{\mathcal{O}}'}}{|\vec{x}|^{2\widehat{\Delta}_{\widehat{\mathcal{O}}}}}~, \quad \widehat{C}_{\widehat{\mathcal{O}}\widehat{\mathcal{O}}'}=\delta_{\widehat{\mathcal{O}}}^{\widehat{\mathcal{O}}'}\widehat{C}_{\widehat{\mathcal{O}}\widehat{\mathcal{O}}}~,
\end{align}
and using that $b_{\mathcal{O}}{}^{\widehat{\mathcal{O}}}\widehat{C}_{\widehat{\mathcal{O}}\widehat{\mathcal{O}}}=b_{\mathcal{O}\widehat{\mathcal{O}}}$. 
We will take unit-normalized boundary two-point functions, except for the protected operators that can appear in the bOPE of the bulk conserved currents $J_\ell$. Such operators, collectively denoted by $\widehat{J}_\ell^{(l)}$ (with $l=0,\dots \ell-1$) have their normalization fixed by the Ward identities \eqref{ward_currents}, and therefore the coefficients in their two-point functions are physical
\begin{align}
\begin{split}\label{JhDtwopt}
\langle \text{D}(\vec{x})\text{D}(0)\rangle &=  \frac{C_{\text{D}}}{|\vec{x}|^{2d}}~,\\
\langle \widehat{{J}}^{(l)}_\ell(\vec{x},z_1)\widehat{{J}}^{(l)}_\ell(0,z_2)\rangle & = {C}_{\widehat{J}^{(l)}_\ell} \frac{(z_1 \cdot I(\hat{x})\cdot z_2)^l}{(\vec{x}^2)ì|^{d+\ell-2}}~.
\end{split}
\end{align}

When the bulk operator $\mathcal{O}$ is a free scalar $\phi$, as we explained in \ref{ss:BDcorr}, the scaling dimensions of its boundary modes $\widehat{O}_i$ are completely determined by the bulk equation of motion. The expression \eqref{bOPEphi2} becomes (compare to \eqref{BOPEphi})
\begin{align}\label{phiO2ptaux}
\phi(\vec{x},y)=\sum_{i=1,2}\sum_n \frac{b_{i}}{y^{\Delta_{\phi} -\widehat{\Delta}_i}}\frac{\left(-\frac{1}{4}y^2   \vec \nabla^2\right)^n }{n! \left(\widehat{\Delta}_i+\frac{3-d}{2}\right)_n}\widehat{O}_i(\vec{x})~.
\end{align}
Conventionally, we choose unit normalization for the boundary modes of $\phi$
\begin{align}
\langle \widehat{O}_i (0)\widehat{O}_j (\infty)\rangle = {\delta_{ij}}{}~,
\end{align}
such that $b_{\phi}^i =b_{\phi i}\equiv b_{i}$.

\subsection{Boundary OPE and physical OPE coefficients}

We denote generic boundary operators as $\widehat{\mathcal{O}}_k$, where the label $k$ indicates collectively all possible indices of the local operator. The OPE between two boundary operators $\widehat{\mathcal{O}}_i$ is (up to boundary descendants)
\begin{align}\label{boundaryOPEgen}
\widehat{\mathcal{O}}_i(\vec{x})\widehat{\mathcal{O}}_j(0)\sim \sum_k\frac{\hat{f}_{ij}{}^{k}}{|\vec{x}|^{\widehat{\Delta}_i+\widehat{\Delta}_j-\widehat{\Delta}_k}}\widehat{\mathcal{O}}_k(0)+\dots
\end{align}
The boundary two-point functions are normalized as in \eqref{Btwopt}. We use the Zamolodchikov metric $\widehat{C}_{\widehat{\mathcal{O}}\widehat{\mathcal{O}}'}$ to raise and lower indices of $\hat{f}_{ij}{}^{k}$s. Concretely (sum over repeated indices)
\begin{align}
\langle \widehat{\mathcal{O}}_i (\vec{x}_1)\widehat{\mathcal{O}}_j (\vec{x}_2)\widehat{\mathcal{O}}_m (\infty)\rangle=\frac{{}\hat{f}_{ij}{}^k \widehat{C}_{km}}{|x_{12}|^{\widehat{\Delta}_i+\widehat{\Delta}_j-\widehat{\Delta}_m}}\equiv\frac{\hat{f}_{ijm}}{|x_{12}|^{\widehat{\Delta}_i+\widehat{\Delta}_j-\widehat{\Delta}_m}}~.
\end{align}
With these conventions, we have that the displacement operator, whose normalization is taken as in \eqref{JhDtwopt}, enters the generic boundary OPE \eqref{boundaryOPEgen} as
	\begin{align}
	\widehat{\mathcal{O}}_i(x) \widehat{\mathcal{O}}_j(0) &\supset \,\frac{{\hat{f}_{ij}{}^{\text{D}}}}{|x|^{\widehat{\Delta}_i+\widehat{\Delta}_j-d}}\text{D}(0)+\dots,
	\end{align}
	and a generic boundary four-point function as
	\begin{align}
	\begin{split}
	\langle \widehat{\mathcal{O}}_i(0) \widehat{\mathcal{O}}_j(x)\widehat{\mathcal{O}}_k(1) \widehat{\mathcal{O}}_m(\infty)\rangle &\supset		 {\hat{f}_{ij}{}^{\text{D}}\hat{f}_{km}{}^{\text{D}}}{}\langle\text{D}(0)\text{D}(\infty)\rangle(1+\dots)\\
	&={\hat{f}_{ij}{}^{\text{D}}\hat{f}_{km\text{D}}}\,\,g_{\text{D}}^{\widehat{\Delta}_{ij},\widehat{\Delta}_{kl}}(u,v)\\
	&=\frac{{\hat{f}_{ij\text{D}}\hat{f}_{km\text{D}}}}{C_{\text{D}}}\,\,g_{\text{D}}^{\widehat{\Delta}_{ij},\widehat{\Delta}_{kl}}(u,v)~.
	\end{split}
	\end{align}
	In the equation above we introduced the conformal blocks, which are normalized as \eqref{blocknorm}. Alternatively we can think of $\text{D}$ to be unit-normalized, such that the physical boundary OPE coefficient is
		\begin{align}
	\langle \widehat{\mathcal{O}}_i(0) \widehat{\mathcal{O}}_j(x)\widehat{\mathcal{O}}_k(1) \widehat{\mathcal{O}}_m(\infty)\rangle &\supset{{\hat{\lambda}_{ij\text{D}}\hat{\lambda}_{km\text{D}}}}{}\,\,g_{\text{D}}^{\widehat{\Delta}_{ij},\widehat{\Delta}_{kl}}(u,v), \qquad \hat{\lambda}_{ij\text{D}}=\frac{\hat{f}_{ij\text{D}}}{\sqrt{C_{\text{D}}}}
	\end{align}
Similar remarks apply for other protected operators that can appear in the bOPE of the bulk conserved currents $J_\ell$.

\section{Three-point function conformal blocks}
\label{app:blocks}

In this appendix we derive the conformal block decomposition of the free field $\phi$ three-point function with a generic boundary operator $\widehat{\mathcal{O}}^{(l)}$
\begin{align}\label{phiphiOgenapp}
\langle \phi(\vec{x}_1,y_1)\phi(\vec{x}_2,y_2){{\widehat{\mathcal{O}}^{(l)}}}(\theta,\infty)\rangle.
\end{align}
We will obtain closed-form expressions for all the conformal blocks exchanged in the boundary channel of this three-point function. We also compute some bulk channel blocks, while leaving a more complete study for the future.

\subsection{Blocks in the boundary channel}\label{app:blocks_defect}
We start from the blocks in the boundary channel. As we explained in the main text (see section \ref{sec:three_pt}), the expansion of the correlator \eqref{phiphiOgenapp} in a basis of boundary conformal blocks can be obtained by acting twice with the bOPE on the generic three-point functions 
\begin{align}\label{3ptsslapp}
\langle \widehat{O}_i (\vec{x}_1)\widehat{O}_j (\vec{x}_2){\widehat{\mathcal{O}}^{(l)}} (\theta,\infty)\rangle=&\,\frac{\hat{f}_{ij{\widehat{\mathcal{O}}^{(l)}}}}{|\vec{x}_{12}|^{\widehat{\Delta}_i+\widehat{\Delta}_j-\widehat{\Delta}}}P^{(l)}_\parallel(\hat{x}_{12},{\theta}),
\end{align}
and then resumming the contributions from boundary descendants. The polynomials $P^{(l)}_\parallel$ are defined in \eqref{parjpoly}. Applying the bOPE \eqref{BOPEphi} and using the identity
\begin{align}
\nabla_{\vec{x}_{12}}^{2n}\left(\frac{(-\vec{x}_{12}\cdot \theta)^l}{|\vec{x}_{12}|^{2t}}\right)=4^n (t)_n \left(1+t-l-\frac{d-1}{2}\right)_n \frac{(-\vec{x}_{12}\cdot \theta)^l}{|\vec{x}_{12}|^{2t+2n}},
\end{align}
we can rewrite \eqref{phiphiOgenapp} as
\begin{align}
\begin{split}\label{summandq}
\langle\phi(\vec{x}_1,y_1)\phi(\vec{x}_2,y_2){{\widehat{\mathcal{O}}^{(l)}}}(\theta,\infty)\rangle&=\sum_{i,j=1,2}b_i b_j \hat{f}_{ij{{\widehat{\mathcal{O}}^{(l)}}}}P^{(l)}_\parallel(\hat{x}_{12},{\theta})\\
&\times \sum_{m,n}\frac{(-1)^{m+n}}{m! n!}\frac{  {y_1}^{{\widehat{\Delta}_i-\Delta_\phi }{}+2n} {y_2}^{{\widehat{\Delta}_j-\Delta_\phi }{}+2m}   }{ |\vec{x}_{12}|^{-2\kappa_{ij}+2m+ 2n-l} 
}\\
& \hspace{-1 cm}\times \frac{(-\kappa_{ij})_m (-\kappa_{ij}+m)_n \left(-\kappa_{ij}-\hat{h}-l\right)_m \left(-\kappa_{ij}+m-\hat{h}-l\right)_n}{\left(\widehat{\Delta}_i-\hat{h}\right)_n\left(\widehat{\Delta}_j-\hat{h}\right)_m }~.
\end{split}
\end{align}
In the above formula we introduced
\begin{align}
\kappa_{ij}\,\equiv- \frac{1}{2}(\widehat{\Delta}_i+\widehat{\Delta}_j-\widehat{\Delta}+l),\quad \hat{h}\,\equiv \frac{d-3}{2}.
\end{align}
The infinite sum in the r.h.s. of \eqref{summandq} can be explicitly performed, and the result takes the form
\begin{align}
\begin{split}\label{phiphiOgenapp2}
\langle \phi(\vec{x}_1,y_1)&\phi(\vec{x}_2,y_2){{\widehat{\mathcal{O}}^{(l)}}}(\theta,\infty)\rangle=\frac{P^{(l)}_\parallel(\hat{x}_{12},{\theta})}{ |\vec{x}_{12}|^{d-2-\widehat{\Delta}}}\\
&\times\left( b_1^2 {}\hat{f}_{11{\widehat{\mathcal{O}}^{(l)}}}\,\hat{\mathcal{F}}^{11}_{\widehat{\Delta},l}({w_+},{w_-})+{b_1 b_2{}\hat{f}_{12{\widehat{\mathcal{O}}^{(l)}}}}{}\hat{\mathcal{F}}^{12}_{\widehat{\Delta},l}({w_+},{w_-})+{b_2^2 {}\hat{f}_{22{\widehat{\mathcal{O}}^{(l)}}}}{}\hat{\mathcal{F}}^{22}_{\widehat{\Delta},l}({w_+},{w_-})\right)~.
\end{split}
\end{align}
The quantities $\hat{\mathcal{F}}^{ij}_{\widehat{\Delta},l}$ are hypergeometric functions of the cross-ratios $w_\pm$ (defined in \eqref{wdef})
\begin{align}
\begin{split}\label{phiphiOblocks_def}
\hat{\mathcal{F}}^{11}_{\widehat{\Delta},l}({w_+},{w_-})& =\frac{1}{2}\left[{}_2F_1\left(\frac{1-\widehat{\Delta}-l}{2} ,\frac{d-2+l-\widehat{\Delta}}{2};\frac{1}{2};-{{w_-}}\right)+({w_+}\leftrightarrow {w_-})\right]~,\\
\hat{\mathcal{F}}^{12}_{\widehat{\Delta},l}({w_+},{w_-})& =\frac{1}{2}\left[\left((-1)^l-1\right) {w_-}^{1/2} \, _2F_1\left(\frac{2-\widehat{\Delta}-l}{2},\frac{d+l-\widehat{\Delta}-1}{2} ;\frac{3}{2};-{w_-}\right)\right.\\
&\,\,\,\,\,\,\,\left.+\left((-1)^l+1\right) {w_+}^{1/2} \, _2F_1\left(\frac{2-l-\widehat{\Delta}}{2},\frac{d+l-\widehat{\Delta}-1}{2} ;\frac{3}{2};-{w_+}\right)\right]~,\\
\hat{\mathcal{F}}^{22}_{\widehat{\Delta},l}({w_+},{w_-})& =\frac{1}{2 (\widehat{\Delta}+l-1) (d-2+l-\widehat{\Delta})}\\
&\,\,\,\,\times\left[\, _2F_1\left(\frac{1-\widehat{\Delta}-l}{2} ,\frac{d-2+l-\widehat{\Delta}}{2};\frac{1}{2};-{{w_-}}\right)-({w_-}\leftrightarrow {w_+})\right]~.\\
\end{split}
\end{align}
Note that in terms of two cross-ratios
\begin{align}
\xi\equiv\frac{|\vec{x}_{12}|^2+(y_1-y_2)^2}{y_1 y_2}, \quad \zeta\equiv \frac{(|\vec{x}_{23}|^2+y_2^2)y_1}{(|\vec{x}_{13}|^2+y_1^2) y_2} \underset{x_3\rightarrow \infty}{\sim} \frac{y_1}{y_2},
\end{align}
the cross-ratios ${w_\pm}$ can be rewritten as
\begin{align}
{w_\pm}=-\frac{(1\pm \zeta)^2}{1-(\xi +2) \zeta+\zeta^2}.
\end{align}

 We have checked that \eqref{phiphiOgenapp2} satisfies the Klein-Gordon equation with the correct conditions. As a further consistency check, we note that the defect channel blocks for the two-point function \eqref{2ptFunctionsDEF} can be recovered from \eqref{phiphiOblocks_def} by setting $\widehat{\Delta}=l=0$ and $\hat{f}_{ij\id}=\delta_{ij}$.

\subsection{Scalar blocks in the bulk channel}\label{app:blocks_bulk}

Next, we will be interested in the bulk conformal block expansion of \eqref{phiphiOgenapp}. For simplicity we will consider only the case where the third operator is a boundary scalar, while leaving the general case for future work. In the bulk channel we plug the $\phi\times \phi$ OPE \eqref{phiphiOPE} to convert \eqref{phiphiOgenapp} into an infinite sum over bulk-to-boundary two-point functions
\begin{align}\label{phiphiOgenappscalar2}
\langle \phi(\vec{x}_1,y_1)\phi(\vec{x}_2,y_2){{\widehat{\mathcal{O}}}}(\infty)\rangle=\sum_{\mathcal{O}_k \subset \phi\times \phi} c_{\phi\phi}{}^{\mathcal{O}_k} \langle \mathcal{O}_k (\vec{x}_{2})\widehat{\mathcal{O}}(\infty)\rangle+\dots
\end{align}
with the ellipsis denoting contributions from bulk descendants, which are fixed by $SO(d+1,1)$ conformal symmetry. As discussed in the main text (see subsection \ref{ss:reg_OPE}), spin selection rules and current conservation imply that the bulk operator $\phi^2$ is the only contribution to the r.h.s of \eqref{phiphiOgenappscalar2} for generic $\widehat{\Delta}$ not equal to the scaling dimension of $J_\ell$. In this more generic case we have
\begin{align}\label{phiphiOgenappscalar3}
\langle \phi(\vec{x}_1,y_1)\phi(\vec{x}_2,y_2){{\widehat{\mathcal{O}}}}(\infty)\rangle=\bigg[\langle \phi^2(x_2)\widehat{\mathcal{O}}(\infty)\rangle+\dots\bigg] \equiv b_{\phi^2 \widehat{\mathcal{O}}}\mathcal{W}_{\phi^2}^{\phi\phi\widehat{\mathcal{O}}}(\vec{x}_{12},y_1,y_2).
\end{align}
Note that when $\widehat{\Delta}$ equals the scaling dimension of $J_\ell$, we should add to the previous expression an additional contribution proportional to $\langle J_\ell(x_2) {{\widehat{\mathcal{O}}}}(\infty)\rangle$ (see e.g. the case of the displacement in \ref{subsec:dispDet}). It is not difficult to compute $\mathcal{W}_{\phi^2}^{\phi\phi\widehat{\mathcal{O}}}$ by plugging the bulk OPE into \eqref{phiphiOspinj} and resumming the bulk descendants. Using the explicit form of the differential operator that controls the scalar exchange (see e.g. \cite{Ferrara:1973yt}), we find the following series expansion
\begin{align}
\begin{split}\label{bulkblockgen}
b_{\phi^2 \widehat{\mathcal{O}}}\mathcal{W}_{\phi^2}^{\phi\phi\widehat{\mathcal{O}}}(\vec{x}_{12},y_1,y_2)=&\frac{b_{\phi^2 \widehat{\mathcal{O}}}}{y_2^{2\Delta_{\phi}-\widehat{\Delta}}}\sum_{n=0}^\infty\frac{(-\frac{\xi}{16}  )^n}{n!}\,\frac{  \Gamma \left(\frac{d-1}{2}\right) \Gamma (2\Delta_{\phi}+ 2n-\widehat{\Delta})}{ \Gamma (2\Delta_{\phi}-\widehat{\Delta}) \Gamma \left(n+\frac{d-1}{2}\right)}\\
&\times {}_2F_1\left(\Delta_\phi+n,d-\widehat{\Delta}+2 n-2;2\Delta_\phi+2 n;1-\frac{y_1}{y_2}\right)~,
\end{split}
\end{align}
where the cross-ratio $\xi$ is defined in \eqref{xidef}. There are various interesting special situations in which the result \eqref{bulkblockgen} produces simple closed-form formulae. In the `cylindrical' configuration $y_1=y_2=y$ the infinite sum gives a simple hypergeometric function
\begin{align}\label{closedbulkblockscalar}
\mathcal{W}_{\phi^2}^{\phi\phi\widehat{\mathcal{O}}}(\vec{x}_{12},y,y)&=\frac{1}{y^{2\Delta_{\phi}-\widehat{\Delta}}}
\, _2F_1\left(\frac{d-\widehat{\Delta}-2}{2} ,\frac{d-\widehat{\Delta}-1}{2} ;\frac{d-1}{2};-\frac{\hat\chi }{4}\right),
\end{align}
where we introduced a cross-ratio $\hat{\chi}$
\begin{align}\label{chidef}
\hat{\chi}=\frac{|\vec{x}_{12}|^2}{y^2},
\end{align}
which is nothing but the restriction of $\xi$ defined in \eqref{xidef} to the `cylindrical' configuration. As explained in appendix \ref{ss:mtchBulk}, this result can be also derived by `inverting' the boundary channel expansion \eqref{phiphiOgenapp2}. As a final comment, we note that the series representation \eqref{bulkblockgen} yields simple closed-form expressions, some of which will be presented in appendix \ref{subsec:dispDet}, when the third operator is of $\text{D}_\ell^{(0)}$ type.

\section{OPE relations and bulk-to-boundary crossing}\label{app:exactrelDer}

In this appendix we discuss in detail the derivation of the main results presented in section \ref{sec:conformalbc}.

\subsection{Derivation of the OPE relations}\label{subsexactrelDer}
In this appendix we derive the OPE relations \eqref{ssspinconstr}. The starting point is the boundary channel expansion for the correlator \eqref{phiphiOgenapp}, given in \eqref{phiphiOgen}. Away from other operator insertions, the $\phi\times \phi$ OPE requires this three-point function to be analytic around $x_1^\mu=x_2^\mu$ (recall that the identity in \eqref{phiphiOPE} decouples). In order to study this limit, it is convenient to place the two bulk operators at the same transverse distance i.e.  $y_1=y_2=y$, such that the expression \eqref{phiphiOgen} simplifies as follows:
\begin{align}
\begin{split}\label{phiphilDiag}
\langle \phi(\vec{x}_1&,y)\phi(\vec{x}_2,y){{\widehat{\mathcal{O}}^{(l)}}}(\theta,\infty)\rangle=\frac{P_\parallel^{(l)}(\hat{x}_{12},\theta)}{y^{d-2-\widehat{\Delta}}}\hat{\chi}^{-\frac{1}{2}(d-2-\widehat{\Delta})}\\
\times&\left\{\frac{1}{2}b_1^2 {}\hat{f}_{11{{\widehat{\mathcal{O}}^{(l)}}}}\left[1+\, _2F_1\left(\frac{1-l-\widehat{\Delta}}{2} ,\frac{d+l-\widehat{\Delta}-2}{2};\frac{1}{2};-\frac{4}{\hat{\chi} }\right)\right]\right.\\
&+b_1 b_2 {}\hat{f}_{12{{\widehat{\mathcal{O}}^{(l)}}}}\left.[1+(-1)^l]{ \,\hat{\chi}^{-\frac{1}{2}}\, _2F_1\left(\frac{2-l-\widehat{\Delta}}{2},\frac{d+l-\widehat{\Delta}-1}{2} ;\frac{3}{2};-\frac{4}{\hat{\chi} }\right)}\right.\\
&\left. + \frac{b_2^2{}\hat{f}_{22{{\widehat{\mathcal{O}}^{(l)}}}}}{2 (\widehat{\Delta}+l-1) (\widehat{\Delta}-l-d+2)}  \left[\, 1-\, _2F_1\left(\frac{1-\widehat{\Delta}-l}{2},\frac{d+l-\widehat{\Delta}-2}{2};\frac{1}{2};-\frac{4}{\hat{\chi} }\right)\right]\right\}~,
\end{split}
\end{align}
where $\hat\chi$ is the cross-ratio defined in \eqref{chidef}. In this configuration with $y_1=y_2$ Bose symmetry \eqref{Bosesymm} implies that this expression vanishes when $l$ is an odd integer, so we first consider even $l$. We then require \eqref{phiphilDiag} to be analytic around $\vec{x}_1=\vec{x}_2$, for finite $y$. For generic values of $d,l,\widehat{\Delta}$, the r.h.s. of \eqref{phiphilDiag} contains unphysical singularities, since
\begin{align}
\begin{split}\label{asymph_blocks}
\langle \phi(\vec{x}_1,y)&\phi(\vec{x}_2,y){{\widehat{\mathcal{O}}^{(l)}}}(\theta,\infty)\rangle\underset{\hat\chi\rightarrow 0}{\sim}\frac{\left(- \vec{x}_{12}\cdot {\theta}\right)^l}{y^{d+l-2-\widehat{\Delta}}}\\
 \times&\left[\frac{1}{2}\hat{\chi}^{1+\frac{\widehat{\Delta}-d-l}{2}} \left(b_1^2\hat{f}_{11{{\widehat{\mathcal{O}}^{(l)}}}}-\frac{b_2^2\hat{f}_{22{{\widehat{\mathcal{O}}^{(l)}}}}}{(\widehat{\Delta}+l-1) (d-\widehat{\Delta}+l-2)}\right)\right.\\
& \left. + \frac{\sqrt{\pi }\,\Gamma \left(\frac{d-3}{2}+l\right)}{ 2^{3-l-\widehat{\Delta}} {\hat\chi}^{\frac{d-3}{2}+l}} \left(\frac{2 b_1b_2\hat{f}_{12{{\widehat{\mathcal{O}}^{(l)}}}}}{\Gamma \left(\frac{l+\widehat{\Delta}+1}{2}\right) \Gamma \left(\frac{d+l-\widehat{\Delta}-1}{2}\right)}\right.\right.\\ & \left.\left.+\frac{b_1^2\hat{f}_{11{{\widehat{\mathcal{O}}^{(l)}}}}(\widehat{\Delta}+l-1)  (d-\widehat{\Delta}+l-2)+b_2^2\hat{f}_{22{{\widehat{\mathcal{O}}^{(l)}}}}}{(\widehat{\Delta}+l-1) \Gamma \left(\frac{l+\widehat{\Delta}}{2}\right) \Gamma \left(\frac{d+l-\widehat{\Delta}}{2}\right)}\right)+\dots\right]~,
\end{split}
\end{align}
up to higher powers of $\hat\chi$. Such unphysical singularities cancel from the r.h.s. of \eqref{phiphilDiag} precisely when the OPE relations \eqref{ssspinconstr} are satisfied, such that the analytic result at $y_1=y_2$ is
\begin{align}
\begin{split}\label{fullresDiageven}
\langle \phi(\vec{x}_1,y)\phi(\vec{x}_2,y)&{{\widehat{\mathcal{O}}^{(l)}}}(\theta,\infty)\rangle=b_1b_2\hat{f}_{12{{\widehat{\mathcal{O}}^{(l)}}}}\frac{P_\parallel^{(l)}(\hat{x}_{12},\theta)}{y^{d-2-\widehat{\Delta}}}\\
&\times\frac{\sqrt{\pi } \Gamma \left(\frac{3-d}{2}-l\right) \left[1-\cot \left(\frac{1}{2} \pi  (\widehat{\Delta}+l)\right) \cot \left(\frac{1}{2} \pi  (d-\widehat{\Delta}+l)\right)\right]}{2^{d-\widehat{\Delta}+l-1} \Gamma \left(1-\frac{l+\widehat{\Delta}}{2}\right) \Gamma \left(\frac{\widehat{\Delta}-d-l+4}{2} \right)}\\
&\times{\hat\chi} ^{l/2}\, _2F_1\left(\frac{d+l-\widehat{\Delta}-2}{2},\frac{d+l-\widehat{\Delta}-1}{2} ;\frac{d-1}{2}+l;-\frac{\hat\chi }{4}\right)~.
\end{split}
\end{align}

When $\widehat{\Delta}$ approaches some special integer dimensions some of the boundary blocks in the r.h.s. of \eqref{phiphilDiag} are themselves regular and \eqref{asymph_blocks} is not valid. This can happen when:
\begin{itemize}
	\item The dimension of $\widehat{\mathcal{O}}$ equals that of a double-twist combination of $\widehat{O}_1$ and $\widehat{O}_1$
	\begin{align}
	\widehat{\Delta}=d+l+2n-2,\quad n \in \mathbb{N},
	\end{align}
	\item The dimension of $\widehat{\mathcal{O}}$ equals that of a double-twist combination of $\widehat{O}_2$ and $\widehat{O}_2$
	\begin{align}
	\widehat{\Delta}=d+l+2n,\quad n \in \mathbb{N},
	\end{align}
	\item The dimension of $\widehat{\mathcal{O}}$ equals that of a double-twist combination of $\widehat{O}_1$ and $\widehat{O}_2$
	\begin{align}
	\widehat{\Delta}=d+l+2n-1,\quad n \in \mathbb{N}.
	\end{align}
\end{itemize}
We then analyse these special cases separately. Requiring the cancellation of any residual singularity on the r.h.s. of \eqref{phiphilDiag}, will again impose certain relations between the boundary OPE coefficients. It is reassuring to see that these special cases are captured by the appropriate limits of the general result \eqref{ssspinconstr}.

We now discuss the case when $l$ is an odd integer. Starting from \eqref{phiphiOgen}, we need to set $\hat{f}_{11{\widehat{\mathcal{O}}^{(l)}}}=\hat{f}_{22{\widehat{\mathcal{O}}^{(l)}}}=0$ (as dictated by Bose symmetry) so that the three-point function is proportional to $\hat{f}_{12{\widehat{\mathcal{O}}^{(l)}}}$. We then study analyticity around $\vec{x}_1=\vec{x}_2$ for finite $y_{12}\equiv y_1-y_2$. For generic values of $d,l,\widehat{\Delta}$ this correlator features unphysical singularities, since
\begin{align}
\begin{split}\label{fullresDiagoddlead}
& \langle \phi(\vec{x}_1,y_1)\phi(\vec{x}_2,y_2)\widehat{\mathcal{O}}^{(l)} (\theta,\infty)\rangle\underset{\vec{x}_{12}\rightarrow 0}{\sim}b_1 b_2\hat{f}_{12\widehat{\mathcal{O}}^{(l)}}\left(- \vec{x}_{12}\cdot {\theta}\right)^l\\
&\hspace{-0.5cm}\times\left(-\frac{\sqrt{\pi}}{2}\right)\left(\frac{y_{12}^{\widehat{\Delta}+l-1}}{|\vec{x}_{12}|^{d+2l-3}} \frac{ \Gamma \left(\frac{d-3}{2}+l\right)}{\Gamma \left(\frac{l+\widehat{\Delta}+1}{2}\right) \Gamma \left(\frac{d+l-\widehat{\Delta}-1}{2} \right)}+\frac{\Gamma \left(\frac{3-d}{2}-l\right) (y_{12})^{2-d+\widehat{\Delta}-l}}{\Gamma \left(1-\frac{l}{2}-\frac{\widehat{\Delta}}{2}\right) \Gamma \left(\frac{4-d-l+\widehat{\Delta}}{2}\right)}+\dots\right)~,
\end{split}
\end{align}
up to subleading terms. Because of the first term in the above expression, which is singular for $d\geq 3$ (for $d=3$ and $l=0$ the singularity is logarithmic), for generic $\widehat{\Delta}$ we must set $\hat{f}_{12{\widehat{\mathcal{O}}^{(l)}}}=0$. On the other hand, the boundary blocks are themselves regular and the \eqref{fullresDiagoddlead} is not valid when the dimension of $\widehat{\mathcal{O}}$ equals
\begin{align}\label{O1O2double}
\widehat{\Delta}=d+l+2n-1,\quad n \in \mathbb{N}.
\end{align}
Again we see that the relations \eqref{ssspinconstr}, together with the constraints from Bose symmetry \eqref{Bosesymm}, promptly capture these special cases. The analytic correlator \eqref{phiphiOgen} on the special dimensions \eqref{O1O2double} then reads
\begin{align}
\begin{split}\label{fullresodd}
\langle& \phi(\vec{x}_1,y_1)\phi(\vec{x}_2,y_2)\widehat{\mathcal{O}}^{(l)} (\theta,\infty)\rangle=b_1 b_2\hat{f}_{12\widehat{\mathcal{O}}^{(l)}}\left(- \vec{x}_{12}\cdot {\theta}\right)^l\\
&\times\left(-\frac{\sqrt{\pi}}{2}\right)\frac{\Gamma \left(\frac{3-d}{2}-l\right)}{\Gamma \left(n+\frac{3}{2}\right) \Gamma \left(\frac{3-d}{2}-l-n\right)} y_{12}^{2n+1}\, _2F_1\left(-n-\frac{1}{2},-n;\frac{d-1}{2}+l;-\frac{1}{w_-}\right)~.
\end{split}
\end{align}

\subsection{Matching with the bulk}\label{ss:mtchBulk}
Owing the results from the previous subsection, we are now ready to discuss the consequences of the bulk-boundary crossing symmetry for the three-point functions \eqref{phiphiOgenapp}. 

The first step is to derive the leading terms in the bulk channel expansion of the correlator \eqref{phiphiOgenapp}. To this end, recall that the $\phi\times \phi$ OPE \eqref{phiphiOPE} contains a scalar $\phi^2$ and infinitely many conserved currents $J_\ell$, with $\ell \in 2\mathbb{N}$ and $\Delta_\ell=d+\ell-2$. The bulk-boundary two-point functions between $J_\ell$ and any boundary operator $\widehat{\mathcal{O}}^{(l)}$ are further constrained by current conservation. The operator $\widehat{\mathcal{O}}^{(l)}$ can appear in the bOPE of $J_\ell$ if 
\begin{align}
\langle \partial_{\mu} J_\ell^{\mu \mu_1\dots \mu_{\ell-1}}(\vec{x},y) \widehat{\mathcal{O}}^{a_1\dots a_l}(0)\rangle =0.
\end{align}
For $l<\ell$, this condition is satisfied only if $\widehat{\Delta}=\Delta_\ell$, so that $\widehat{\mathcal{O}}$ is a protected boundary primary. On the other hand, for $l=\ell$, conservation is automatically ensured with no extra conditions on $\widehat{\Delta}$ \footnote{Indeed, $\langle (\partial J_\ell)\, \widehat{\mathcal{O}}_\ell\rangle=\langle {J}_{\ell-1} \widehat{\mathcal{O}}_\ell\rangle$ must vanish since ${J}_{\ell-1}$ does not contain any spin $\ell$ component in its bOPE.}, so $\widehat{\mathcal{O}}$ is unprotected. This is of course compatible with the Ward identities \eqref{ward_currents}.

 We now plug the $\phi\times \phi$ OPE into \eqref{phiphiOgenapp}, impose the selection rules from conservation in order to figure out which bulk primary can couple to $\widehat{\mathcal{O}}^{(l)}$ and finally compare to the boundary channel expansion. We conclude that:
\begin{itemize}
	\item When $\widehat{\Delta}\neq d+\ell-2$ and $l$ is odd, $\widehat{\mathcal{O}}^{(l)}$ cannot couple to any bulk operator in the $\phi \times \phi$, and the three-point function must vanish. This perfectly matches with the expectations from the boundary channel.

	\item When $\widehat{\Delta}\neq d+\ell-2$ and $l$ is even, $\widehat{\mathcal{O}}^{(l)}$ can only couple to a spin $l$ bulk current $J_l$ (or to $\phi^2$ when $l=0$). This is consistent with what we expect from the the boundary channel, where we are left with only one unknown OPE coefficient $\hat{f}_{12{\widehat{\mathcal{O}}^{(l)}}}$. In either case, from the leading bulk OPE we have
	\begin{align}\label{bulkOPEspinlgen}
	\langle \phi(\vec{x}_1,y)\phi(\vec{x}_2,y){{\widehat{\mathcal{O}}^{(l)}}} (\theta,\infty)\rangle& \underset{\vec{x}_{12}\rightarrow 0}{\sim}\frac{c_{\phi\phi l}\,b_{J_l\widehat{\mathcal{O}}^{(l)}}}{C_{J_l}}P_\parallel^{(l)}(\hat{x}_{12},\theta)\frac{{\hat\chi}^{{l}/{2}}}{y^{d-2-\widehat{\Delta}}}+\dots,
	\end{align}
	and, after comparing to \eqref{fullresDiageven} we find
	\begin{align}\label{aux1}
	\hspace{-0.5cm}\frac{c_{\phi\phi l}b_{{J_l}\widehat{\mathcal{O}}^{(l)}}}{C_{J_l}}={b_1 b_2 {}\hat{f}_{12{\widehat{\mathcal{O}}^{(l)}}}}{}\frac{\sqrt{\pi }\, \Gamma \left(\frac{3-d}{2}-l\right) \left[1-\cot \left(\frac{1}{2} \pi  (\widehat{\Delta}+l)\right) \cot \left(\frac{1}{2} \pi  (d-\widehat{\Delta}+l)\right)\right]}{2^{d-1-\widehat{\Delta}+l}\Gamma \left(1-\frac{l+\widehat{\Delta}}{2}\right) \Gamma \left(\frac{\widehat{\Delta}-l-d+4}{2}\right)}~.
	\end{align}
	The result for a scalar ($l=0$) operator $\widehat{\mathcal{O}}$ is simply obtained from the former by setting $c_{\phi\phi 0}=C_{J_0}=1$ and $b_{{J_0}\widehat{\mathcal{O}}^{(0)}}\equiv b_{\phi^2\widehat{\mathcal{O}}}$.

We can use the result above in order to re-interpret the expression for \eqref{phiphiOgenapp} obtained using the boundary OPE in terms of the bulk OPE, and compute the corresponding bulk block. This procedure is unambiguous, since in both channels there is just one undetermined OPE coefficient. In practice, we solve \eqref{aux1} for $\hat{f}_{12 \widehat{\mathcal{O}}^{(l)}}$ and plug the result into \eqref{fullresDiageven}. We find
	\begin{align}
	\begin{split}\label{bulkblockinvert}
	& \mathcal{W}_{J_l}^{\phi\phi\widehat{\mathcal{O}}^{(l)}}(\vec{x}_{12},y,y) \\&=\frac{P_\parallel^{(l)}(\hat{x}_{12},\theta)\hat{\chi}^{{l}/{2}}}{y^{2\Delta_{\phi}-\widehat{\Delta}}}
	\, _2F_1\left(\frac{d+l-\widehat{\Delta}-2}{2} ,\frac{d+l-\widehat{\Delta}-1}{2} ;\frac{d+2l-1}{2};-\frac{\hat{\chi} }{4}\right)~.
	\end{split}
	\end{align}
	As a consistency check, note that for $l=0$ the above expression reproduces the block $\mathcal{W}_{\phi^2}^{\phi\phi\widehat{\mathcal{O}}}(\vec{x}_{12},y,y)$, which was computed explicitly in \eqref{closedbulkblockscalar}. 
The same logic can be applied to compute the bulk blocks at generic transverse positions $y_1, y_2$, starting from the boundary channel decomposition \eqref{phiphiOgen}.

	\item When $\widehat{\Delta}=d+\ell-2$ and $l$ is even there are two cases. For $\ell>l$, the primary 
	$\widehat{\mathcal{O}}^{(l)}$ can couple to both $J_l$ and $J_\ell$. The number of undetermined bulk OPE coefficients then matches that of the boundary ones ($\hat{f}_{11\widehat{\mathcal{O}}^{(l)}}$ and $\hat{f}_{22\widehat{\mathcal{O}}^{(l)}}$). As an example, in section \ref{subsec:dispDet} we explicitly solve the bulk-to-boundary bootstrap for the case of $\ell=2$ with $l=0$, but similar results can be obtained for the more general case of $\text{D}^{(l)}_\ell$. When $\ell=l$ the operator $\widehat{\mathcal{O}}^{(l)}$ can only couple to $J_l$, and this matches with the number of undetermined boundary OPE coefficients ($\hat{f}_{11\widehat{\mathcal{O}}^{(l)}}$).

	\item When $\widehat{\Delta}=d+\ell-2$ and $l$ is odd the only possible bulk contribution comes from the spin $\ell$ currents $J_{\ell}$. From the leading bulk OPE limit at $|\vec{x}_{12}|=0$ we have
	\begin{align}
	\langle \phi(\vec{x}_1,y_1)\phi(\vec{x}_2,y_2){\text{V}_\ell^{(l)}} (\theta,\infty)\rangle&\underset{{x}_{12}\rightarrow 0}{\sim}\frac{c_{\phi\phi \ell}}{C_{J_\ell}}b_{J_\ell\text{V}_\ell^{(l)}}(y_{12})^{\ell-l}\left( \vec{x}_{12}\cdot {\theta}\right)^l+\dots,
	\end{align}
	where $y_{12}=y_1-y_2$. So, after comparing to \eqref{fullresodd} (note that $\ell-l=2n+1$) we find
	\begin{align}\label{aux2}
	\frac{c_{\phi\phi \ell}b_{J_\ell\text{V}_\ell^{(l)}}}{C_{J_\ell}}={b_1 b_2 {}\hat{f}_{12{\text{V}_\ell^{(l)}}}}{}\frac{\sqrt{\pi } \Gamma \left(\frac{3-d}{2}-l\right)}{2 \Gamma \left(\frac{\ell-l+2}{2}\right) \Gamma \left(\frac{4-d-\ell-l}{2}\right)}.
	\end{align}
	The Ward identity \eqref{ward_currents} further relates the coefficient $b_{J_\ell\text{V}_\ell^{(l)}}$ to the coefficient in the two-point function of $\text{V}_\ell^{(l)}$, e.g. for the flux operator $\text{V}_2^{(1)}\equiv \text{V}^{(1)}$ the coefficient in eq. \eqref{VV}. In the case of $\text{V}^{(1)}$, upon plugging the value of $c_{\phi\phi T}$ and $C_T$ given in eq. \eqref{cphiphiT2}, the result \eqref{aux2} gives precisely the first equality of eq. \eqref{bTVlambda_rel}. The second equality is obtained upon using that $b_{T\text{V}^{(1)}}=2 \widehat{C}_{\text{V}^{(1)}}$, as dictated by the Ward identity \eqref{ward_currents}.

\end{itemize}

\subsection{Displacement Ward identity}\label{subsec:dispDet}
In this appendix we derive the displacement Ward identity presented in section \ref{ss:displ_flux}. The starting point is the three-point function of the displacement operator $\text{D}$ with the free bulk scalar. From \eqref{phiphiOgen} and after imposing the OPE relations \eqref{ssspinconstr}, this reads
\begin{align}\label{Displ3pt}
\langle \phi(\vec{x}_1,y_1)\phi(\vec{x}_2,y_2)\text{D}(\infty)\rangle= y_1 y_2 b_2^2 {}\hat{f}_{22\text{D}}+b_1^2 {}\hat{f}_{11\text{D}} \left[|\vec{x}_{12}|^2-(d-1) \left(y_1^2+y_2^2\right)\right].
\end{align}
We want to match this expression against the bulk OPE channel expansion. As we discussed in the main text -- see around \eqref{phiphiDblockexpalead} -- this receives a contribution from the $\phi^2$ as well as from the stress tensor. The complete expression, i.e. including contributions from bulk descendants, is
\begin{align}\label{phiphiDblockexpa}
\langle \phi(x_1)\phi(x_2)\text{D} (\infty)\rangle &=b_{\phi^2 \text{D}}\mathcal{W}_{\phi^2}^{\phi\phi\text{D}}(\vec{x}_{12},y_1,y_2)+\frac{c_{\phi\phi T}}{C_T}\,{x}_{12}^\mu {x}_{12}^\nu\langle T_{\mu\nu}(x_2)\text{D}(\infty)\rangle.
\end{align}
The first term in the r.h.s. of the above equation is the $\langle\phi^2\text{D}\rangle$ bulk block, which is computed by \eqref{bulkblockgen}
\begin{align}\label{Displphi2block}
\mathcal{W}_{\phi^2}^{\phi\phi\text{D}}(\vec{x}_{12},y_1,y_2)&=\frac{(d-1)(y_1+y_2)^2-|\vec{x}_{12}|^2}{4(d-1)}.
\end{align}
The second term is the contribution from the bulk stress tensor and reads \cite{McAvity:1993ue,McAvity:1995zd}
\begin{align}\label{TD2pt}
\langle T_{\mu\nu}(x)\text{D}(\infty)\rangle =&{b_{T\text{D}}}{}\left(\delta_{\mu y}\delta_{\nu y}-\frac{1}{d}\delta_{\mu\nu}\right), \quad b_{T\text{D}}=\frac{d\, C_{\text{D}}}{d-1}.
\end{align}
Note bulk descendant operators of $T^{\mu\nu}$ do not enter into \eqref{phiphiDblockexpa}, since \eqref{TD2pt} is a constant. One can further use the Ward identities for the displacement operator \cite{McAvity:1993ue,McAvity:1995zd} to relate the bOPE coefficient $b_{\phi^2 \text{D}}$ to the one-point function of $\phi^2$:
\begin{align}\label{cphiphiT}
b_{\phi^2 \text{D}}=-a_{\phi^2}\,\, \frac{2^d (d-2)}{S_d}, \quad S_d\equiv\text{Vol}(S^{d-1})=\frac{2 \pi ^{d/2}}{\Gamma \left(\frac{d}{2}\right)}.
\end{align}
We can now equate \eqref{Displ3pt} to \eqref{phiphiDblockexpa} and solve for $\hat{f}_{11\text{D}}$ and $\hat{f}_{22\text{D}}$. The result is
\begin{align}\label{displRel2}
{}\hat{f}_{11\text{D}}=\frac{a_{\phi^2} 2^d C_T (d-2)-4  C_{\text{D}}c_{\phi\phi T} S_d}{4C_T (d-1) S_d b_1^2},\qquad 
{}\hat{f}_{22\text{D}}=-\frac{a_{\phi^2} 2^d C_T (d-2)+4 C_{\text{D}} c_{\phi\phi T} S_d}{2C_T S_d b_2^2}.
\end{align}
The final formula \eqref{displRel} is obtained by plugging in the above expression the values \eqref{cphiphiT2} of $c_{\phi\phi T}$ and $C_T$ corresponding to a $d$-dimensional free scalar field. It is pleasant to see that the final result \eqref{displRel} is consistent with the Ward identity \cite{McAvity:1993ue}
\begin{align}
\int\,\mathrm{d}^{d-1}\vec{x}\, \langle \phi(x_1)\phi(x_2)\text{D}(\vec{x})\rangle=(\partial_{y_1}+\partial_{y_2})\langle \phi(x_1)\phi(x_2)\rangle.
\end{align}

\section{Crossing equations in a vectorial form}\label{app:crossing}
In section \ref{ssec:crossingeq} we derived the system of 7 independent crossing equations \eqref{cross_eq_red}. The latter can be rewritten in a vectorial form by introducing the 7-component vectors  of $3\times 3$ matrices $\vec{V}_{\widehat{\Delta},l}$
\begin{align}\label{crossEqVecGen}
0=\sum_{\widehat{\mathcal{O}},l}\begin{pmatrix}
{}\hat{f}_{11}{}^{{\widehat{\mathcal{O}}^{(l)}}} &
{}\hat{f}_{12}{}^{{\widehat{\mathcal{O}}^{(l)}}} & {}\hat{f}_{22}{}^{{\widehat{\mathcal{O}}^{(l)}}}
\end{pmatrix} \vec{V}_{\widehat{\Delta},l}\begin{pmatrix}
{}\hat{f}_{11{\widehat{\mathcal{O}}^{(l)}}} \\
{}\hat{f}_{12{\widehat{\mathcal{O}}^{(l)}}} \\
{}\hat{f}_{22{\widehat{\mathcal{O}}^{(l)}}} \\
\end{pmatrix}.
\end{align}
The explicit form of $\vec{V}_{\widehat{\Delta},l}$ is given in \eqref{Vdef}. The odd $l$ terms in the above expression are subjected to further restrictions. Firstly, Bose symmetry implies that ${}\hat{f}_{11{\widehat{\mathcal{O}}^{(l)}}}={}\hat{f}_{22{\widehat{\mathcal{O}}^{(l)}}}=0$. Secondly, the odd-$l$ primaries must have scaling dimensions $\widehat{\Delta}=d+l+2n-1$, with $n \in \mathbb{N}$, as follows from the exact relations \eqref{ssspinconstr}. It is then convenient to rewrite \eqref{crossEqVecGen} as
\begin{align}
\begin{split}\label{crossEqVecGen2}
0=& \sum_{\widehat{\mathcal{O}},l= \text{even}}\begin{pmatrix}
{}\hat{f}_{11}{}^{{\widehat{\mathcal{O}}^{(l)}}} &
{}\hat{f}_{12}{}^{{\widehat{\mathcal{O}}^{(l)}}} & {}\hat{f}_{22}{}^{{\widehat{\mathcal{O}}^{(l)}}}
\end{pmatrix} \vec{V}_{\widehat{\Delta},l}\begin{pmatrix}
{}\hat{f}_{11{\widehat{\mathcal{O}}^{(l)}}} \\
{}\hat{f}_{12{\widehat{\mathcal{O}}^{(l)}}} \\
{}\hat{f}_{22{\widehat{\mathcal{O}}^{(l)}}} \\
\end{pmatrix} \\ & +\sum_{\underset{n=0,1,\dots}{\underset{\widehat{\Delta}=d+l+2n-1}{l=\text{odd}}}}{}\hat{f}_{12}{}^{{\widehat{\mathcal{O}}^{(l)}}}\hat{f}_{12{\widehat{\mathcal{O}}^{(l)}}}\begin{pmatrix}
0 &
1 & 0
\end{pmatrix} \vec{V}_{\widehat{\Delta},l}\begin{pmatrix}
0\\
1 \\
0 \\
\end{pmatrix}~.
\end{split}
\end{align}
For numerical purposes it is convenient to isolate, in the crossing equations above, the contributions from the primaries with fixed dimensions from those in the continuum. Such special primaries are the identity $\id$ (for which \eqref{ssspinconstr} implies ${\hat{f}}_{12\id}=0$ and we choose the normalization ${\hat{f}}_{ii\,\id}\equiv {\hat{f}}_{ii}=1$), as well as the boundary modes of the bulk higher-spin currents, $\text{D}_\ell^{(l)}$ and $\text{V}_\ell^{(l+1)}$ of spin $l$ and $l+1$ in the notation of section \ref{ss:reg_OPE}. Upon implementing the exact relations \eqref{ssspinconstr}, we rewrite \eqref{crossEqVecGen2} as follows
\begin{align}
\begin{split}\label{crossEqVecGen3}
0=&\underbrace{\begin{pmatrix}
	1 &
	0&
	1
	\end{pmatrix} \vec{V}_{0,0}\begin{pmatrix}
	1 \\
	0\\
	1 \\
	\end{pmatrix}}_{\equiv\vec{V}_{\id}}+\sum_{\widehat{\mathcal{O}},l= \text{even}}{}{\hat{f}_{12{\widehat{\mathcal{O}}^{(l)}}}\hat{f}_{12}{}^{\widehat{\mathcal{O}}^{(l)}}}{}\underbrace{\begin{pmatrix}
	\kappa_1(\widehat{\Delta},l) &
	1 & \kappa_2(\widehat{\Delta},l)
	\end{pmatrix} \vec{V}_{\widehat{\Delta},l}\begin{pmatrix}
	\kappa_1(\widehat{\Delta},l) \\
	1 \\
	\kappa_2(\widehat{\Delta},l) \\
	\end{pmatrix}}_{\equiv\vec{V}_{+,\widehat{\Delta},l}}\\
+&\sum_{\underset{n=0,1,\dots}{\underset{\widehat{\Delta}=d+l+2n-1}{l=\text{odd}}}}{}{\hat{f}_{12{\widehat{\mathcal{O}}^{(l)}}}\hat{f}_{12}{}^{{\widehat{\mathcal{O}}^{(l)}}}}\underbrace{\begin{pmatrix}
	0 &
	1 & 0
	\end{pmatrix} \vec{V}_{\widehat{\Delta},l}\begin{pmatrix}
	0 \\
	1 \\
	0 \\
	\end{pmatrix}}_{\equiv \vec{V}_{-,\widehat{\Delta},l}}\\
+&\sum_{{\ell \in 2\mathbb{N}}{} }\,\sum_{{}{\underset{\widehat{\Delta}=d+\ell-2}{l<\ell, \text{even}}}}\underbrace{\begin{pmatrix}
	{}\hat{f}_{11{\widehat{\mathcal{O}}^{(l)}}} &
	0&
	{}\hat{f}_{22{\widehat{\mathcal{O}}^{(l)}}}
	\end{pmatrix} \vec{V}_{\widehat{\Delta},l}\begin{pmatrix}
	{}\hat{f}_{11}{}^{{\widehat{\mathcal{O}}^{(l)}}} \\
	0\\
	{}\hat{f}_{22}{}^{{\widehat{\mathcal{O}}^{(l)}}} \\
	\end{pmatrix}}_{\equiv\begin{pmatrix}
	{}\hat{f}_{11{\widehat{\mathcal{O}}^{(l)}}} &
	{}\hat{f}_{22{\widehat{\mathcal{O}}^{(l)}}}
	\end{pmatrix} \vec{V}_{\textbf{0},\widehat{\Delta},l}\begin{pmatrix}
	{}\hat{f}_{11}{}^{{\widehat{\mathcal{O}}^{(l)}}} \\
	{}\hat{f}_{22}{}^{{\widehat{\mathcal{O}}^{(l)}}} \\
	\end{pmatrix}}~.
\end{split}	
\end{align}
The first line in the expression above accounts for the identity $\id$ as well as for the unprotected, even-spin operators. The second line accounts for the odd-spin operators i.e. belonging to the family $\text{V}_\ell^{(l)}$. The third line contains even-spin protected operators i.e. belonging to the family $\text{D}_\ell^{(l)}$. The 7-component vectors $\vec{V}_{\pm,\widehat{\Delta},l},\vec{V}_{\id,\widehat{\Delta},l}$ are defined in \eqref{VminusDefIgen},\eqref{VplusDefgen}. The quantities $\vec{V}_{\textbf{0}\widehat{\Delta},l}$ are 7-component vectors of $2\times 2$ matrices defined in \eqref{Vodefgen}. 

\begin{align}\footnotesize\label{Vdef}
\vec{V}_{\widehat{\Delta},l}=\left(
\begin{array}{c}
\left(
\begin{array}{ccc}
F^{11,11}_{-,\widehat{\Delta},l}(u,v) & 0 & 0 \\
0 & 0 & 0 \\
0 & 0 & 0 \\
\end{array}
\right)\\
\left(
\begin{array}{ccc}
0 & 0 & 0 \\
0 & 0 & 0 \\
0 & 0 & F^{22,22}_{-,\widehat{\Delta},l}(u,v) \\
\end{array}
\right)\\
\left(
\begin{array}{ccc}
0 & \frac{1}{2} F^{11,12}_{-,\widehat{\Delta},l}(u,v) & 0 \\
\frac{1}{2} F^{11,12}_{-,\widehat{\Delta},l}(u,v) & 0 & 0 \\
0 & 0 & 0 \\
\end{array}
\right)\\
\left(
\begin{array}{ccc}
0 & 0 & 0 \\
0 & 0 & \frac{1}{2} F^{12,22}_{-,\widehat{\Delta},l}(u,v) \\
0 & \frac{1}{2} F^{12,22}_{-,\widehat{\Delta},l}(u,v) & 0 \\
\end{array}
\right)\\
\left(
\begin{array}{ccc}
0 & 0 & 0 \\
0 & F^{12,12}_{-,\widehat{\Delta},l}(u,v) & 0 \\
0 & 0 & 0 \\
\end{array}
\right)\\
\left(
\begin{array}{ccc}
0 & 0 & \frac{1}{2} F^{11,22}_{-,\widehat{\Delta},l}(u,v) \\
0 & (-1)^s F^{12,21}_{-,\widehat{\Delta},l}(u,v) & 0 \\
\frac{1}{2} F^{11,22}_{-,\widehat{\Delta},l}(u,v) & 0 & 0 \\
\end{array}
\right)\\
\left(
\begin{array}{ccc}
0 & 0 & \frac{1}{2} F^{11,22}_{+,\widehat{\Delta},l}(u,v) \\
0 & -(-1)^s F^{12,21}_{+,\widehat{\Delta},l}(u,v) & 0 \\
\frac{1}{2} F^{11,22}_{+,\widehat{\Delta},l}(u,v) & 0 & 0 \\
\end{array}
\right)\\
\end{array}
\right),
\end{align}

\begin{align}\label{VminusDefIgen}
\footnotesize
\vec{V}_{\id}=\left(
\begin{array}{c}
F^{11,11}_{-,\widehat{\Delta},l}(u,v) \\
F^{22,22}_{-,\widehat{\Delta},l}(u,v) \\
0 \\
0 \\
0 \\
F^{11,22}_{-,\widehat{\Delta},l}(u,v) \\
F^{11,22}_{+,\widehat{\Delta},l}(u,v) \\
\end{array}
\right),&\qquad\vec{V}_{-,\widehat{\Delta},l}=\left(
\begin{array}{c}
0 \\
0 \\
0 \\
0 \\
F^{12,12}_{-,\widehat{\Delta},l}(u,v) \\
- F^{12,21}_{-,\widehat{\Delta},l}(u,v) \\
F^{12,21}_{+,\widehat{\Delta},l}(u,v) \\
\end{array}
\right),
\end{align}

\begin{align}\label{VplusDefgen}
\footnotesize
\vec{V}_{+,\widehat{\Delta},l}=
\left(
\begin{array}{c}
\kappa_1(\widehat{\Delta},l)^2 F^{11,11}_{-,\widehat{\Delta},l}(u,v) \\
\kappa_2(\widehat{\Delta},l)^2 F^{22,22}_{-,\widehat{\Delta},l}(u,v) \\
\kappa_1(\widehat{\Delta},l) F^{11,12}_{-,\widehat{\Delta},l}(u,v) \\
\kappa_2(\widehat{\Delta},l) F^{12,22}_{-,\widehat{\Delta},l}(u,v) \\
F^{12,12}_{-,\widehat{\Delta},l}(u,v) \\
\kappa_1(\widehat{\Delta},l) \kappa_2(\widehat{\Delta},l) F^{11,22}_{-,\widehat{\Delta},l}(u,v)+F^{12,21}_{-,\widehat{\Delta},l}(u,v) \\
\kappa_1(\widehat{\Delta},l) \kappa_2(\widehat{\Delta},l) F^{11,22}_{+,\widehat{\Delta},l}(u,v)-F^{12,21}_{+,\widehat{\Delta},l}(u,v) \\
\end{array}
\right),
\end{align}

\begin{align}\label{Vodefgen}
\footnotesize
\vec{V}_{\textbf{0},{\widehat{\Delta},l}}=\left(
\begin{array}{c}
\left(
\begin{array}{cc}
F^{11,11}_{-,\widehat{\Delta},l}(u,v) & 0 \\
0 & 0 \\
\end{array}
\right)\\
\left(
\begin{array}{cc}
0 & 0 \\
0 & F^{22,22}_{-,\widehat{\Delta},l}(u,v) \\
\end{array}
\right)\\
\left(
\begin{array}{cc}
0 & 0 \\
0 & 0 \\
\end{array}
\right)\\
\left(
\begin{array}{cc}
0 & 0 \\
0 & 0 \\
\end{array}
\right)\\
\left(
\begin{array}{cc}
0 & 0 \\
0 & 0 \\
\end{array}
\right)\\
\left(
\begin{array}{cc}
0 & \frac{1}{2} F^{11,22}_{-,\widehat{\Delta},l}(u,v) \\
\frac{1}{2} F^{11,22}_{-,\widehat{\Delta},l}(u,v) & 0 \\
\end{array}
\right)\\
\left(
\begin{array}{cc}
0 & \frac{1}{2} F^{11,22}_{+,\widehat{\Delta},l}(u,v) \\
\frac{1}{2} F^{11,22}_{+,\widehat{\Delta},l}(u,v) & 0 \\
\end{array}
\right)
\end{array}
\right).
\end{align}

\bibliography{bib}

\providecommand{\href}[2]{#2}\begingroup\raggedright\begin{thebibliography}{10}

\bibitem{Cardy:2004hm}
J.~L. Cardy, {\it {Boundary conformal field theory}},
  \href{http://arxiv.org/abs/hep-th/0411189}{{\tt hep-th/0411189}}.

\bibitem{Rattazzi:2008pe}
R.~Rattazzi, V.~S. Rychkov, E.~Tonni, and A.~Vichi, {\it {Bounding scalar
  operator dimensions in 4D CFT}},  {\em JHEP} {\bf 12} (2008) 031,
  [\href{http://arxiv.org/abs/0807.0004}{{\tt arXiv:0807.0004}}].

\bibitem{Poland:2018epd}
D.~Poland, S.~Rychkov, and A.~Vichi, {\it {The Conformal Bootstrap: Theory,
  Numerical Techniques, and Applications}},  {\em Rev. Mod. Phys.} {\bf 91}
  (2019) 015002, [\href{http://arxiv.org/abs/1805.04405}{{\tt
  arXiv:1805.04405}}].

\bibitem{Liendo:2012hy}
P.~Liendo, L.~Rastelli, and B.~C. van Rees, {\it {The Bootstrap Program for
  Boundary CFT$_d$}},  {\em JHEP} {\bf 07} (2013) 113,
  [\href{http://arxiv.org/abs/1210.4258}{{\tt arXiv:1210.4258}}].

\bibitem{McAvity:1993ue}
D.~M. McAvity and H.~Osborn, {\it {Energy momentum tensor in conformal field
  theories near a boundary}},  {\em Nucl. Phys.} {\bf B406} (1993) 655--680,
  [\href{http://arxiv.org/abs/hep-th/9302068}{{\tt hep-th/9302068}}].

\bibitem{McAvity:1995zd}
D.~McAvity and H.~Osborn, {\it {Conformal field theories near a boundary in
  general dimensions}},  {\em Nucl. Phys. B} {\bf 455} (1995) 522--576,
  [\href{http://arxiv.org/abs/cond-mat/9505127}{{\tt cond-mat/9505127}}].

\bibitem{Witten:2003ya}
E.~Witten, {\it {SL(2,Z) action on three-dimensional conformal field theories
  with Abelian symmetry}},  \href{http://arxiv.org/abs/hep-th/0307041}{{\tt
  hep-th/0307041}}.

\bibitem{Gaiotto:2008ak}
D.~Gaiotto and E.~Witten, {\it {S-Duality of Boundary Conditions In N=4 Super
  Yang-Mills Theory}},  {\em Adv.\ Theor.\ Math.\ Phys.} {\bf 13} (2009), no.~3
  721--896, [\href{http://arxiv.org/abs/0807.3720}{{\tt arXiv:0807.3720}}].

\bibitem{Seiberg:2016gmd}
N.~Seiberg, T.~Senthil, C.~Wang, and E.~Witten, {\it {A Duality Web in 2+1
  Dimensions and Condensed Matter Physics}},  {\em Annals Phys.} {\bf 374}
  (2016) 395--433, [\href{http://arxiv.org/abs/1606.01989}{{\tt
  arXiv:1606.01989}}].

\bibitem{DiPietro:2019hqe}
L.~Di~Pietro, D.~Gaiotto, E.~Lauria, and J.~Wu, {\it {3d Abelian Gauge Theories
  at the Boundary}},  {\em JHEP} {\bf 05} (2019) 091,
  [\href{http://arxiv.org/abs/1902.09567}{{\tt arXiv:1902.09567}}].

\bibitem{Ferrara:1972xe}
S.~Ferrara and G.~Parisi, {\it {Conformal covariant correlation functions}},
  {\em Nucl. Phys. B} {\bf 42} (1972) 281--290.

\bibitem{Ferrara:1972ay}
S.~Ferrara, A.~Grillo, and G.~Parisi, {\it {Nonequivalence between conformal
  covariant Wilson expansion in Euclidean and Minkowski space}},  {\em Lett.
  Nuovo Cim.} {\bf 5S2} (1972) 147--151.

\bibitem{Ferrara:1972uq}
S.~Ferrara, A.~Grillo, G.~Parisi, and R.~Gatto, {\it {The shadow operator
  formalism for conformal algebra. Vacuum expectation values and operator
  products}},  {\em Lett. Nuovo Cim.} {\bf 4S2} (1972) 115--120.

\bibitem{Ferrara:1973vz}
S.~Ferrara, A.~Grillo, G.~Parisi, and R.~Gatto, {\it {Covariant expansion of
  the conformal four-point function}},  {\em Nucl. Phys. B} {\bf 49} (1972)
  77--98. [Erratum: Nucl.Phys.B 53, 643--643 (1973)].

\bibitem{Lauria:2020emq}
E.~Lauria, P.~Liendo, B.~C. van Rees, and X.~Zhao, {\it {Line and surface
  defects for the free scalar field}},
  \href{http://arxiv.org/abs/2005.02413}{{\tt arXiv:2005.02413}}.

\bibitem{Paulos:2015jfa}
M.~F. Paulos, S.~Rychkov, B.~C. van Rees, and B.~Zan, {\it {Conformal
  Invariance in the Long-Range Ising Model}},  {\em Nucl.\ Phys.\ B} {\bf 902}
  (2016) 246--291, [\href{http://arxiv.org/abs/1509.00008}{{\tt
  arXiv:1509.00008}}].

\bibitem{Behan:2017emf}
C.~Behan, L.~Rastelli, S.~Rychkov, and B.~Zan, {\it {A scaling theory for the
  long-range to short-range crossover and an infrared duality}},  {\em J.
  Phys.} {\bf A50} (2017), no.~35 354002,
  [\href{http://arxiv.org/abs/1703.05325}{{\tt arXiv:1703.05325}}].

\bibitem{Behan:2017dwr}
C.~Behan, L.~Rastelli, S.~Rychkov, and B.~Zan, {\it {Long-range critical
  exponents near the short-range crossover}},  {\em Phys.\ Rev.\ Lett.} {\bf
  118} (2017), no.~24 241601, [\href{http://arxiv.org/abs/1703.03430}{{\tt
  arXiv:1703.03430}}].

\bibitem{Behan:2018hfx}
C.~Behan, {\it {Bootstrapping the long-range Ising model in three dimensions}},
   {\em J. Phys.} {\bf A52} (2019), no.~7 075401,
  [\href{http://arxiv.org/abs/1810.07199}{{\tt arXiv:1810.07199}}].

\bibitem{Poland:2011}
D.~Poland and D.~Simmons-Duffin, {\it Bounds on 4d conformal and superconformal
  field theories},  {\em JHEP} {\bf 05} (2011) 017,
  [\href{http://arxiv.org/abs/1009.2087}{{\tt arXiv:1009.2087}}].

\bibitem{ElShowk:2013}
S.~El-Showk and M.~F. Paulos, {\it Bootstrapping conformal field theories with
  the extremal functional method},  {\em Phys. Rev. Lett.} {\bf 111} (Dec,
  2013) 241601, [\href{http://arxiv.org/abs/1211.2810}{{\tt arXiv:1211.2810}}].

\bibitem{Dimofte:2012pd}
T.~Dimofte and D.~Gaiotto, {\it {An E7 Surprise}},  {\em JHEP} {\bf 10} (2012)
  129, [\href{http://arxiv.org/abs/1209.1404}{{\tt arXiv:1209.1404}}].

\bibitem{Gaiotto:2014gha}
D.~Gaiotto, {\it {Boundary F-maximization}},
  \href{http://arxiv.org/abs/1403.8052}{{\tt arXiv:1403.8052}}.

\bibitem{Gliozzi:2015qsa}
F.~Gliozzi, P.~Liendo, M.~Meineri, and A.~Rago, {\it {Boundary and Interface
  CFTs from the Conformal Bootstrap}},  {\em JHEP} {\bf 05} (2015) 036,
  [\href{http://arxiv.org/abs/1502.07217}{{\tt arXiv:1502.07217}}].

\bibitem{Giombi:2019enr}
S.~Giombi and H.~Khanchandani, {\it {$O(N)$ Models with Boundary Interactions
  and their Long Range Generalizations}},
  \href{http://arxiv.org/abs/1912.08169}{{\tt arXiv:1912.08169}}.

\bibitem{Osborn:1993cr}
H.~Osborn and A.~C. Petkou, {\it {Implications of conformal invariance in field
  theories for general dimensions}},  {\em Annals Phys.} {\bf 231} (1994)
  311--362, [\href{http://arxiv.org/abs/hep-th/9307010}{{\tt hep-th/9307010}}].

\bibitem{Costa:2011mg}
M.~S. Costa, J.~Penedones, D.~Poland, and S.~Rychkov, {\it {Spinning Conformal
  Correlators}},  {\em JHEP} {\bf 11} (2011) 071,
  [\href{http://arxiv.org/abs/1107.3554}{{\tt arXiv:1107.3554}}].

\bibitem{Billo:2016cpy}
M.~Bill\`o, V.~Gon\c{c}alves, E.~Lauria, and M.~Meineri, {\it {Defects in
  conformal field theory}},  {\em JHEP} {\bf 04} (2016) 091,
  [\href{http://arxiv.org/abs/1601.02883}{{\tt arXiv:1601.02883}}].

\bibitem{Komargodski:2016auf}
Z.~Komargodski and D.~Simmons-Duffin, {\it {The Random-Bond Ising Model in 2.01
  and 3 Dimensions}},  {\em J. Phys.} {\bf A50} (2017), no.~15 154001,
  [\href{http://arxiv.org/abs/1603.04444}{{\tt arXiv:1603.04444}}].

\bibitem{Behan:2017mwi}
C.~Behan, {\it {Conformal manifolds: ODEs from OPEs}},  {\em JHEP} {\bf 03}
  (2018) 127, [\href{http://arxiv.org/abs/1709.03967}{{\tt arXiv:1709.03967}}].

\bibitem{Stanic}
R.~Stanic, {\it {Renormalization group flows in quantum field theories with a
  boundary}},  Master's thesis, EPFL, 2019.

\bibitem{Herzog:2017xha}
C.~P. Herzog and K.-W. Huang, {\it {Boundary Conformal Field Theory and a
  Boundary Central Charge}},  {\em JHEP} {\bf 10} (2017) 189,
  [\href{http://arxiv.org/abs/1707.06224}{{\tt arXiv:1707.06224}}].

\bibitem{Prochazka:2019fah}
V.~Procházka and A.~Söderberg, {\it {Composite operators near the boundary}},
   {\em JHEP} {\bf 03} (2020) 114, [\href{http://arxiv.org/abs/1912.07505}{{\tt
  arXiv:1912.07505}}].

\bibitem{Kos:2014bka}
F.~Kos, D.~Poland, and D.~Simmons-Duffin, {\it {Bootstrapping Mixed Correlators
  in the 3D Ising Model}},  {\em JHEP} {\bf 11} (2014) 109,
  [\href{http://arxiv.org/abs/1406.4858}{{\tt arXiv:1406.4858}}].

\bibitem{Dolan:2000ut}
F.~A. Dolan and H.~Osborn, {\it {Conformal four point functions and the
  operator product expansion}},  {\em Nucl. Phys.} {\bf B599} (2001) 459--496,
  [\href{http://arxiv.org/abs/hep-th/0011040}{{\tt hep-th/0011040}}].

\bibitem{Dolan:2011dv}
F.~A. Dolan and H.~Osborn, {\it {Conformal Partial Waves: Further Mathematical
  Results}},  \href{http://arxiv.org/abs/1108.6194}{{\tt arXiv:1108.6194}}.

\bibitem{SimmonsDuffin:2015}
D.~Simmons-Duffin, {\it A semidefinite program solver for the conformal
  bootstrap},  {\em JHEP} {\bf 06} (2015) 174,
  [\href{http://arxiv.org/abs/1502.02033}{{\tt arXiv:1502.02033}}].

\bibitem{Landry:2019}
W.~Landry and D.~Simmons-Duffin, {\it {Scaling the semidefinite program solver
  SDPB}},  \href{http://arxiv.org/abs/1909.09745}{{\tt arXiv:1909.09745}}.

\bibitem{Go:2019}
M.~Go and Y.~Tachikawa, {\it {Autoboot: A generator of bootstrap equations with
  global symmetry}},  {\em JHEP} {\bf 06} (2019) 084,
  [\href{http://arxiv.org/abs/1903.10522}{{\tt arXiv:1903.10522}}].

\bibitem{Mortici:2010}
C.~Mortici, {\it New approximation formulas for evaluating the ratio of gamma
  functions},  {\em Math. Comput. Model.} {\bf 52} (2010), no.~1 425 -- 433.

\bibitem{Behan:2016}
C.~Behan, {\it {PyCFTBoot: A Flexible Interface for the Conformal Bootstrap}},
  {\em Commun. Comput. Phys.} {\bf 22} (2017), no.~1 1–38,
  [\href{http://arxiv.org/abs/1602.02810}{{\tt arXiv:1602.02810}}].

\bibitem{Go:2020}
M.~Go, {\it {An Automated Generation of Bootstrap Equations for Numerical Study
  of Critical Phenomena}},  \href{http://arxiv.org/abs/2006.04173}{{\tt
  arXiv:2006.04173}}.

\bibitem{Behan:2018}
C.~Behan, {\it Unitary subsector of generalized minimal models},  {\em Phys.
  Rev. D} {\bf 97} (May, 2018) 094020,
  [\href{http://arxiv.org/abs/1712.06622}{{\tt arXiv:1712.06622}}].

\bibitem{Fuente:2019}
A.~de~la Fuente, {\it Bootstrapping mixed correlators in the 2d ising model},
  \href{http://arxiv.org/abs/1904.09801}{{\tt arXiv:1904.09801}}.

\bibitem{Esterlis:2016}
I.~Esterlis, A.~L. Fitzpatrick, and D.~Ramirez, {\it Closure of the operator
  product expansion in the non-unitary bootstrap},  {\em JHEP} {\bf 11} (2016)
  030, [\href{http://arxiv.org/abs/1606.07458}{{\tt arXiv:1606.07458}}].

\bibitem{Karateev:2019}
D.~Karateev, P.~Kravchuk, M.~Serone, and A.~Vichi, {\it Fermion conformal
  bootstrap in 4d},  {\em JHEP} {\bf 06} (2019) 088,
  [\href{http://arxiv.org/abs/1902.05969}{{\tt arXiv:1902.05969}}].

\bibitem{Atanasov:2018}
A.~Atanasov, A.~Hillman, and D.~Poland, {\it {Bootstrapping the minimal 3D
  SCFT}},  {\em JHEP} {\bf 11} (2018) 140,
  [\href{http://arxiv.org/abs/1807.05702}{{\tt arXiv:1807.05702}}].

\bibitem{Nakayama:2016}
Y.~Nakayama, {\it {Bootstrapping Critical Ising Model on Three Dimensional Real
  Projective Space}},  {\em Phys. Rev. Lett.} {\bf 116} (Apr, 2016) 141602,
  [\href{http://arxiv.org/abs/1601.06851}{{\tt arXiv:1601.06851}}].

\bibitem{Cappelli:2019}
A.~Cappelli, L.~Maffi, and S.~Okuda, {\it {Critical Ising model in varying
  dimension by conformal bootstrap}},  {\em JHEP} {\bf 01} (2019) 161,
  [\href{http://arxiv.org/abs/1811.07751}{{\tt arXiv:1811.07751}}].

\bibitem{ElShowk:2012}
S.~El-Showk, M.~F. Paulos, D.~Poland, S.~Rychkov, D.~Simmons-Duffin, and
  A.~Vichi, {\it {Solving the 3D Ising model with the conformal bootstrap}},
  {\em Phys. Rev. D} {\bf 86} (Jul, 2012) 025022,
  [\href{http://arxiv.org/abs/1203.6064}{{\tt arXiv:1203.6064}}].

\bibitem{ElShowk:2014}
S.~El-Showk, M.~F. Paulos, D.~Poland, S.~Rychkov, D.~Simmons-Duffin, and
  A.~Vichi, {\it {Solving the 3d Ising Model with the Conformal Bootstrap II.
  c-Minimization and Precise Critical Exponents}},  {\em J. Stat. Phys.} {\bf
  157} (2014) 869--914, [\href{http://arxiv.org/abs/1403.4545}{{\tt
  arXiv:1403.4545}}].

\bibitem{Dymarsky:2018}
A.~Dymarsky, F.~Kos, P.~Kravchuk, D.~Poland, and D.~Simmons-Duffin, {\it The 3d
  stress-tensor bootstrap},  {\em JHEP} {\bf 02} (2018) 164,
  [\href{http://arxiv.org/abs/1708.05718}{{\tt arXiv:1708.05718}}].

\bibitem{SimmonsDuffin:2017}
D.~Simmons-Duffin, {\it {The lightcone bootstrap and the spectrum of the 3d
  Ising CFT}},  {\em JHEP} {\bf 03} (2017) 086,
  [\href{http://arxiv.org/abs/1612.08471}{{\tt arXiv:1612.08471}}].

\bibitem{Kos:2016}
F.~Kos, D.~Poland, D.~Simmons-Duffin, and A.~Vichi, {\it {Precision islands in
  the Ising and $O(N)$ models}},  {\em JHEP} {\bf 08} (2016) 036,
  [\href{http://arxiv.org/abs/1603.04436}{{\tt arXiv:1603.04436}}].

\bibitem{Caracciolo:2010}
F.~Caracciolo and S.~Rychkov, {\it Rigorous limits on the interaction strength
  in quantum field theory},  {\em Phys. Rev. D} {\bf 81} (Apr, 2010) 085037,
  [\href{http://arxiv.org/abs/0912.2726}{{\tt arXiv:0912.2726}}].

\bibitem{QFTinAdS}
M.~F. Paulos, J.~Penedones, J.~Toledo, B.~C. van Rees, and P.~Vieira, {\it {The
  S-matrix bootstrap. Part I: QFT in AdS}},  {\em JHEP} {\bf 11} (2017) 133,
  [\href{http://arxiv.org/abs/1607.06109}{{\tt arXiv:1607.06109}}].

\bibitem{Baggio:2017mas}
M.~Baggio, N.~Bobev, S.~M. Chester, E.~Lauria, and S.~S. Pufu, {\it {Decoding a
  Three-Dimensional Conformal Manifold}},  {\em JHEP} {\bf 02} (2018) 062,
  [\href{http://arxiv.org/abs/1712.02698}{{\tt arXiv:1712.02698}}].

\bibitem{Richards:2015}
A.~Richards, {\it {University of Oxford Advanced Research Computing}},  {\em
  Zenodo} (2015).

\bibitem{Ferrara:1973yt}
S.~Ferrara, A.~Grillo, and R.~Gatto, {\it {Tensor representations of conformal
  algebra and conformally covariant operator product expansion}},  {\em Annals
  Phys.} {\bf 76} (1973) 161--188.

\end{thebibliography}\endgroup
\bibliographystyle{JHEP}

\end{document}